\documentclass[preprint,12pt,authoryear]{elsarticle}

\usepackage{amssymb}
\usepackage[noend]{algpseudocode}
\usepackage{algorithm}
\usepackage{xcolor}
\usepackage{multirow}
\usepackage{subcaption}
\NewDocumentCommand{\rot}{O{45} O{1em} m}{\makebox[#2][l]{\rotatebox{#1}{#3}}}

\begin{document}

\begin{frontmatter}



\title{ Navigating the corporate disclosure gap: Modelling of \textit{Missing Not at Random} Carbon Data}


\author[inst1,inst2]{Malgorzata Paulina Olesiewicz}

\affiliation[inst1]{organization={ Humboldt Universit\"at zu Berlin},
            city={Berlin},
            country={Germany}}

\author[inst2]{ Jaakko Kooroshy }
\author[inst1]{ Sonja Greven}

\affiliation[inst2]{organization={London Stock Exchange Group (LSEG)},
            city={London},
            country={United Kingdom}}

\begin{abstract}
Corporate carbon emissions data is disclosed by approximately 65\% of large and mid-sized companies globally, despite being a key indicator of corporate climate performance. With investors increasingly looking to integrate climate risk into their investment strategies and risk reporting, this creates demand for robust prediction models that can generate reliable estimates for missing carbon disclosures. However, these estimates lack transparency and are frequently used in the investment decisions process with the same confidence as corporate reported data. As disclosures remain mostly voluntary and the propensity to disclose is shaped by several factors (e.g. size, sector, geography), missing emissions data should be assumed to be \textit{‘missing not at random’} (MNAR). However, widely used estimation methods (e.g. linear regression models) typically do not correct for MNAR bias and do not accurately reflect the uncertainty of estimated data. The objective of this paper is to address these issues:  (1) account for the uncertainty of the missing data and thus obtain regression coefficients by multiple imputation (MI) (2) correct for potential bias by using MI algorithms based on Heckman's sample selection model introduced by Galimard et al. (3) estimate missing carbon disclosures with  linear models based on MI and report on the uncertainty of predicted values, measured as the length of the prediction interval. In the simulation, our approach resulted in an accuracy gain based on root mean squared error of up to 30\%, and up to a 40\% higher coverage rate than the existing models. When applied to commercial data, the results suggested up to 20\% higher coverage for proposed methods. 
\end{abstract}


\begin{highlights}
\setlength{\emergencystretch}{3em}

\item Only around 65\% of companies report their carbon emissions and the undisclosed data is most likely missing not at random (MNAR) due to voluntary act of reporting. 

\item We use multiple imputation with Heckman's model approach to improve the prediction of corporate Scope 1 and Scope 2 emissions.

\item Multiple imputation with Heckman's model performs best among tested approaches if data is missing not at random.

\item We achieve accuracy gains (Root mean squared error (RMSE)) of up to 30\% and up to 40\% higher coverage (for 95\% confidence intervals) relative to existing models (median, linear regression) if data is missing not at random.
\end{highlights}

\end{frontmatter}


\setlength{\emergencystretch}{3em}

Sustainable investment directs investment capital to companies that manage climate-related risks and opportunities effectively by including non-financial indicators (e.g., carbon emissions) in the investment decision process \citep{schoenmaker2018principles}. Mounting pressure from asset owners, governments and society more widely has contributed to the rapid proliferation of sustainable investment strategies, with a record \$155bn raised globally in the third quarter of 2020 \citep{Refinitive2020}. This has created a growing demand for high-quality sustainability data to underpin these strategies. However, non-financial reporting is not yet standardized and remains largely voluntary, resulting in incomplete, "noisy" datasets (see e.g \citeauthor{li2020difference}, \citeyear{li2020difference}). 
\vspace{2.5mm}

Therefore, for investors to meet their sustainability and risk management objectives, it is important that accurate data is available for all companies in the investible universe. However, corporate carbon emissions, which is a crucial climate-related indicator, is only disclosed by approximately 65\%  of listed companies\footnote{Based on All Word Index 2018.  Non-financial reporting is most advanced for listed equities and similar issues are present across other asset classes.}. To fill the gap between the available and required data, it is common practice to replace missing emissions values with estimates. The methods used to predict such missing emissions vary across data providers, but tend to fall under one of three categories:
\begin{enumerate}
    \item sector median or median carbon intensity estimation, where missing values are predicted based on the reported emissions within the respective peer group \citep{busch2018consistency}. 
    \item regression analysis, where missing values are predicted using various linear models with predictors based on the financial, energy consumption, and historical emissions data  \citep{kalesnik2020green} (the corresponding regression coefficients are typically obtained using complete case analysis (CCA) or single imputation (SI) techniques), and
    \item input-output models, where the sectors' total direct emissions and energy consumption are used to derive their carbon intensities, reflecting the emissions associated with one dollar of gross output \citep{liu2017value}.
\end{enumerate}

\vspace{2.5mm}

Data providers typically do not disclose the details of their models' assumptions and corresponding uncertainty, which can lead to a misplaced perception that estimated and reported emissions are equally accurate \citep{kalesnik2020green}. In particular, quantitative investment strategies - such as investment against specific climate-focused indices - typically treat estimated and disclosed data in the same way. This is concerning, given some of the discrepancies in datasets across different data providers that have been identified in several studies (e.g. \citeauthor{busch2018consistency}, \citeyear{busch2018consistency} and \citeauthor{berg2019aggregate}, \citeyear{berg2019aggregate}). Consequently, without a good understanding of the estimation process and associated uncertainty, investors may struggle to make fully informed decisions when choosing a data provider, and to select the one whose estimation model is most suitable for their purposes \citep{busch2018consistency}.

\vspace{2.5mm}

Another concern with the current emissions estimation models is that they are often built based on only the observed data within the sample, i.e., without accounting for the uncertainty of missing values.  This matter has been highlighted in a recent publication by \cite{hoepner2021emissions}, in which the authors recommended that emissions estimation models embed a precautionary principle to discourage the lack of disclosure. In practice, not accounting for missing data during a model estimation may lead to biased and inefficient estimates of model parameters such as regression coefficients and an underestimation of standard errors, resulting in incorrect confidence intervals \citep{white2011multiple}. In this paper, we focus on linear regression, which can perform well as a prediction tool if obtained model parameters are unbiased, and efficient (\citeauthor{greene2003econometric}, \citeyear{greene2003econometric}, Section 6.6). Therefore, if missing data are not handled properly, the resulting linear prediction model may produce inaccurate estimates.

\vspace{2.5mm}

Incomplete datasets are the norm rather than the exception in social, behavioural and medical sciences. To address aforementioned issues related to missing data, imputation techniques, which replace missing observations with plausible values, have been the subject of extensive research in the statistical literature since the 1970s (see e.g \citeauthor{rubin1976inference}, \citeyear{rubin1976inference} and \citeauthor{gleason1975proposal}, \citeyear{gleason1975proposal}). However, imputation may lead to a situation where an imputed dataset is treated as if it was complete, which makes imputation "seductive and dangerous" (\citeauthor{madow1983incomplete}, \citeyear{madow1983incomplete}, p.8), since the inherent uncertainty of the imputation process may be overlooked. Therefore, multiple imputation (MI), where several imputed datasets are created, should be the preferred technique as it accounts for the uncertainty of the missing values. 

\vspace{2.5mm}

According to Rubin's framework, missing data can be classified into three categories: 

\begin{enumerate}
\item missing completely at random (MCAR), 
\item missing at random (MAR), where the probability of the data being missing depends on the observed data, and 
\item missing not at random (MNAR), where the probability of the data being missing depends on both the observed and the unobserved data. 

\end{enumerate}

Carbon emissions data cannot be classified as MCAR, since there is extensive research on the (observed) drivers that compel companies to disclose their emissions. This research suggests that the biggest motivator for disclosure tends to be regulatory pressure (for detailed discussion see \citeauthor{prado2009stakeholder}, \citeyear{prado2009stakeholder}, and \citeauthor{venturelli2019state}, \citeyear{venturelli2019state}). Several countries such as Brazil  have succeeded in increasing disclosure rates by providing a wide range of incentives – either through legislation or changes in the regulatory regime \citep{araya2006exploring}. The positive impact on disclosure has also been credited to stakeholder activism, especially effective if initiated by investors \citep{flammer2019shareholder}, the industry's environmental sensitivity and companies' international operations \citep{araya2006exploring}. Moreover, highly carbon-intensive industries tend to be more regulated and publicly pressured to report their data, leading to higher disclosure rates within those industries \citep{noronha2013corporate}. 

\vspace{2.5mm}

 However, the exact determinants of corporate carbon disclosures within each sector are difficult to identify, and therefore the majority of related studies considers the carbon emission data to be MNAR (e.g. \citeauthor{allen2017}, \citeyear{allen2017} and \citeauthor{saka2014disclosure}, \citeyear{saka2014disclosure}). The voluntary decision to disclose is a form of self-selection, which could lead to bias, since companies with a relatively high level of emissions might be reluctant to disclose this information (\citeauthor{matsumura2014firm}, \citeyear{matsumura2014firm} and \citeauthor{clarkson2008revisiting}, \citeyear{clarkson2008revisiting}). Companies are also less likely to report their emissions if the cost of measuring and collecting this information exceeds the potential benefit (often the case for low carbon-intensive industries or smaller out-of-public-eye firms) \citep{matsumura2014firm}. 

\vspace{2.5mm}

When dealing with MNAR data, it is crucial to account for potential bias during the prediction process \citep{heckman1976common}. However, the current methods used by data providers when estimating missing emissions data typically do not systematically correct for this and the predicted data risk being inaccurate. Moreover, most commercially available datasets rely on CCA or SI techniques to estimate prediction model parameters, which do not account for the uncertainty caused by the data incompleteness. This results in prediction intervals that do not adequately reflect estimates’ uncertainty. 

\vspace{2.5mm}

To address these challenges, the objective of this paper is to predict the missing carbon emissions values with a regression model using parameters obtained with multiple imputation (MI) techniques. To account for the potential non-randomness of the observed sample, we use MI algorithms based on Heckman's sample selection model introduced in \cite{galimard2016multiple} and extended in \cite{galimard2018heckman}. Finally, to reflect the uncertainty of our estimation, we report the length of the prediction interval for each predicted value.

 \vspace{2.5mm}

The remainder of this paper is organized as follows. Section 2 presents the theoretical foundations related to Heckman's model and multiple imputation. The methods proposed by \citeauthor{galimard2016multiple} are discussed in detail. Section 3 introduces the 2018 carbon emissions data (n = 6547) used for this study, which was curated by the London Stock Exchange Group for use in investment solutions (LSEG data).  In section 4, we describe the design of a simulation study and evaluation methods that were applied. To further assess the accuracy of our prediction models, we apply the proposed methods to LSEG data, and, for companies which started reporting their emissions data the following year (n = 423 for Scope 1 and n = 447 for Scope 2), we compare the estimated values with the reported 2019 emissions. In section 5, we present results for the simulation and the empirical study.  In the last section, we discuss our findings.

\section{Methods}

We introduce the applied methods by first giving an overview of Heckman's model, also known as the sample selection method, which corrects for the bias in not randomly selected samples \citep{heckman1976common}. We then discuss the basic multiple imputation algorithm and its extensions proposed by \citeauthor{galimard2016multiple} (\citeyear{galimard2016multiple}, \citeyear{galimard2018heckman}). Finally, the \textit{predict then combine} (PC) procedure \citep{miles2016obtaining} is presented, which we later use to obtain the final estimates of the missing values and their standard errors. 

\subsection{Heckman's model}

Let $Y_{i}, i = 1, \dots, n,$ be an incomplete set of continuous outcome variables and $R'_{i}$ the associated selection latent variable, where the outcome is missing ($R_{ i} = 0$) if $R'_{i} < 0 $ and observed ($R_{i} = 1$) otherwise. Thus, vector $Y$ can be divided into two parts: $Y_{miss}$ ($n_{0}\times1$) containing all $Y_{i}$ for which $R_{i} = 0$ and $Y_{obs}$ ($n_{1}\times1$) with all the values of  $Y_{i}$ where $R_{i} = 1$. The joint outcome and selection model is defined as
  \vspace{2.5mm}
  
 \begin{equation}
\begin{array}{l}
R_{i}^{\prime}=X_{i}^{s} \beta^{s}+\varepsilon_{i}^{s} \\
Y_{i}=X_{i} \beta+\varepsilon_{i}
\end{array}
\begin{array}{c}
\mbox{with} 
\end{array}
\begin{array}{r}
\left(\begin{array}{l}
\varepsilon^{s} \\
\varepsilon
\end{array}\right) \sim N\left(\left(\begin{array}{c}
0 \\
0
\end{array}\right),\left(\begin{array}{cc}
1 & \rho \sigma_{\varepsilon} \\
\rho \sigma_{\varepsilon} & \sigma^{2}_{\varepsilon}
\end{array}\right)\right),
\end{array}
\label{eq:joint_model}
 \end{equation}
 \vspace{2.5mm}
 
 where  $X_{i}^{s}$ is a vector ($1 \times q$)  of covariates associated with the missingness, $X_{i}$ is a vector ($1 \times p$)   of outcome covariates, $\beta^{s}$ is a ($q\times 1$) vector of selection coefficients, $\beta$ is a ($p\times 1$) vector of outcome coefficients, $\rho$ is  thecorrelation between the error terms, $\varepsilon_{i}^{s}$ is the error term of the selection model, $\sigma^{2}_{\varepsilon}$ is its variance  and $\varepsilon_{i}$ is the error of the outcome model. The variance of the selection latent variable $\sigma^{2}_{\varepsilon^{s}}$ is set to 1 since only the sign of $R'_{i}$ is observed (\citeauthor{cameron_trivedi_2005}, \citeyear{cameron_trivedi_2005}, Section 16.5). The model requires $X_{i} \neq X_{i}^{s}$, which is known as exclusion restriction (see Section(\ref{sec:H2Step})). By the properties of the bivariate normal distribution
 
 \begin{equation}
     E\left[R'_{i} \mid Y_{i}\right]=X_{i}^{s} \beta^{s}+\rho \frac{\sigma_{\varepsilon^{s}}}{\sigma_{\varepsilon}}(Y_{i}-X_{i} \beta),
 \end{equation}
 
which implies that if the correlation $\rho$ is not 0, the missingness depends on the outcome and the observed data (MNAR mechanism). If $\rho = 0$, the missingness depends on the observed data only and the corresponding mechanism is MAR. As $\rho$ further deviates from 0, bias becomes stronger \citep{galimard2016multiple}. The probability of a value being observed is estimated using a probit model and is given by 
\begin{equation}
\begin{array}{cc}
        P\left(R_{i}=1 \mid X_{i}^{s}\right)=\Phi\left(X_{i}^{s} \beta^{s}\right), \\
\end{array}
    \label{eq:selection}
\end{equation}
  
where $\Phi$ is the standard normal cumulative distribution function and $\phi$ its density. Consequently, Heckman derives the conditional mean and the variance of the outcome variable as  

  \begin{equation}
  \setlength{\emergencystretch}{3em}
     E\left(Y_{i} \mid X_{i}, X_{i}^{s}, R_{y i}=1\right)=X_{i} \beta+\rho \sigma_{\varepsilon} \lambda_{i}  
     \label{eq:conditionalmeant}
  \end{equation}
  \begin{equation}
  \setlength{\emergencystretch}{3em}
      \begin{array}{cc}
         Var\left(Y_{i} \mid X_{i}, X_{i}^{s}, R_{y i}=1\right)=\sigma_{\varepsilon}^{2}\left(1-\rho^{2} \delta_{i}\right) \quad \mbox{with}  \ \delta_{i}=\lambda_{i}\left(\lambda_{i}+X_{i}^{s} \beta^{s}\right)
         \label{eq:varianceHeckman}
      \end{array}
  \end{equation}
\vspace{2.5mm}

  where $\lambda_{i}$ is defined as inverse Mills ratio (IMR) with
\begin{equation}
   \lambda_{i}=\frac{\phi\left(X_{i}^{s} \beta^{s}\right)}{\Phi\left(X_{i}^{s} \beta^{s}\right)} \quad .
   \label{eq:imr} 
\end{equation} 

\vspace{2.5 mm}
 It is important to note that the $Var (Y_{i})$ in Eq.(\ref{eq:varianceHeckman}) is homoskedastic, which needs to be corrected for during the variance estimation process. 
 To obtain the model parameters, Heckman proposed two-step estimation and maximum-likelihood (ML) methods which are presented in the sections below.

 \subsubsection{Two-step (2-Step) Estimation} \label{sec:H2Step}
 
The process of obtaining the two-step estimator is as follows: 
 \begin{enumerate}
     \item From the probit selection Equation (\ref{eq:selection}) estimate $\widehat{\beta}^{s}$ using ML. For each observation obtain IMR ($\hat{\lambda_{i}}$) as defined in Equation (\ref{eq:imr}).
     
     \item Estimate $\hat{\beta}$ and $\hat{\beta}_{\lambda}$ from the following linear equation:
\begin{equation}
    Y_{i}=X_{i} \beta+\hat{\lambda}_{i} \beta_{\lambda}+\eta_{i}, \eta_{i} \sim N\left(0, \sigma_{\eta}^{2}\right) \label{eq:secondstep}
\end{equation}
 \end{enumerate}
 
where $\beta_{\lambda} = \rho \sigma_{\varepsilon}$ is the coefficient of Mills ratio (i.e, 'correction term' associated with probability of observing the value) and $\eta_{i}$ is the error term  with variance $\sigma_{\eta}^{2} = \sigma_{\varepsilon}^{2}\left(1-\rho^{2} \delta_{i}\right)$. 

\vspace{2.5mm}

 Since $\hat{\lambda_{i}}$ is obtained using covariates $X_{i}^{s}$, to avoid collinearity issues in the second step, the model requires exclusion restrictions where a different set of covariates is used in selection and outcome equation (\citeauthor{cameron_trivedi_2005}, \citeyear{cameron_trivedi_2005}, Section 16.5). In practice, it is often suggested to use at least one more variable ($q=p+1$) in the selection equation, which has no significant relation with the outcome \citep{galimard2016multiple}.  However, finding such a variable in the real world has been shown to be challenging \citep{liu2020comparison}. For further details see \cite{cameron_trivedi_2005}.

\subsubsection{Maximum Likelihood (ML)}\label{sec:ml}

Since Heckman's model is fully parametric, it is possible to derive ML estimates of the joint model parameters $\theta = (\beta, \beta^{s}$, $\rho$) from the bivariate normal distribution. The bivariate selection model has the likelihood function
\begin{equation}
   L =\prod_{i=1}^{n}\left\{Pr \left[R'_{i} \leq 0\right]\right\}^{1-R_{i}}\left\{f\left(Y_{i} \mid R'_{i}>0\right) \times Pr\left[R'_{i}>0\right]\right\}^{R_{i}}
\end{equation}

\vspace{2.5mm}

where the first component corresponds to the observations with the unobserved outcome variable ($Y_{miss}$) and the second to the complete observations ($Y_{obs}$) \citep{cameron_trivedi_2005}. 
The ML is more efficient than the two-step estimator but computationally more expensive. Today, computational cost is not an issue but the two-step estimator is considered more robust, since the ML may not converge at all or converge to a local rather than a global maximum \citep{henningsen2019package}.

\subsection{Multiple Imputation}

 \vspace{3mm}

Multiple imputation techniques are motivated by Bayes' theory and their standard implementation provides unbiased inference if the missing data are MAR \citep{liu2015methods}. Let $\theta$ be a parameter of interest, which could be estimated if the entire population observed. To obtain an unbiased and efficient parameter of interest $\hat{\theta}$, despite the incomplete sample, the MI  approximates the complete-data posterior mean with an average, using a small number of imputed data sets.

\vspace{2.5mm}

Consequently, any multiple imputation algorithm consists of three steps: \begin{enumerate}
    \item generation of $m$ multiply-imputed data sets,
    \item estimation of the parameters of interest ($\theta$) for each data set separately and
    \item calculation of the aggregated estimates.
\end{enumerate}
The final estimate of the parameter of interest  is an approximation of the mean in Eq.(\ref{eq:conditionalmeant}) with an average as 

\begin{equation}
    \widehat{\theta}=\frac{1}{m} \sum_{j=1}^{m} \widehat{\theta}_{j}
    \label{eq:finalest}
\end{equation}

where $m$ is the total number of the imputed data sets and $\hat{\theta}_{j}$ is the estimate obtained from the $j^{th}$ imputation. Using Rubin's rules one can approximate the total variance of $\widehat{\theta}$ such that:
\begin{equation}
    Var(\hat{\theta})=W+\left(1+\frac{1}{m}\right) B
    \label{eq:finalvar}
\end{equation}

where $W$ is the within-imputation variance $W=(1 / m) \sum_{j=1}^{m} Var(\hat{\theta_{j}})$ and $B$ is the between-imputation variance $B=(1 /(m-1)) \sum_{j=1}^{m}\left(\hat{\theta}_{j}-\hat{\theta}\right)^{2}$. The total variance accounts for three sources of uncertainty: the variance caused by observing the sample rather than the entire population ($W$), the additional variance caused by the missing values ($B$) and the extra simulation variance ($B/m$) caused by the fact that $\hat{\theta}$ is estimated for a finite $m$ \citep{van2018flexible}. The single imputation techniques do not capture the between-imputation part of the variance and therefore often result in too small standard errors \citep{white2011multiple}. It is important to note that for the model used in the first step of the MI algorithm to result in proper imputation, it is required that both the imputed values and imputation model's parameters are drawn from a distribution. Consequently, each set of imputed data will be based on a different set of model parameters, which are drawn at each iteration from their respective Bayesian posterior \citep{carpenter2007missing}. The imputed data should be then drawn from the Bayesian posterior distribution with uninformative priors for the model parameters \citep{white2011multiple}.

\subsubsection{Multiple Imputation using Heckman's Model}

Selection of an imputation model strongly depends on the type of data at hand.  If the data is MNAR, the standard MI techniques will not be satisfactory, as the obtained estimates tend to be biased. To address this issue, in their first paper \cite{galimard2016multiple} proposed a MI algorithm using Heckman's two-step estimation. Algorithm is build out of three main steps: $\beta^{s}$ and $\lambda_{i}$ estimation, $\beta$, $\beta_{\lambda}$ estimation and imputation step. In the first step, the algorithm follows the 2-Step estimation procedure to obtain estimates of $\beta^{s}$ and $\lambda_{i}$. As shown in Eq.(\ref{eq:varianceHeckman}), the variance of the outcome equation in the second step is not homeostatic and the algorithm corrects for this using a Cholesky decomposition (for details see \citeauthor{galimard2016multiple} \citeyear{galimard2016multiple}, Annex A). In the second step, to obtain model parameters ($\beta$, $\beta_{\lambda}$) for each imputation model the algorithm follows steps of the initial MI algorithm under the normal linear model proposed by Rubin (see \citeauthor{rubin2004multiple} \citeyear{rubin2004multiple}, p.167). Thirdly, the imputation is then performed as specified in the second Step of the Eq.(\ref{eq:secondstep}) with the parameters obtained in the previous steps.

\vspace{2.5mm}

As stated earlier, the advantage of the two-step approach is its computational efficiency. However, this method is often susceptible to collinearity, and additional exclusion restrictions need to be implemented \citep{ogundimu2019robust}. Moreover, the uncertainty of the first step $\hat{\beta^{s}}$ is not accounted for in the second step and may lead to biased estimates \citep{galimard2018heckman}. Consequently, in their second paper \cite{galimard2018heckman} followed up with an algorithm for  MI based on  maximum likelihood estimation. 
In the first step of the algorithm, Heckman's model parameters $\theta = (\beta, \beta^{s},\sigma_{\varepsilon}, \rho)$ are estimated using ML. In the second step, the parameters for each imputation model are drawn from normal distribution. in the third step, the imputation is performed,following this same logic as in 2-Step estimation. 

\vspace{2.5mm}

In our study, we apply both of the proposed algorithms since each offers a different set of advantages. To compare their performance with other MI  methods, we also implemented  MI using predictive mean matching \citep{van2018flexible} and random forest \citep{doove2014recursive}, which implicitly assume a MAR mechanism.  Both of these methods are \textit{hot deck} methods, meaning that the imputed values are limited only to the already observed data. 

\subsection{Final Prediction}\label{sec:fp} 

The performance of an imputation technique should not be evaluated based on the proximity of the imputed and true values since the imputation of missing data is not the same as their prediction. The aim of the imputation is to account for the uncertainty resulting from the incomplete data,  whereas prediction tries to minimise the distance between the true and predicted values \citep{van2018flexible}. Therefore, in practice, the final estimates rather than imputed missing values are used to replace the missing observations. The final estimates can be easily obtained from the models fitted to multiply-imputed data \citep{miles2016obtaining}. For each missing value $Y_{i}$, the final predicted value is estimated as the average of the predicted values $\hat{Y}_{ij}$ across the $m$ imputed data sets with 
    
    \begin{equation}
      \hat{Y_{i}} =  \frac{1}{m} \sum_{j=1}^{m} \hat{Y}_{ij} = \frac{1}{m} \sum_{j=1}^{m} X_{i}\hat{\beta_{j}} = X_{i}\hat{\beta}
    \end{equation}

where $\hat{\beta_{j}}$ is the vector of coefficients obtain for the data set $j$, $j = 1,\dots,m$. In the case of multiple imputation, the standard error of prediction should account for both; the within and between-variance of the prediction. To combine the estimation of the standard errors from multiply imputed data \cite{miles2016obtaining} proposed the \textit{predict then combine} (PC) procedure which follows Rubin's rule. The total variance of the prediction can be calculated
as:
\begin{equation}
    V_{\hat{Y_{i}}}=V_{W_{i}}+V_{B_{i}}\left(1+\frac{1}{m}\right)
\end{equation}

with the within ($V_{W_{i}}$) and the between-variance ($V_{B_{i}}$) given as

    \begin{equation}
    V_{W_{i}}=\frac{1}{m} \sum_{j=1}^{m} \widehat{se}_{ij}^{2}
    \label{eq:within}
    \end{equation}
    
    \begin{equation}
        V_{B_{i}}=\frac{1}{m-1} \sum_{j=1}^{m}\left(\hat{Y}_{ij}-\hat{Y_{i}}\right)^{2}
    \end{equation}

The estimated standard error of the prediction $\hat{se}_{ij}$ from imputed data set $j$ in Eq.(\ref{eq:within}) is calculated as in the standard linear model:
    
    \begin{equation}
    se_{ij}^{2} = \sqrt{\sigma_{\varepsilon}^2 + \left( X_{i}[\sigma_{\varepsilon}^2(\mathbf{X}_{obs}\mathbf{X}^{\top}_{obs})^{-1}]X^{\top}_{i}\right).}
    \end{equation}
where $\mathbf{X}_{obs}$ is a ($n_{1}\times p$) matrix of outcome covariates for the observed part of the sample.

\vspace{2.5mm}

Finally, the prediction interval (PI) is obtained using a t-distribution with 
    
    \begin{equation}
      PI= \hat{Y_{i}} \pm t_{(\lambda / 2, df)} \times \sqrt{V_{\hat{Y_{i}}}}
        \label{eq:PI}
    \end{equation}

\vspace{2.5mm}

where  $t_{\lambda / 2}$  is the $(1- \frac{\lambda}{2})$-quantile of t distribution with $df$ degrees of freedom.
\vspace{2.5mm}

\section{Data}

\subsection{Data Overview}

 For this study, we use the 2018 data set of 6547 companies, which was curated by the London Stock Exchange Group (LSEG) for use in investment solutions. We focus on $Scope\ 1$ and $Scope \ 2 $ emissions, which refer to emissions resulting directly from company activity and  emissions resulting from its energy consumption, respectively. Companies with reported emissions value constituted 39\% of the sample \footnote{The sample includes additional micro companies, which are not included in All World Index}.  For each of the companies, the following covariates are available:  $Region$, $Size$, $Industry$, $SuperSector$, $Sector$, $SubSector$, $RevenueUSD$ and $FirstActivity$. All the covariates are categorical except $RevenueUSD$ (for details see Annex Table \ref{tab:data}). Prior to the analysis, we checked whether the common support criterium is satisfied, meaning that for each category within categorical variables, observations with reported and missing emission values were present. The observations without common support have been excluded from the analysis (less than 2\% of the total sample).
 
 \vspace{2.5mm}
 
 The probability of the emissions being observed varies significantly depending on the $Size$ of the company. Therefore, to investigate the impact of the level of missingness in the sample on the precision of the prediction, we also analysed data for each size category separately. The summary of the analysed data sets can be found in Table (\ref{tab:breakdown}) where the size of the company is denoted by $L$ for large, $M$ for medium, $S$ for small and $U$ for micro.

\begin{table}[h!]
\caption{Overview of the data and in particular reported and missing values according to the $Size$ of the company. }
\centering
\begin{tabular}{|c|c|c|c|c|c|c|c|}
\hline
 &  & \multicolumn{3}{c|}{Scope 1} &  \multicolumn{3}{c|}{Scope 2}\\
 \cline{3-8}
 Size  &	$n$ & $n_{1}$ & $n_{0}$ & \% reported & $n_{1}$ & $n_{0}$ & \% reported \\
 \hline
 L &	1343 &	950 &	393	& 71\% & 948 & 395 & 71\% \\
 M &	1694 &	927	& 767 &	55\% & 923 & 771 & 55\% \\
S  &	2085 &	508 & 1577 & 24\% & 504 & 1581 & 24\% \\
U  &	1425 &	169	& 1256 & 12\% & 167 & 1258 & 12\% \\
\hline
 Total & 6547 & 2554 & 3993 & 39\% & 2542 & 4005 & 39\% \\

   \hline
\end{tabular}
\vspace{1.5mm}
\label{tab:breakdown}
\end{table}

\subsection{Variable Selection Process} \label{sec:varselection}

\vspace{2.5mm}

For each analysed data set, we considered all the available covariates as predictor variables in the outcome and selection equation. For this purpose, we used the stepwise regression (stepwise selection) approach, where predictors are iteratively added and/or removed from the prediction model. The process is terminated when further improvements in the model fit are no longer statistically significant \citep{efron2016computer}. In this study, we applied both-directions stepwise selection, which is a hybrid of forward and backward selection. 

\vspace{2.5mm}

The selection was performed with the \texttt{stepAIC} function from the \texttt{MASS} package, which uses the AIC as the model performance evaluation criterion. 
For each analysed dataset, we performed variable selection separately for the outcome and selection equation (emissions model and disclosure probability model). For the outcome equation, we conducted a variable selection based on a linear model. For the selection equation, we fit  a probit model to the data at every step of the selection. The process has been repeated for both Scope 1 and Scope 2 separately. 

\vspace{2.5mm}

The stepwise selection based on the AIC criterion does not guarantee that all the selected variables will be statistically significant. Therefore, the authors of the \texttt{stepAIC} function, \citeauthor{venables2013modern} (\citeyear{venables2013modern}, section 6.8), recommend manual removal of insignificant variables. Selected in this ways outcome and selection variables for all datasets met the exclusion restriction criterion required for the 2-Step Heckman model. 

 \section{Simulation Study}

The objective of this simulation study is to evaluate the performance of the selected and alternative imputation techniques in various scenarios for the MAR and MNAR mechanisms. In each scenario, $N$ independent data sets were generated based on the LSEG data. 

\subsection{Data-generating process}

\vspace{2.5mm}

In the LSEG dataset, covariates are complete and were used in the simulation without any modifications.  The jointly distributed error terms, $\varepsilon$, $\varepsilon^{s}$ were generated from the bivariate normal distribution specified in Eq.({\ref{eq:joint_model}})  with $\rho$ equal to 0, -0.3 and -0.6, corresponding to a MAR, Light MNAR and Heavy MNAR mechanism respectively. We assume a negative correlation between the selection and outcome equations to reflect the empirical research suggesting that companies with a higher level of emissions are less likely to report their data. To evaluate the performance of the methods when the missingness mechanism is different from specified in Heckman's model, we use the approach proposed by \cite{ganjali2008comparison}, which uses a Bernoulli distribution for $R_{i}$ with a function of $Y_{i}$ as its parameter $P(R_{i} = 1)$. This fourth scenario is named Non-Heckman in our notation. In the Results section and Discussion, we collectively refer to the Light MNAR, Heavy MN AR and Non-Heckman scenario as MNAR scenarios.

 \vspace{2.5mm}
 
 For each subset and the $Total$ dataset, we fitted  Heckman's model (ML estimation) and obtained vectors of estimates $\hat{\beta}$ and $\hat{\beta^{s}}$. The estimated $\hat{\sigma}_{\varepsilon}$ was between 1.44 and 1.64 for all the datasets. For each missingness mechanism, we generated data under three scenarios of $\sigma_{\varepsilon}^{2}$: 1, 2.5 and 4. The $\hat{\beta}^{s}$ were used directly from the ML Heckman's estimation to keep the same missingness level and distribution within each sample,  and the binary missingness variable $R_{i}$ was estimated accordingly with $R_{i} = 1$ if $P(R_{i}= 1) >= 0.5$  . For $\hat{\beta}$, we  drew  a vector from $\beta \sim N(\hat{\beta}, \hat{se}(\hat{\beta})^{2})$ and defined it as $\beta^{*}$. 

\vspace{2.5 mm}

The outcome variable $Y_{i}$ was generated according to the outcome equation $Y_{i}=X_{i} \beta^{*}+\varepsilon_{i}$. The missingness indicator $R_{i}$ was generated according to the following selection equation: $ X_{i}^{s} \hat{\beta}^{s} + \varepsilon_{i}^{s} \ge 0 $, then $R_{i} = 1$; $R_{i} = 0$ otherwise. The predictors in $X_{i}$ and $X^{s}_{i}$ were selected as specified in Section(\ref{sec:varselection}) and Annex in Table(\ref{tab:methods}), except for the $Total$ dataset where SubSector (113 categories) was replaced by the Sector (40 categories) variable for computational efficiency of the simulation (in the empirical study SubSector is used). For each model, we also added an intercept.  For the \textit{Non-Heckman} scenario, the missingness indicator $R_{i}$ was generated using Bernoulli distribution with parameter $P(R_{i}= 1) = \Phi(c + 0.45 * \Delta Y_{i})$ for $i = 1,\ldots,n$ where $\Delta Y_{i} = \bar{Y}_{sector(Y_{i})} - Y_{i}$. Here, $\bar{Y}_{sector(Y_{i})}$ is the average emission for the Sector to which observation $Y_{i}$ belongs. To obtain a similar level of missingness as in the original data $c$ for $Total$ dataset  is equal to $-0.5$, $0.62$ for $L$, $0.15$ for $M$, $-0.9$ for $S$ and $-1.6$ for $U$. Generating the probability of data being observed as a function of $\Delta Y_{i}$ allowed us to ensure that within each Sector observations with emissions above the average level are less likely to be observed, following the same hypothesis as the other implemented MNAR mechanisms.    

\vspace{2.5mm}

 A total of N = 500 datasets were generated under each missingness scenario and value of $\sigma_{\varepsilon}^{2}$. The entire data-generating process was repeated separately for Scope 1 and Scope 2.

\subsection{Computational Setting}

The analysis was performed with the R statistical software, version 4.0.3, including the following packages: \texttt{mice} \citep{van2015package}, \texttt{sample- Selection} \citep{toomet2008sample} and \texttt{miceMNAR} \citep{Galimard2018package}. 

\subsection{ Analysis Methods} \label{sec:methods}

 The complete simulated data were first saved, before the deletion of some values, to later serve as a benchmark. The incomplete data were then analysed with the following methods: 

\begin{enumerate}
\setlength{\emergencystretch}{5em}
    \item Single Imputation (SI): 
    \begin{itemize}
        \item  Linear model (LM) consisting of imputation based on the linear model prediction, with the least squares estimates calculated from the observed data. Imputations were obtained with the \texttt{norm.predict} function from the \texttt{mice} package.
        \item  Heckman's model with maximum likelihood estimation (Hml) as described in \citep{galimard2016multiple}. Imputations were obtained with the \texttt{hecknorm} function from the \texttt{miceMNAR} package.
    \end{itemize}
    \item Multiple Imputation (MI) with imputation performed five times for each missing value: 
    \begin{itemize}
    \setlength{\emergencystretch}{6em}
        \item  MI using Predictive Mean Matching (MIPmm) consisted of imputations from  a small list of candidates, which are chosen based on their proximity to the \textit{predictive mean} (for details see \citeauthor{van2018flexible}, \citeyear{van2018flexible}). Imputations were obtained with the \texttt{pmm} function from the \texttt{mice} package.
        \item MI using Random Forest (MIRF) consists of imputation, where the $predictive \ mean$ is now calculated by a tree model instead of a regression model (for details see \citeauthor{doove2014recursive}, \citeyear{doove2014recursive}). Imputations were obtained with the \texttt{rf} function from the \texttt{mice} package with the default number of 10 trees in each random forest.
        \item MI using Heckman's model with maximum likelihood estimation (MIHml) based on\citep{galimard2018heckman}. Imputations were obtained with the \texttt{hecknorm} function from the \texttt{miceMNAR} package.
        \item MI using Heckman's two-step model estimation (MIH2step) based on \citep{galimard2016multiple}. Imputations were obtained with the \\ \texttt{hecknorm2step} function from the \texttt{miceMNAR} package.
        
    \end{itemize}
\end{enumerate}

The missing values were then predicted using linear models with coefficients obtained from an analysis of the imputed datasets. For the MI methods, the final coefficient estimates were obtained using Rubin's rule. In the Annex, Table (\ref{tab:methods}) summarises the predictors used for each of the models, which were selected according to the procedure described in Section(\ref{sec:varselection}). The final estimates of the missing values based on the MI methods were obtained with the $PC$ procedure described in the Methods section. As stated earlier, sector median estimation (Median)  and LM are currently widely used approaches for the estimation of missing emissions.  Therefore,  we additionally implemented median estimation, where the missing values were estimated with the median of the observed emissions within the (Sub-) Sector peer group. Based on the coefficients obtained from the imputation analysis, the same set of predictors were used for the imputation for the linear prediction models.  We then evaluated the performance of each method.

\vspace{2.5mm}

First, we evaluated the above methods in terms of accuracy of the estimated parameters $\theta$ (coefficients). For this purpose, we computed the empirical mean of the parameter estimates ($\bar{\theta}$), relative bias (Rbias \%),  the root mean square of estimated (model) standard errors ($SE_{m}$), empirical standard error ($SE_{e}$), coverage rate (CR \%) and the root mean square error (RMSE). To calculate these metrics, we use formulas provided in  \cite{morris2019using}. The coverage rate was estimated for a $\alpha =0.05$ confidence level, which implies that, according to the property of randomisation validity, exactly $95\%$ of intervals should cover the true $\theta$ \citep{morris2019using}. The root mean of the squared model standard errors is often also called "average model SE" and should be close to $SE_{e}$ since  $\mathrm{E}[SE_{m}^{2}]=SE_{e}^{2}$. If the $SE_{m}$ is systematically underestimated, this suggests a bias, which may lead to undercoverage, meaning that the true value of the parameter is not significantly often covered by the confidence interval. In the case when $SE_{m} < SE_{e}$, \cite{morris2019using} recommends using the relative error in the average model SE ($RE_{SE_{m}}$) as an informative performance measure.  

\vspace{2.5mm}

The accuracy of the models’ predictions in each data-generating setting was evaluated with the coverage rate of the nominal $95 \%$ confidence intervals (CR\%), the average length of the prediction interval (PI), the mean relative error (RE), and the root mean square error (RMSE). The prediction intervals for each missing variable was calculated according to Eq.(\ref{eq:PI}). Since RE is a measure relative to the true value, it reports much higher values for the observations with very small true values, i.e. RE is sensitive to small true values. In contrast, RMSE penalises heavily high absolute variation between the estimate and its true values, regardless of the true values. Therefore, it is often used to report the overall error of the prediction model.

\section{Results}

In this section, we present the results of the simulation study for the  Scope 1 $Total$ dataset, including all data-generating scenarios. To test our approach beyond the simulation, we report results obtained from the application of the proposed methods to the original 2018 LSEG data. The results for the data sub-sets (Scope 1), based on company size, and Scope 2 $Total$ dataset can be found in the Annex.

\subsection{Simulation}
 
 \vspace{2.5mm}
 
  \noindent\textit{Estimation of Coefficients} 
  
  \vspace{2.5mm}
  
\begin{figure}[H]

\centering

   \caption{$Total$ Scope 1: Boxplot over simulations for $\hat{\beta}_{Revenue}$.
   Red dotted lines denote the true value of the  
   coefficient.}
   
\begin{subfigure}{\textwidth}
\centering
    \includegraphics[width=0.7\textwidth]{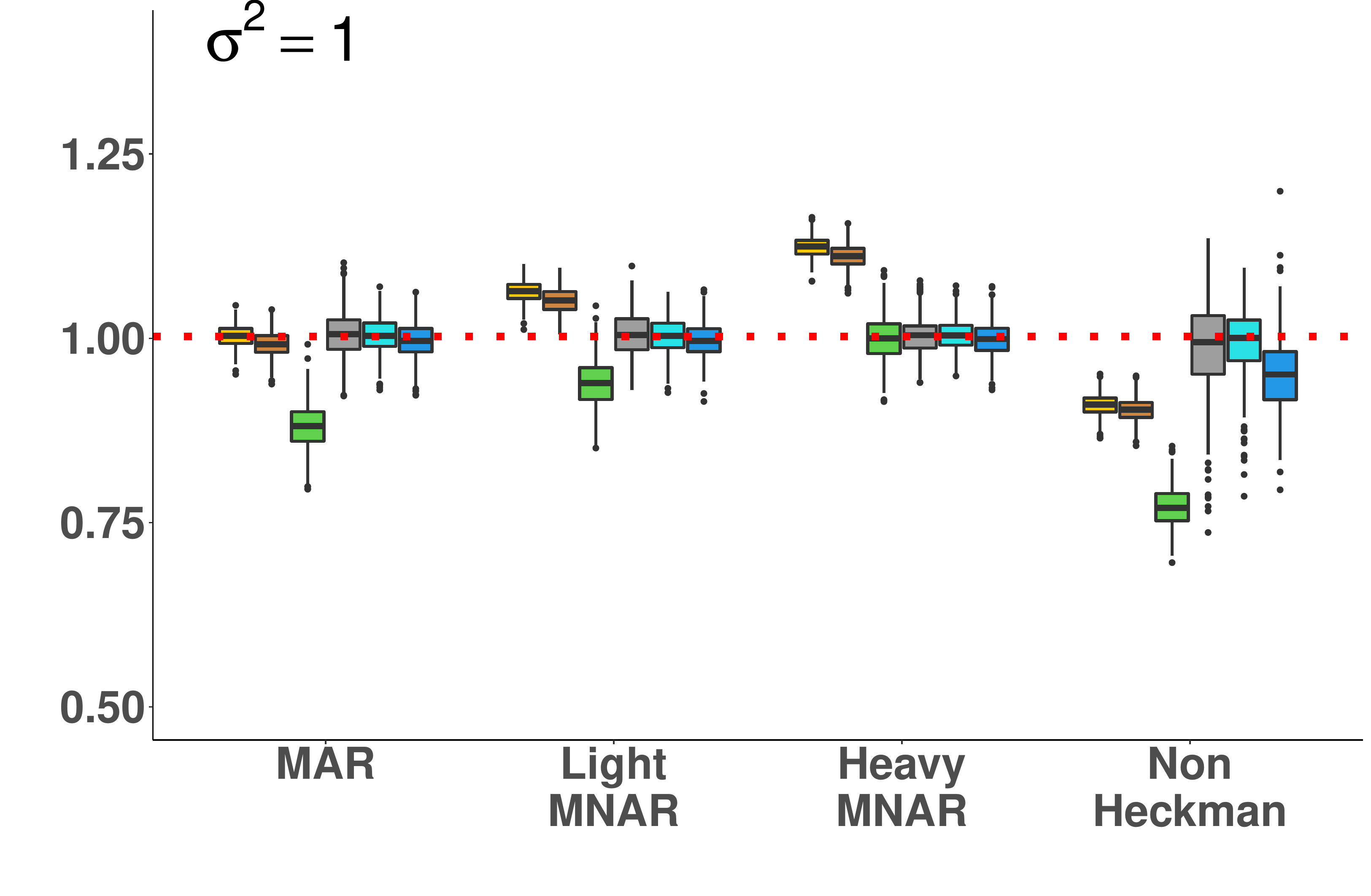}
  \end{subfigure}
  \begin{subfigure}{\textwidth}
  \centering
   \includegraphics[width=0.7\textwidth]{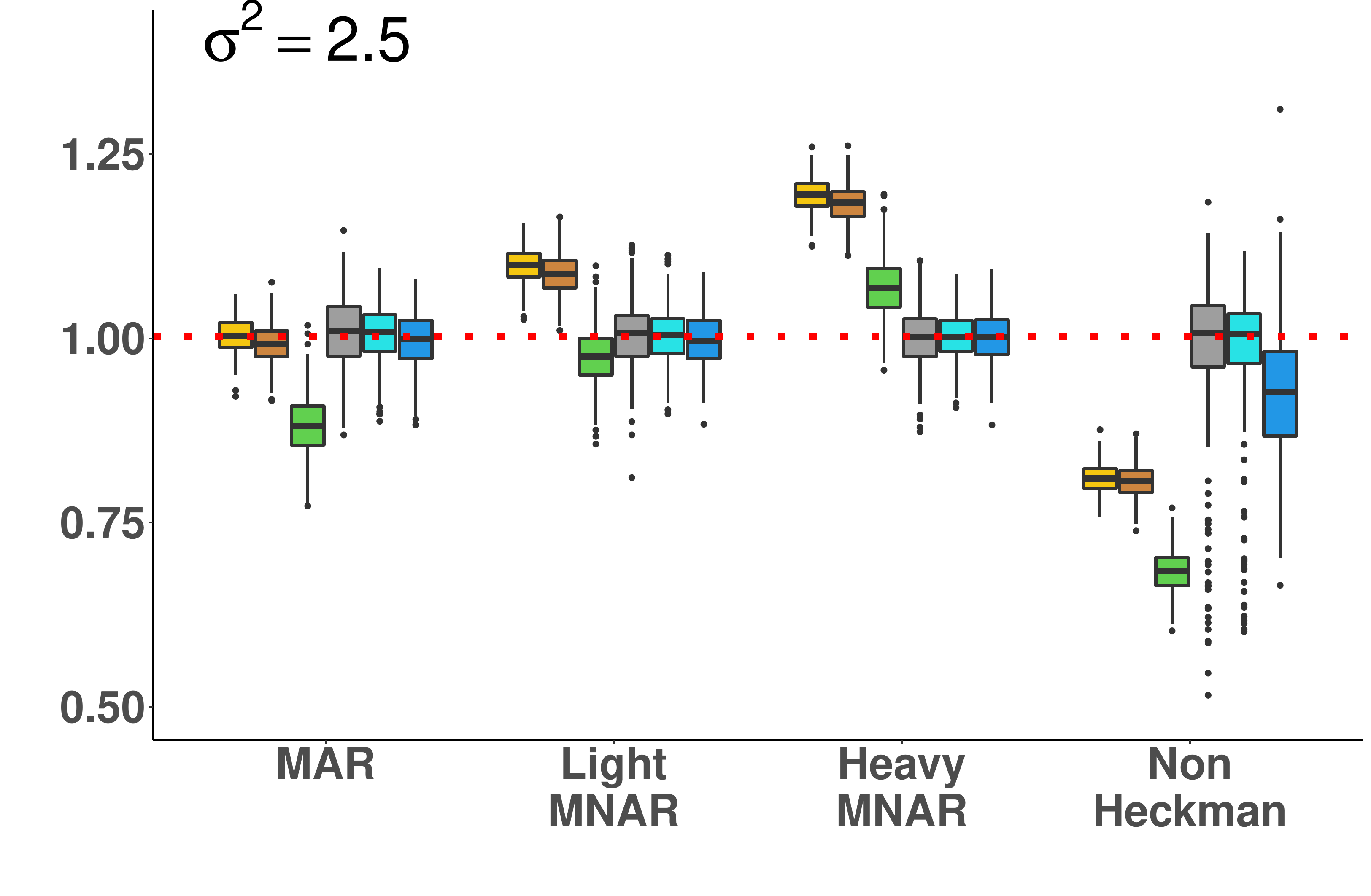}
 \end{subfigure}
  \begin{subfigure}{\textwidth}
  \centering
    \includegraphics[width=0.7\textwidth]{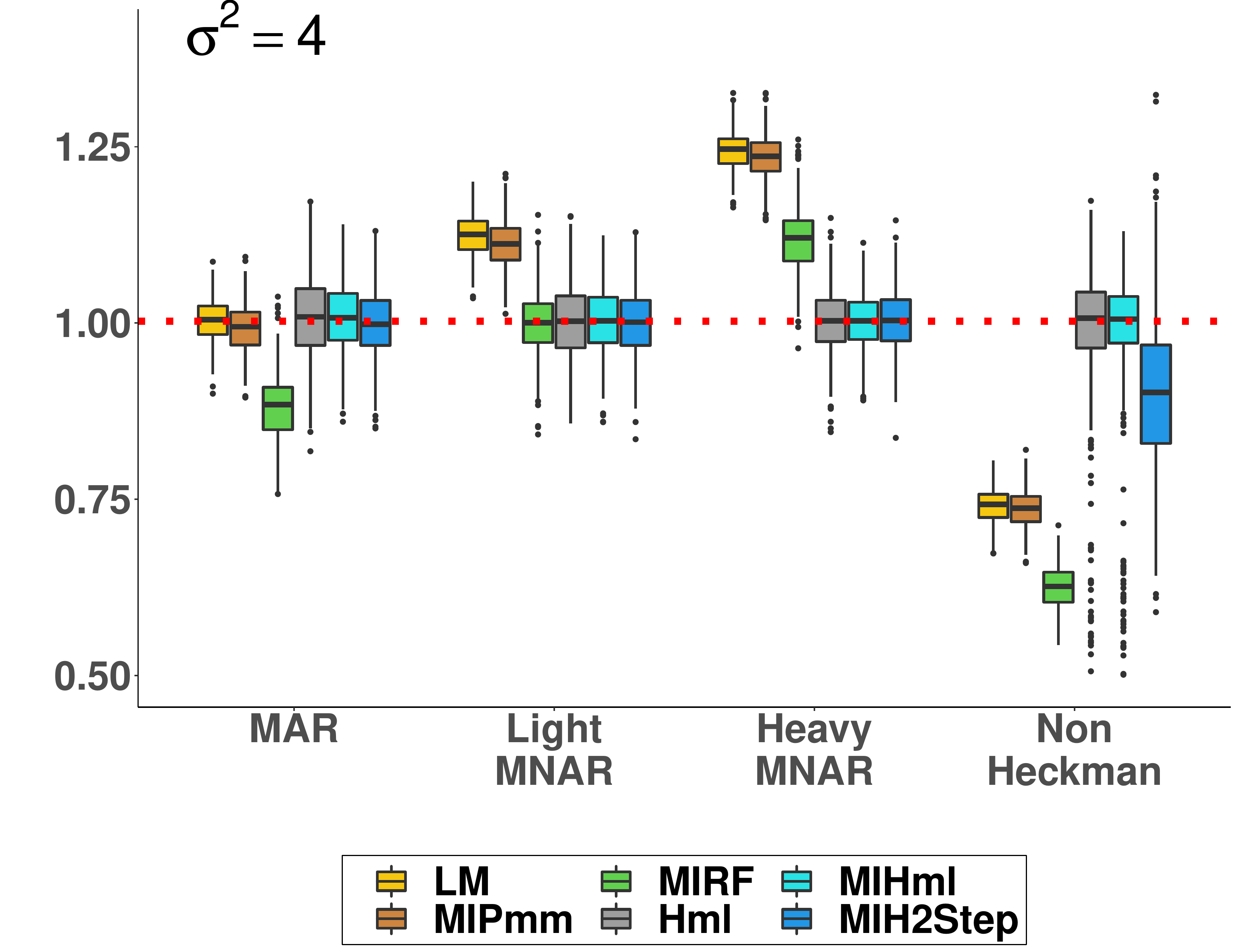}
    \end{subfigure}
    
 \label{fig:betast}
  \end{figure}

 Table \ref{tab:BetaTotal} and Fig \ref{fig:betast} present results of $\beta_{Revenue}$ estimation for the $Total$ dataset. With MAR mechanism ($\rho=0$), all methods except MIRF appeared to give unbiased estimates of $\beta_{Revenue}$ (with relative bias (Rbias) less than 2\%). In Light MNAR ($\rho=-0.3$) and Heavy MNAR ($\rho=-0.6$), only Hml, MIHml and MI2Step appeared to consistently give unbiased estimates across all the values of $\sigma^{2}_{\varepsilon}$. The estimates obtained with methods not based on Heckman's model appeared to be biased and their bias increased with $\rho$ and $\sigma^{2}_{\varepsilon}$. In Non-Heckman scenario, Hml and MIHml appeared to be unbiased (or almost unbiased i.e., $Rbias$ less than 2.5\% ), whereas bias of MI2Step was much lower than of the other methods not based on Heckman's model.  As expected, the MIHml was more efficient (smaller $SE_{m}$) than MIH2Step. To sum up, if data is MNAR, proposed approaches estimated coefficients with lower bias than alternatives.

 \vspace{2.5mm}

 For all scenarios, due to the exclusion of between-variance, the $SE_{m}$ for LM and Hml was smaller than that for MI methods. $SE_{m}$ was smaller than $SE_{e}$ for all the methods, suggesting that $SE_{m}$ of all models was underestimated. The relative error ($RE_{SE}$) was higher for single imputation methods. 
 
 \vspace{2.5mm}
 
 The coverage rate (CR) was below its nominal value (95\%) for all the methods, with especially low value for all methods except MIHml and MI2Step in  Heavy MNAR and Non-Heckman scenarios ($CR \leq 56\% $) when $\sigma^{2}_{\varepsilon} > 1$. The proposed MIHml and MI2Step performed best with relatively high coverage rate across all scenarios in comparison to other tested methods. 

 \noindent  

\vspace{2.5mm}

In the MAR scenario, the RMSE was the lowest for LM or MIPmm for all values of $\sigma^{2}_{\varepsilon}$, suggesting a smaller estimation error across all 500 datasets. In non-Heckman scenarios, the RMSE of methods based on Heckman's model was lower than the other methods and MIHml tended to have the lowest RMSE among all the methods and across all values of $\sigma^{2}_{\varepsilon}$.  In general, higher value of $\sigma^{2}_{\varepsilon}$ brought an increase in $SE_{m}$, $SE_{e}$ and RMSE across all the methods. The results for the sub-samples are detailed in the Annex.

 \vspace{2.5mm}
 
 For the $Total$ dataset, we obtained the coefficient estimates for all 42 variables. For each categorical covariate with a total of $k$ categories, $k-1$ coefficients were estimated. To summarise results for all the categories, Table\ref{tab:allbetas} presents the median of the relative bias (median of Rbias) across all the categories of each covariate. As before, Hml, MIHml and MI2Step performed better than other methods for most covariates across MNAR scenarios, resulting in lower median relative bias. The results for coefficients estimation for Scope 2 emissions were very similar to Scope 1 and  similar trends were observed for the sub-sets. The results are detailed in the Annex.

\begin{table}[H]
\caption{$Total$ Scope 1: Simulation results for $\beta_{Revenue} = 1.00 $ estimates }
\label{tab:BetaTotal}
\centering
\small{ 1.a) \quad $\sigma^{2}_{\varepsilon}$ =1 }

\vspace{1mm}

\resizebox{0.65\textwidth}{!}{\begin{tabular}{|ll|rrrrrrr|}
  \hline

 & Methods & Mean & Rbias & $SE_{m}$ & $SE_{e}$ & $RE_{SE}$ &  CR   & {\small RMSE } \\ 
\hline

 & LM & 1.00 & -0.25 & 0.52 & 5.26 & -90.11 & 48.80 & 1.50 \\  
 &  MIPmm & 0.99 & -1.24 & 1.48 & 5.00 & -70.40 & 80.80 & 2.04 \\  
\textbf{MAR} &  MIRF & 0.88 & -12.22 & 3.09 & 11.38 & -72.85 & 4.00 & 12.64 \\  
   &  Hml & 1.01 & 0.75 & 0.84 & 5.95 & -85.88 & 44.00 & 3.02 \\  
  &  MIHml & 1.00 & -0.25 & 2.27 & 5.68 & -60.04 & 85.00 & 2.44 \\  
   &  MIH2Step & 1.00 & -0.25 & 2.26 & 5.50 & -58.91 & 89.60 & 2.44 \\
\hline
   & LM & 1.06 & 5.74 & 0.51 & 6.77 & -92.47 & 0.20 & 6.29 \\  
   & MIPmm & 1.05 & 4.74 & 1.43 & 5.94 & -75.93 & 9.80 & 5.16 \\  
 \textbf{ Light} &  MIRF & 0.94 & -6.23 & 3.27 & 8.54 & -61.71 & 48.00 & 7.04 \\  
 \textbf{ MNAR }& Hml & 1.01 & 0.75 & 0.84 & 4.80 & -82.50 & 42.20 & 2.87 \\  
   &  MIHml & 1.00 & -0.25 & 2.13 & 4.50 & -52.67 & 88.20 & 2.25 \\  
   &  MIH2Step & 1.00 & -0.25 & 2.23 & 4.76 & -53.15 & 88.00 & 2.35 \\  
\hline

   &  LM & 1.12 & 11.73 & 0.48 & 10.20 & -95.29 & 0.00 & 12.20 \\  
   &  MIPmm & 1.11 & 10.73 & 1.33 & 9.21 & -85.56 & 0.00 & 10.98 \\  
 \textbf{ Heavy} &  MIRF & 1.00 & -0.25 & 3.28 & 7.32 & -55.19 & 91.20 & 3.10 \\  
   \textbf{ MNAR } &  Hml & 1.00 & -0.25 & 0.87 & 6.91 & -87.41 & 56.00 & 2.32 \\  
   &  MIHml & 1.00 & -0.25 & 1.92 & 6.76 & -71.60 & 87.80 & 2.02 \\  
   &  MIH2Step & 1.00 & -0.25 & 2.20 & 7.01 & -68.62 & 89.20 & 2.25 \\  
   
\hline 

   &  LM & 0.91 & -9.22 & 0.48 & 7.50 & -93.60 & 0.00 & 9.39 \\  
   &  MIPmm & 0.90 & -10.22 & 1.38 & 7.81 & -82.33 & 0.00 & 10.06 \\  
  \textbf{Non-} &  MIRF & 0.77 & -23.19 & 2.71 & 16.85 & -83.92 & 0.00 & 23.31 \\  
  \textbf{Heckman} &  Hml & 0.99 & -1.24 & 0.85 & 11.94 & -92.88 & 21.20 & 6.40 \\  
   &  MIHml & 0.99 & -1.24 & 4.50 & 11.78 & -61.80 & 85.20 & 4.78 \\  
   &  MIH2Step & 0.95 & -5.23 & 5.32 & 9.47 & -43.82 & 76.60 & 7.20 \\  
\hline  
  
\end{tabular}}

\centering
{\small Mean = Average estimate $\hat{\beta}_{Revenue}$ (rounded to 2 digits), Rbias =  Average relative bias (\%), $SE_{m}$ = Root mean square of the estimated standard errors $\times 10^{2}$, $SE_{e}$ = Empirical standard error $\times  10^{2}$, CR = Coverage rate (based on 95\% confidence interval), $RE_{SE}$ = relative error of $SE_{m}$(\%), RMSE = Root mean square error $\times 10^{2}$.}
\end{table}

\begin{table}
\centering

\small{ 1.b) \quad $\sigma^{2}_{\varepsilon}$ =2.5 }

\vspace{1mm}

\resizebox{0.65 \textwidth}{!}{\begin{tabular}{|ll|rrrrrrr|}
  \hline

 & Methods & Mean & Rbias & $SE_{m}$ & $SE_{e}$ & $RE_{SE}$ &  CR   & {\small RMSE } \\ 
\hline

   &  LM & 1.00 & -0.25 & 0.82 & 5.71 & -85.64 & 46.80 & 2.41 \\  
   &  MIPmm & 0.99 & -1.24 & 2.28 & 5.41 & -57.86 & 85.40 & 2.89 \\  
 \textbf{MAR}  &  MIRF & 0.88 & -12.22 & 3.51 & 11.74 & -70.10 & 10.80 & 12.63 \\  
   &  Hml & 1.01 & 0.75 & 1.16 & 7.18 & -83.84 & 36.40 & 4.82 \\  
   &  MIHml & 1.01 & 0.75 & 3.57 & 6.59 & -45.83 & 85.00 & 3.84 \\  
   &  MIH2Step & 1.00 & -0.25 & 3.57 & 6.21 & -42.51 & 90.20 & 3.64 \\  
\hline  

   &  LM & 1.10 & 9.73 & 0.80 & 9.08 & -91.19 & 0.00 & 9.96 \\  
   &  MIPmm & 1.09 & 8.73 & 2.21 & 8.24 & -73.18 & 6.40 & 8.84 \\  
  \textbf{ Light}  &  MIRF & 0.97 & -3.24 & 3.58 & 7.95 & -54.97 & 79.40 & 4.65 \\  
  \textbf{ MNAR }  &  Hml & 1.00 & -0.25 & 1.17 & 6.94 & -83.14 & 41.80 & 4.37 \\  
   &  MIHml & 1.00 & -0.25 & 3.39 & 6.46 & -47.52 & 88.60 & 3.52 \\  
   &  MIH2Step & 1.00 & -0.25 & 3.51 & 6.69 & -47.53 & 89.60 & 3.61 \\  
\hline  

   &  LM & 1.19 & 18.71 & 0.75 & 14.83 & -94.94 & 0.00 & 19.31 \\  
   &  MIPmm & 1.18 & 17.71 & 2.08 & 13.97 & -85.11 & 0.00 & 18.19 \\  
 \textbf{ Heavy}  &  MIRF & 1.07 & 6.74 & 3.82 & 9.01 & -57.60 & 54.20 & 7.61 \\  
   \textbf{ MNAR } &  Hml & 1.00 & -0.25 & 1.21 & 11.70 & -89.66 & 46.00 & 3.81 \\  
   &  MIHml & 1.00 & -0.25 & 2.84 & 11.37 & -75.02 & 87.80 & 3.13 \\  
   &  MIH2Step & 1.00 & -0.25 & 3.38 & 11.33 & -70.17 & 93.00 & 3.35 \\  
\hline  

   &  LM & 0.81 & -19.20 & 0.73 & 12.61 & -94.21 & 0.00 & 19.29 \\  
   &  MIPmm & 0.81 & -19.20 & 2.00 & 13.20 & -84.85 & 0.00 & 19.83 \\  
 \textbf{Non-}   &  MIRF & 0.68 & -32.17 & 2.59 & 21.81 & -88.12 & 0.00 & 31.96 \\  
   \textbf{Heckman}  &  Hml & 0.99 & -1.24 & 1.22 & 18.84 & -93.52 & 30.80 & 9.47 \\  
   &  MIHml & 0.99 & -1.24 & 4.88 & 18.95 & -74.25 & 83.20 & 8.59 \\  
   &  MIH2Step & 0.93 & -7.23 & 8.90 & 14.73 & -39.58 & 78.40 & 11.39 \\  
\hline  
 
\end{tabular}}

\vspace{3mm}
\small{
1.c) \quad $\sigma^{2}_{\varepsilon}$ =4 }

\vspace{1mm}

\resizebox{0.65\textwidth}{!}{\begin{tabular}{|ll|rrrrrrr|}
  \hline

 & Methods & Mean & Rbias & $SE_{m}$ & $SE_{e}$ & $RE_{SE}$ &  CR   & {\small RMSE } \\ 
\hline

   &  LM & 1.00 & -0.25 & 1.03 & 5.87 & -82.45 & 48.80 & 3.01 \\  
   &  MIPmm & 0.99 & -1.24 & 2.92 & 5.86 & -50.17 & 84.80 & 3.53 \\  
 \textbf{MAR}  &  MIRF & 0.88 & -12.22 & 3.71 & 12.06 & -69.24 & 15.00 & 13.02 \\  
   &  Hml & 1.01 & 0.75 & 1.39 & 8.01 & -82.65 & 35.80 & 6.06 \\  
   &  MIHml & 1.01 & 0.75 & 4.42 & 7.25 & -39.03 & 85.40 & 4.91 \\  
   &  MIH2Step & 1.00 & -0.25 & 4.52 & 7.03 & -35.70 & 89.80 & 4.75 \\  
\hline  

    & LM & 1.12 & 11.73 & 1.01 & 10.60 & -90.47 & 0.00 & 12.57 \\  
   &  MIPmm & 1.11 & 10.73 & 2.84 & 9.77 & -70.93 & 5.40 & 11.47 \\  
 \textbf{ Light}    &  MIRF & 1.00 & -0.25 & 3.95 & 7.80 & -49.36 & 88.00 & 4.34 \\  
    \textbf{ MNAR } &  Hml & 1.00 & -0.25 & 1.40 & 8.80 & -84.09 & 36.00 & 5.63 \\  
    & MIHml & 1.00 & -0.25 & 4.15 & 7.86 & -47.20 & 86.60 & 4.50 \\  
    & MIH2Step & 1.00 & -0.25 & 4.48 & 7.97 & -43.79 & 91.40 & 4.55 \\  
\hline  

   &  LM & 1.25 & 24.69 & 0.95 & 18.28 & -94.80 & 0.00 & 24.42 \\  
    & MIPmm & 1.23 & 22.70 & 2.63 & 17.58 & -85.04 & 0.00 & 23.45 \\  
    \textbf{ Heavy} &  MIRF & 1.12 & 11.73 & 4.06 & 11.77 & -65.51 & 19.00 & 12.49 \\  
  \textbf{ MNAR }    & Hml & 1.00 & -0.25 & 1.44 & 15.41 & -90.66 & 46.80 & 4.74 \\  
   &  MIHml & 1.00 & -0.25 & 3.46 & 15.13 & -77.13 & 87.00 & 3.94 \\  
     & MIH2Step & 1.00 & -0.25 & 4.34 & 14.99 & -71.05 & 91.80 & 4.41 \\  
\hline  

    &  LM & 0.74 & -26.18 & 0.89 & 15.86 & -94.39 & 0.00 & 26.21 \\  
  & MIPmm & 0.74 & -26.18 & 2.34 & 16.45 & -85.78 & 0.00 & 26.74 \\  
   \textbf{Non-} &  MIRF & 0.63 & -37.15 & 2.66 & 24.45 & -89.12 & 0.00 & 37.78 \\  
     \textbf{Heckman}  &  Hml & 0.98 & -2.24 & 1.46 & 23.46 & -93.78 & 37.20 & 11.83 \\  
     & MIHml & 0.98 & -2.24 & 5.21 & 23.46 & -77.79 & 82.00 & 11.63 \\  
     & MIH2Step & 0.90 & -10.22 & 11.63 & 17.90 & -35.03 & 74.80 & 14.81 \\  
   \hline 
\end{tabular}}

\centering
{\small Mean = Average estimate $\hat{\beta}_{Revenue}$ (rounded to 2 digits), Rbias =  Average relative bias (\%), $SE_{m}$ = Root mean square of the estimated standard errors $\times 10^{2}$, $SE_{e}$ = Empirical standard error $\times  10^{2}$, CR = Coverage rate (based on 95\% confidence interval), $RE_{SE}$ = relative error of $SE_{m}$(\%), RMSE = Root mean square error $\times 10^{2}$.}
\end{table}

\vspace{2.5mm}

\begin{table}[H]
\centering
\caption{\textit{Total} Scope 1: Median relative bias (\% Rbias) for variable coefficients estimates across all the categories of selected predictors}
 $\sigma^{2}_{\varepsilon}$ =2.5 
\resizebox{\textwidth}{!}{\begin{tabular}{|l|rrrr|rrrr|rrrr|}
  \hline
  & \multicolumn{4}{c|}{Sector (40 categories)} & \multicolumn{4}{c|}{Region (8 categories)}& \multicolumn{4}{c|}{FirstActivity (4 categories)}\\
  \cline{2-13}
 & MAR &\multicolumn{1}{c}{Light} & Heavy & \multicolumn{1}{c|}{Non-} & MAR &\multicolumn{1}{c}{Light} & Heavy &  \multicolumn{1}{c|}{Non-} & MAR &\multicolumn{1}{c}{Light} & Heavy &  \multicolumn{1}{c|}{Non-} \\
 & & MNAR & MNAR & Heckman & & MNAR & MNAR & Heckman & & MNAR & MNAR & Heckman\\
  \hline
LM & 0.04 & 1.31 & 2.44 & -1.98 & -1.39 & -16.36 & -32.40 & -23.91 & 0.84 & 16.19 & 24.48 & -18.86 \\ 
  MIPmm & -0.24 & 1.98 & 3.40 & -2.72 & -2.10 & -17.21 & -32.49 & -23.65 & 2.05 & 15.47 & 24.45 & -18.32 \\ 
  MIRF & -21.16 & -18.59 & -20.32 & -11.89 & -34.47 & -44.50 & -61.39 & -46.37 & -40.89 & -36.09 & -30.89 & -40.40 \\ 
  Hml & 0.01 & 0.06 & -0.09 & 0.10 & -1.83 & -0.94 & -0.94 & -3.40 & 3.29 & 1.95 & 2.37 & 1.48 \\ 
  MIHml & 0.00 & 0.11 & -0.18 & -0.66 & -1.90 & -1.69 & -1.33 & -2.04 & 2.27 & 2.15 & 4.80 & 0.35 \\ 
  MIH2Step & -0.34 & -0.45 & -0.37 & -1.35 & -1.39 & -0.22 & -0.52 & -7.77 & 2.57 & 2.55 & 1.78 & -4.40 \\ 
   \hline
\end{tabular}}
\label{tab:allbetas}
\footnotesize{Results for  $\sigma^{2}_{\varepsilon}$ =1 and  $\sigma^{2}_{\varepsilon}$ =4 are enclosed in the Annex}
\end{table}

\noindent\textit{Prediction Accuracy}

\vspace{2.5mm}

 Table \ref{tab:accuracyt} presents the results for the prediction accuracy for the $Total$ dataset. As expected, the average length of prediction interval (PI) based on MI methods was wider than for single imputation methods. Moreover, PI tended to increase with $\sigma^{2}_{\varepsilon}$ for all methods. The average prediction interval length for LM was the lowest across all the methods and scenarios, whereas the highest RE and RMSE appeared to be associated with the Median based estimation. The MIRF had the highest CR across all the methods for most scenarios (except Non-Heckman); however, its relatively high RMSE suggested that the high CR was caused by the length of the prediction interval and not high prediction accuracy. 
 
\vspace{2.5mm}
 
  With the MAR mechanism, LM showed the lowest CR (72.5-87.7\%) within each scenario, which was well below the nominal coverage rate (95\%). The coverage of the MI methods fluctuated between approx. 90\% and 99\% conditional on $\sigma^{2}_{\varepsilon}$. Across all values of $\sigma^{2}_{\varepsilon}$, Median and MIRF had the highest RE and RMSE and all the other methods reported similar values.

\begin{table}[H]
\centering
\caption{$Total$:  Simulation results for the predicted missing Scope 1 emissions values.}
\label{tab:accuracyt}
\resizebox{\textwidth}{!}{\begin{tabular}{|ll|rrrr|rrrr|rrrr|rrrr|}
  \hline
  && \multicolumn{4}{c|}{MAR} & \multicolumn{4}{c|}{Light MNAR} & \multicolumn{4}{c|}{Heavy MNAR} & \multicolumn{4}{c|}{NonHeckman}  \\

$\sigma^{2}_{\varepsilon}$  & Method & \small{RE} & \small{RMSE} & \small{CR} & \small{PI} & \small{RE} & \small{RMSE} & \small{CR} & \small{PI}& \small{RE} & \small{RMSE} & \small{CR} & \small{PI} & \small{RE} & \small{RMSE} & \small{CR} & \small{PI}  \\ 
  \hline
  
\multirow{7}{*}{1} & Median & 68.03 & 2.57 & - & - & 60.67 & 2.44 & - & - & 61.27 & 2.33 & - & - & 25.06 & 2.89 & - &  -\\ 
 & LM & 17.77 & 1.00 & 87.72 & 3.09 & 17.07 & 1.09 & 83.91 & 3.06 & 18.40 & 1.32 & 72.25 & 2.97 & 10.85 & 1.19 & 78.75 & 3.00 \\ 
&  MIPmm & 17.98 & 1.00 & 95.19 & 3.96 & 17.19 & 1.08 & 92.68 & 3.88 & 18.15 & 1.31 & 83.15 & 3.64 & 10.88 & 1.19 & 88.32 & 3.74 \\ 
&  MIRF & 23.63 & 1.09 & 99.00 & 5.59 & 20.04 & 1.06 & 99.22 & 5.64 & 18.29 & 1.18 & 98.41 & 5.60 & 12.65 & 1.42 & 93.80 & 5.14 \\ 
&  Hml & 17.66 & 1.00 & 95.06 & 3.93 & 17.91 & 1.00 & 95.24 & 3.95 & 17.16 & 0.98 & 95.79 & 4.01 & 9.53 & 1.01 & 95.08 & 3.97 \\ 
&  MIHml & 17.71 & 1.00 & 95.25 & 3.97 & 17.93 & 1.00 & 95.48 & 4.00 & 17.14 & 0.99 & 96.21 & 4.10 & 9.48 & 1.00 & 96.13 & 4.16 \\ 
 & MIH2Step & 17.89 & 1.00 & 95.29 & 3.98 & 18.11 & 0.99 & 95.43 & 3.98 & 17.27 & 0.98 & 95.68 & 3.99 & 9.93 & 1.07 & 94.05 & 4.05 \\ 
 \hline
\multirow{7}{*}{2.5}  & Median & 104.21 & 2.85 & - & - & 148.31 & 2.68 & - & - & 171.89 & 2.56 & - & - & 30.06 & 3.46 & - & - \\ 
  & LM & 43.44 & 1.58 & 78.03 & 3.88 & 36.60 & 1.72 & 73.41 & 3.85 & 47.13 & 2.08 & 60.02 & 3.73 & 18.79 & 2.12 & 57.66 & 3.68 \\ 
  & MIPmm & 44.05 & 1.58 & 95.11 & 6.24 & 36.37 & 1.72 & 92.50 & 6.11 & 46.12 & 2.08 & 82.67 & 5.72 & 18.82 & 2.12 & 79.96 & 5.54 \\ 
  & MIRF & 52.91 & 1.64 & 97.63 & 7.42 & 48.08 & 1.66 & 97.54 & 7.45 & 40.88 & 1.91 & 94.77 & 7.30 & 20.43 & 2.32 & 82.56 & 6.35 \\ 
  & Hml & 43.15 & 1.58 & 85.77 & 4.64 & 44.65 & 1.58 & 86.00 & 4.66 & 59.46 & 1.57 & 86.76 & 4.72 & 14.83 & 1.60 & 86.13 & 4.74 \\ 
  & MIHml & 43.19 & 1.58 & 91.98 & 5.54 & 44.61 & 1.58 & 92.21 & 5.58 & 58.57 & 1.57 & 93.01 & 5.70 & 14.83 & 1.60 & 92.94 & 5.81 \\ 
  & MIH2Step & 43.87 & 1.58 & 95.28 & 6.29 & 45.75 & 1.57 & 95.39 & 6.29 & 60.63 & 1.56 & 95.59 & 6.29 & 15.48 & 1.71 & 93.42 & 6.35 \\ 
  \hline
 \multirow{7}{*}{4} & Median & 145.56 & 3.11 & - & - & 138.90 & 2.92 & - & - & 94.58 & 2.82 & - & - & 34.13 & 3.97 & - & - \\ 
  & LM & 68.19 & 2.00 & 72.49 & 4.37 & 60.23 & 2.18 & 67.73 & 4.33 & 42.92 & 2.64 & 54.06 & 4.19 & 24.82 & 2.85 & 45.47 & 4.05 \\ 
  & MIPmm & 68.97 & 2.00 & 95.10 & 7.89 & 60.50 & 2.17 & 92.46 & 7.73 & 42.82 & 2.63 & 82.49 & 7.21 & 24.83 & 2.85 & 73.65 & 6.67 \\ 
  & MIRF & 81.07 & 2.05 & 96.93 & 8.86 & 71.88 & 2.10 & 96.54 & 8.87 & 44.98 & 2.44 & 92.57 & 8.62 & 26.26 & 3.03 & 74.76 & 7.22 \\ 
  & Hml & 67.79 & 2.00 & 79.53 & 5.08 & 72.55 & 2.00 & 79.78 & 5.10 & 49.67 & 2.00 & 80.37 & 5.15 & 18.56 & 2.06 & 79.20 & 5.19 \\ 
  & MIHml & 67.68 & 2.00 & 90.28 & 6.64 & 72.45 & 2.00 & 90.48 & 6.68 & 49.78 & 2.00 & 91.19 & 6.81 & 18.58 & 2.06 & 90.78 & 6.98 \\ 
 &  MIH2Step & 68.68 & 2.00 & 95.29 & 7.95 & 73.53 & 1.99 & 95.37 & 7.95 & 50.54 & 1.97 & 95.58 & 7.95 & 19.45 & 2.21 & 92.42 & 7.95 \\ 
 \hline
\end{tabular}}
{\small RE =  Average relative error for all $Y_{miss}$ (\%) , PI = Average length of the prediction interval, RMSE = Root mean square error, CR = Coverage rate (\%)}
\end{table}

  \vspace{2.5mm}
  
 With the MNAR mechanisms, for each value of $\sigma^{2} _{\varepsilon}$, MIHml, MIH2Step and Hml had lower RMSE than the other methods. The PI of MIH2Step was higher than that of MIHml across all the MNAR scenarios and values of $\sigma^{2}_{\varepsilon}$, which tended to result in a higher coverage rate for MIH2Step. The MIPmm had the lowest coverage rate among all MI methods. For larger variance in Heavy MNAR and Non-Heckman scenario, MIH2Step had higher coverage than MIRF but also higher PI, suggesting inefficiency in the estimation. In summary, MI methods based on Heckman's model tend to have relatively high coverage and small PI in comparison to other tested methods if data is MNAR.   

\vspace{2.5mm}

The results for the data sub-sets were similar and are detailed in the Annex. In general, a higher level of data missingness did not increase the RMSE of the estimates. Only for LM, the CR significantly decreased with the level of missingness.

\subsection{Empirical Study Results: 2018 LSEG Data}
For the empirical part of our study, we have applied the methods described in Section \ref{sec:methods} to the original 2018 LSEG  Scope 1 and Scope 2 data. It is important to note that the MAR mechanism cannot be ruled out for LSEG data, based on testing $\hat{\rho} = 0$ (for Scope 1 p-value = 0.99  and for Scope 2 p-value = 0.48 , $\hat{\rho}$ obtained with  Heckman's ML estimation). 

\vspace{2.5mm}

Fig.2 presents a  boxplot of the relative prediction interval length (relative PI) (100 \%  $\times \frac{PI length}{Predicted value}$) for the Scope 1 and Scope 2 estimates of all missing values in the LSEG dataset. As expected,  the relative PI was smaller for single imputation methods than for MI methods with its average value for Scope 1 of 47.75\% (Scope 2: 36.42\%) for LM, 57.29\% (Scope 2: 46.08\%) for Hml, 93.58\%  (Scope 2: 63.40\%) for MIRF, 76.15\% (Scope 2: 56.21\%) for MIPmm, 71.02\% (Scope 2: 50.06\%) for MIHml and  75.45\% (Scope 2: 54.46\%) for MIH2step. Among MI methods,  MIHml had the lowest relative PI for both Scope 1 and Scope 2. The variation in relative prediction interval values was smaller for Scope 2 than Scope 1 across all the methods, which can be explained by smaller variation of the reported values.

\begin{figure}[H]
  \label{fig:pibox2}
 \caption{Empirical results: Boxplots of the relative prediction interval length according to the method for the $Total$ data set.}
 \setlength{\emergencystretch}{3em}
 
 \begin{subfigure}{0.5\textwidth}
      \centering
    \includegraphics[width=\textwidth]{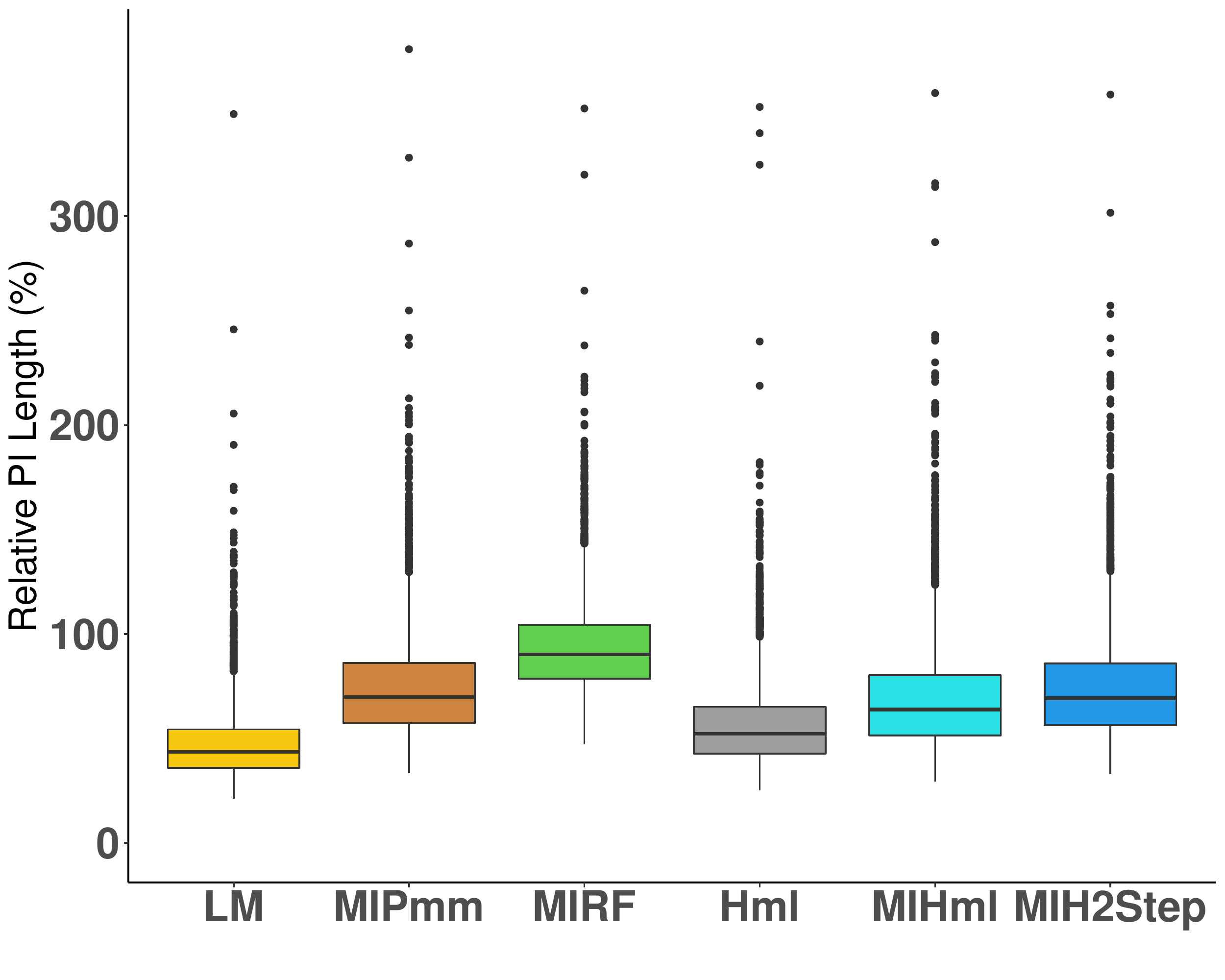} 
    \caption{Scope 1}
 \end{subfigure}%
 \begin{subfigure}{0.5\textwidth}
      \centering
    \includegraphics[width=\textwidth]{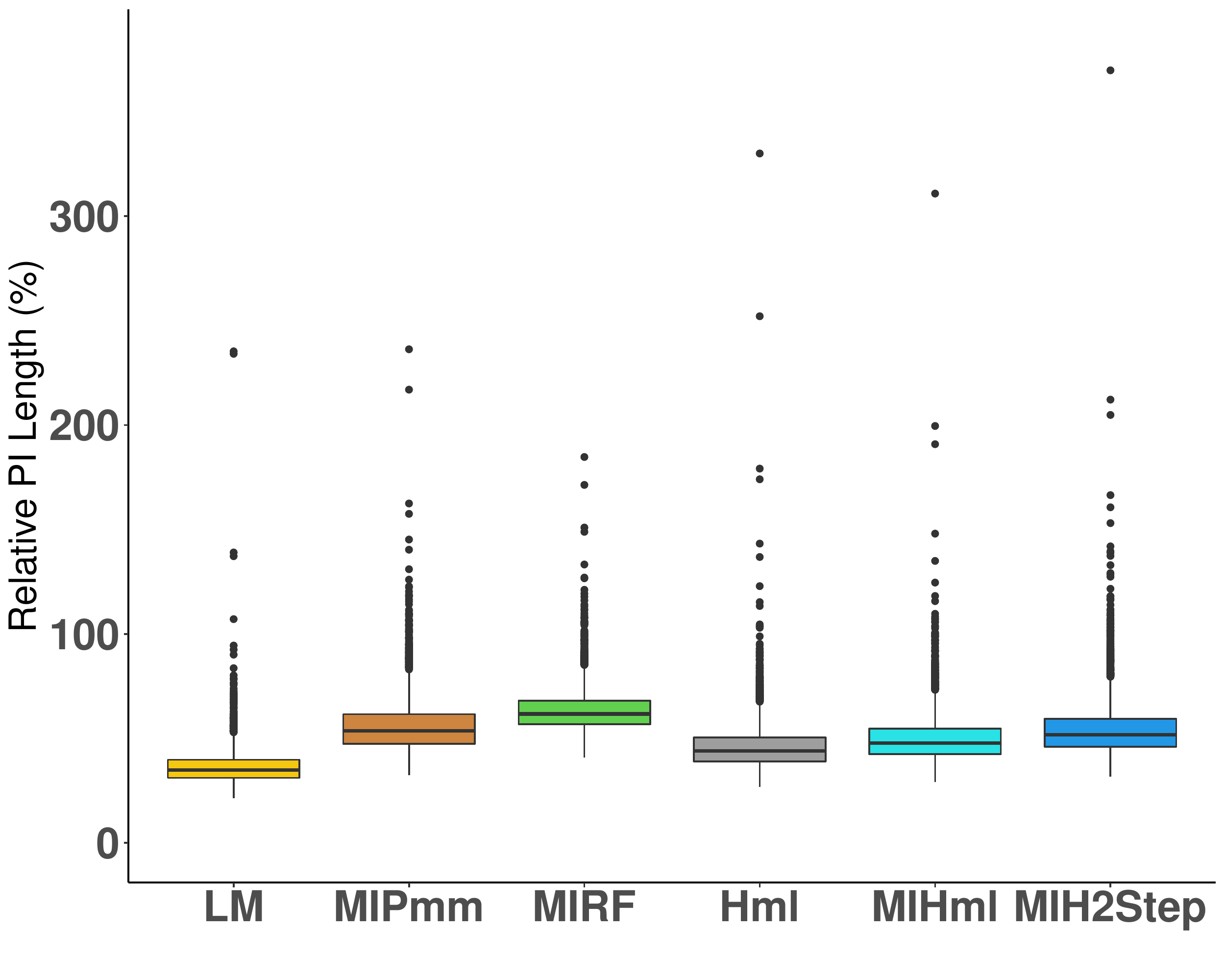} 
        \caption{Scope 2}
 \end{subfigure}
   \end{figure}

   \vspace{2.5mm}
   
In practice, the evaluation of the accuracy of emissions prediction models is often performed retrospectively, after a company starts to reports its data (e.g. \citeauthor{kalesnik2020green}, \citeyear{kalesnik2020green}). Thus, for companies that disclosed their emissions for the first time in 2019 ("first-time reporters"), we compared the reported values (2019) with the obtained estimates (2018). The sample size ($n_{2019}$) for Scope 1 was 423 and for Scope 2 was 447. In our analysis, we did not adjust for a year effect to avoid inclusion of any additional bias. Due to a small sample size available for this exercise, the following results should not be considered conclusive.

\vspace{2.5mm}

\begin{figure}[H]
 \caption{Empirical results. Estimated 2018 values vs reported 2019 values for the $Total$ data set.}
 \label{fig:pibox}
\begin{subfigure}{\textwidth}

  \centering
    \includegraphics[width= 0.75\linewidth]{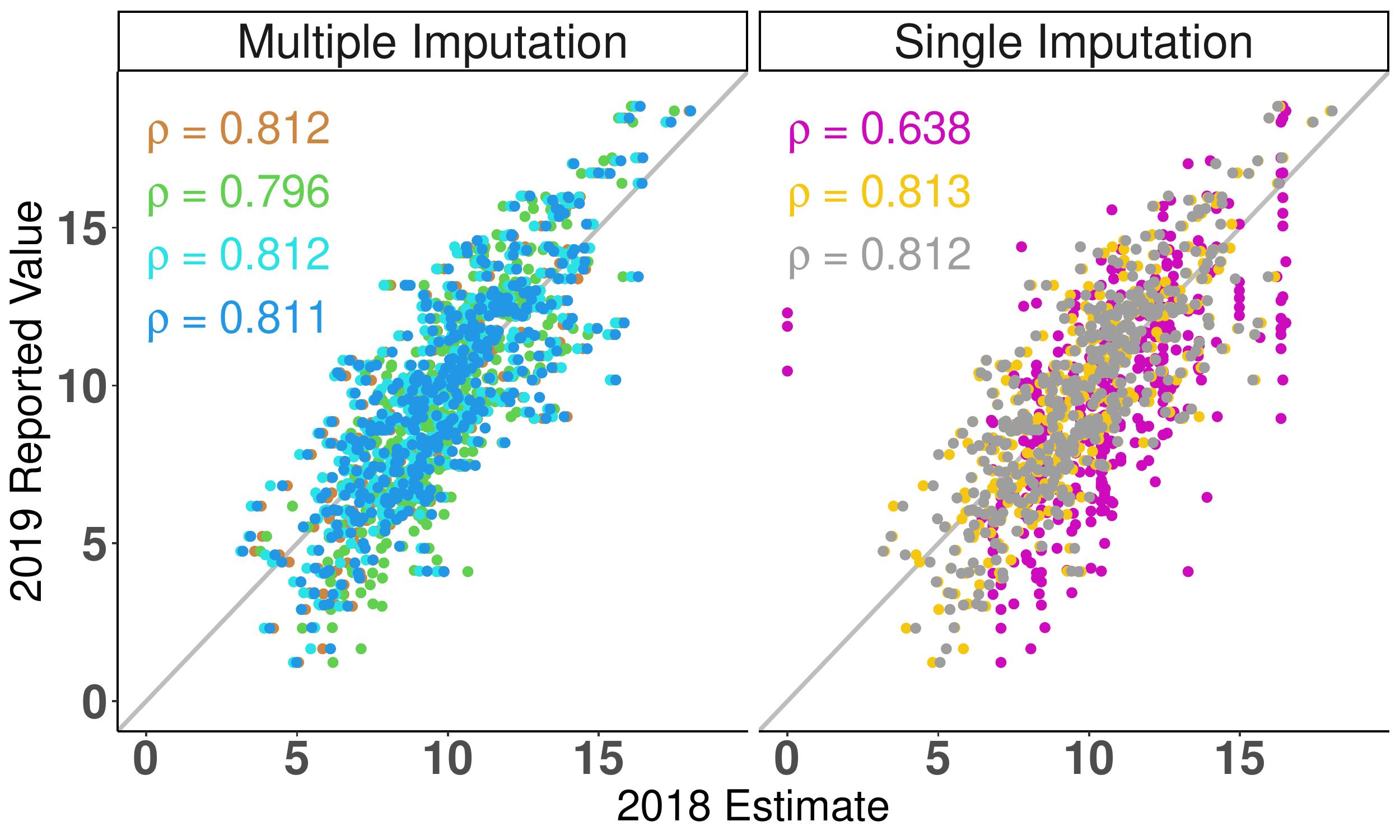} 
    \caption{Scope 1}
    
\end{subfigure}

\vspace{2.5mm}
\begin{subfigure}{\textwidth}

      \centering
    \includegraphics[width=0.75\linewidth]{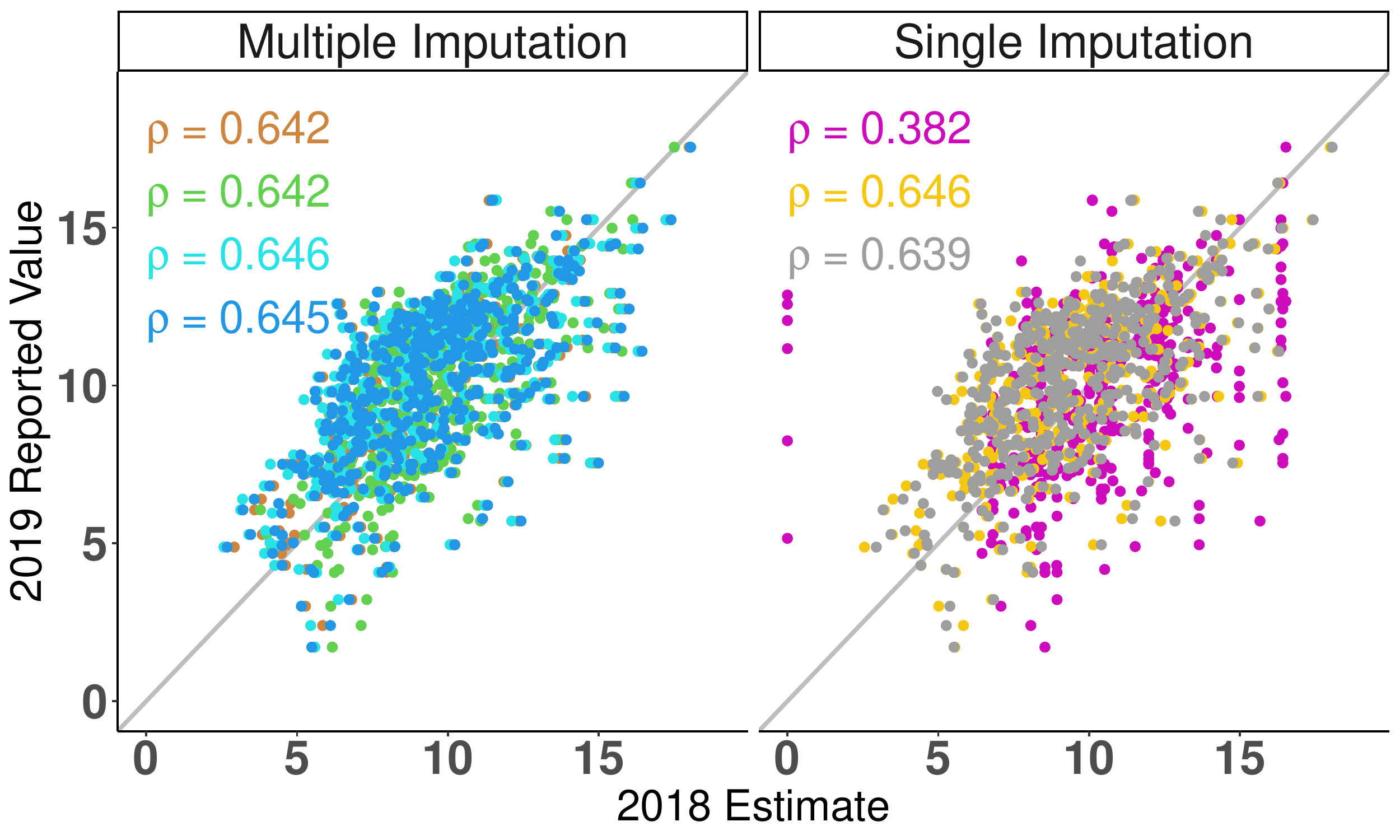} 
        \caption{Scope 2}

    \end{subfigure}
    
    \begin{subfigure}{\textwidth}

      \centering
    \includegraphics[width=0.55 \linewidth]{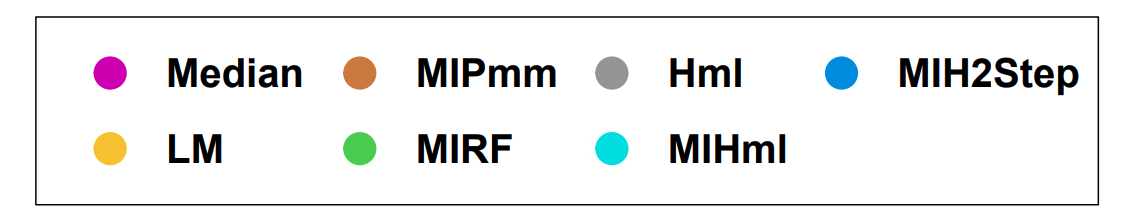} 

    \end{subfigure}
     \label{fig:scatter}
\end{figure}

Fig.\ref{fig:scatter} shows the 2018 estimated values against the corresponding 2019 reported values. It can be also seen that estimates based on multiple imputation methods are distributed more closely to the diagonal line with slope equal to 1 (2018 estimate = 2019 reported value), suggesting higher estimation accuracy. For Scope 1, LM had the highest correlation ($\rho$ Pearson correlation) among all the methods.  For Scope 1 among MI methods, MIPmm and MIHml had the highest correlation $\rho$ = 0.812 and MIHml had the highest $\rho$ = 0.646 for Scope 2. 
 
 \vspace{2.5mm}
 
The detailed results of prediction accuracy for the first-time reporters are presented in Table \ref{tbs:empirical}. We have also included the results for prediction accuracy of the joint Scope 1 and Scope 2 estimates (Scope $1+2$). In terms of coverage rate, MIPmm performed better than MIHml, whereas MIRF performed better than both MI methods based on Heckman's model for all estimates.

\vspace{2.5mm}

Across all the methods, only MIRF had CR above (Scope 1, Scope $1+2$) or close to (Scope 2) the nominal rate (95\%). Similar to what we have seen in simulation, the comparatively high coverage of MIRF was caused by its wide prediction  interval but not higher accuracy, since its RMSE was higher than for other methods. All the methods except MIRF and Median had very similar RMSE values for Scope 1, 2 and $1+2$. The LM exhibited substantial undercoverage, but a relatively low RMSE, which indicated that its prediction errors were not much higher than for the other methods.  The value of Spearman correlation behaved similarly to Pearson and was almost the same for all the methods except MIRF and Median for Scope 1, 2 and $1+2$. It is important to note that the 2019 data may vary from the true 2018 values and thus the coverage rate could be reflected incorrectly. Scope $1+2$ estimates tend to have higher CR and lower RMSE than Scope 1 and Scope 2 estimates.

\vspace{2mm}

\vspace{2.5mm}
In the Annex in Table \ref{tbs:empiricalsub}, we present results for estimation  accuracy within the subsets. For each of the subsets, the RMSE of all methods was higher in comparison to their value in the $Total$ sample, with Median having the highest RMSE value across all the samples. Similarly to results in the simulation study, the CR of LM decreased significantly with the level of missingness in the sample since  the model was fitted to a smaller number of observations, causing the estimated SE and, consequently PI to decrease. CR varied much less across the subsets for the methods other than LM and their PI tended to increase with the level of missingness. The higher level of missingness did not necessarily lead to an increase in  RE across the methods. The sample with the highest RMSE was $U$ (12\% of the data observed); however, the coverage rate across the MI methods in that sample was still pretty high, varying between 77.55\% for MIHml in Scope 1 and 98.04\% for MIRF in Scope 2. Across all methods and samples, CR for Scope 2 tended to be higher than for Scope 1 whereas RMSE, Pearson and Spearman Correlations tended to be lower.    

\vspace{2.5mm}

\begin{figure}[H]
 \caption{Empirical results. Relative Error per Industry for the $Total$ data set.}
 \label{fig:ind}
\resizebox{\textwidth}{!}{ \begin{tabular}{rl}
  \begin{subfigure}{\textwidth}
    \includegraphics[width=0.99\linewidth]{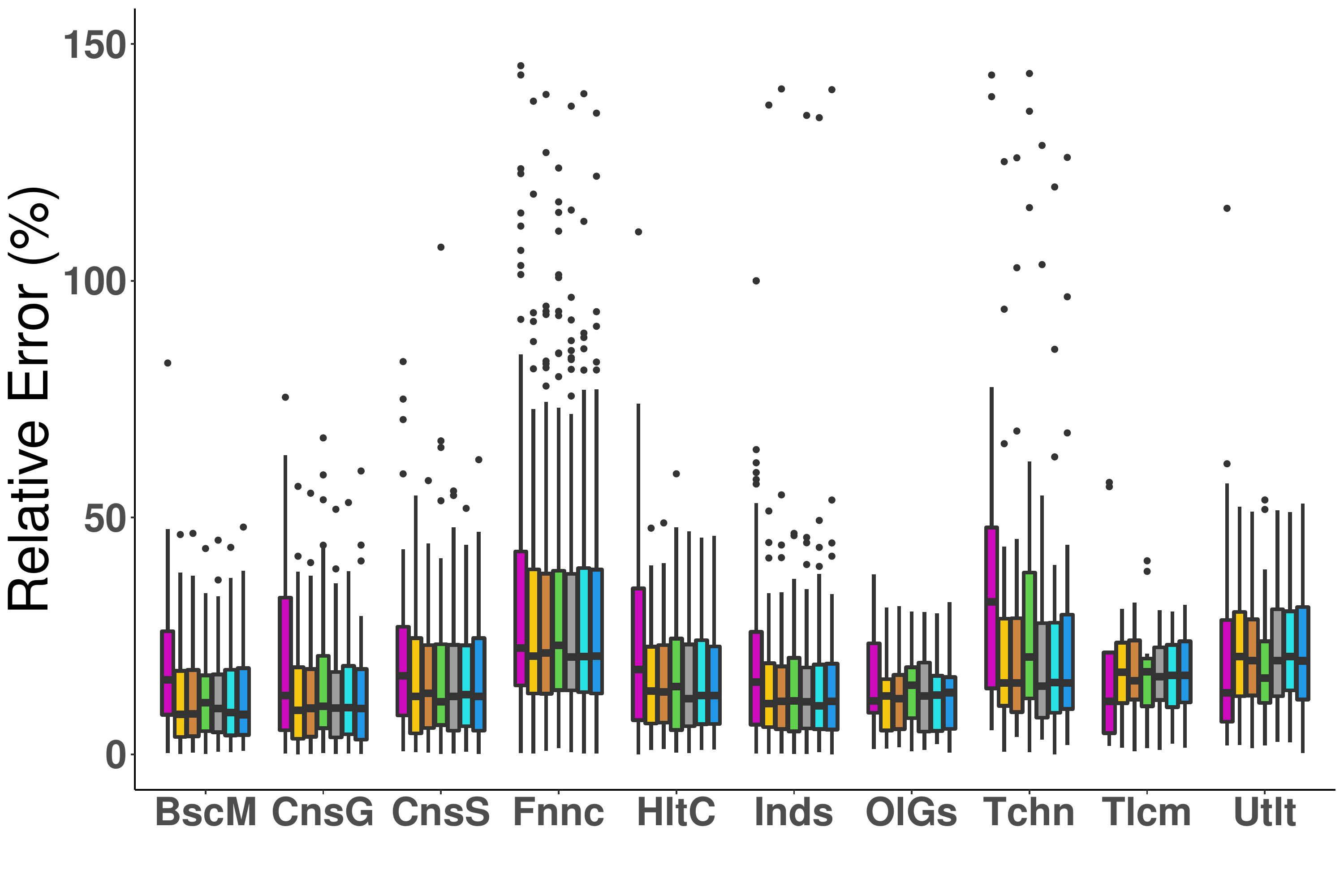}
    \caption{Scope 1 ($n_{2019}$ = 423)}
\end{subfigure}    &     \footnotesize{ \begin{tabular}{|c|r|c|r|}
\hline
Industry  &    Obs. \% & $n_{2019}$ & mean \\ 
\hline
 BscM &    41.6 & 51 & 13.8\\
 CnsG &    40.1 & 62  & 11.2 \\
 CnsS & 33.6 & 44 & 10.5\\
 Fnnc &  36.4 & 78 & 7.9\\
HltC &  34.8 & 24 & 10.4 \\
 Inds & 39.0 & 87 & 11.4 \\
 Ol\&G & 53.2 & 14 & 14.1 \\
 Tchn &  34.7 & 32 & 9.1\\
Tlcm & 59.8 & 11 & 10.2 \\ 
Utlt &  59.0 & 20 & 15.0 \\
\hline
\end{tabular}}

 \end{tabular}}

\vspace{1.5mm}

\resizebox{\textwidth}{!}{ \begin{tabular}{rl}
  \begin{subfigure}{\textwidth}
    \includegraphics[width=0.99\linewidth]{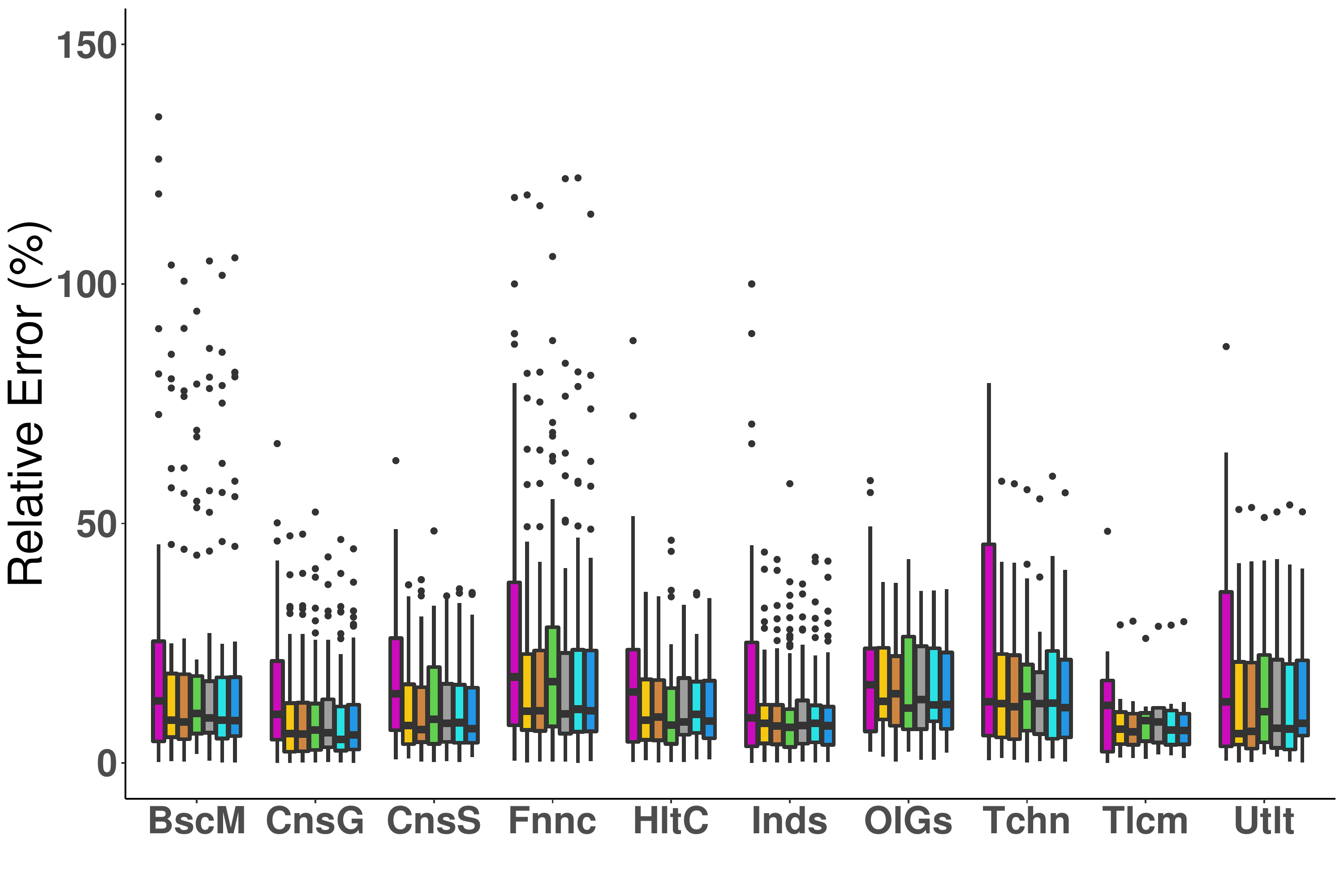}
    \caption{Scope 1 ($n_{2019}$ = 423)}
\end{subfigure}    &     \footnotesize{ \begin{tabular}{|c|r|c|r|}
\hline
Industry  &    Obs. \% & $n_{2019}$ & mean \\ 

\hline

 BscM &    40.3 & 51 & 13.0 \\
 CnsG &    39.4 & 63 & 11.7\\
 CnsS & 34.0 & 45 & 11.1\\
 Fnnc &  38.0 & 79 & 9.6\\
HltC &  35.2 & 27 & 10.9 \\
 Inds & 39.1 & 96 & 11.3 \\
 Ol\&G & 46.0 & 17 & 11.8 \\
 Tchn &  35.9 & 36 & 11.2 \\
Tlcm & 59.8 & 12 & 12.7 \\ 
Utlt &  53.8 & 21 & 11.9 \\
\hline
\end{tabular}}

 \end{tabular}}

\begin{subfigure}{\textwidth}

        \centering
    \includegraphics[width= 0.55 \linewidth]{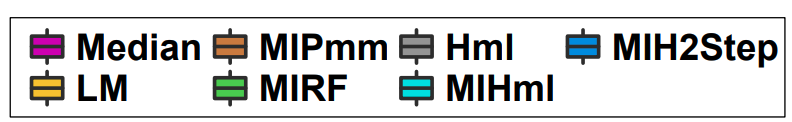} 
    
\footnotesize{ BscM: Basic Materials, CnsG: Consumer Goods, 
CnsS: Consumer Services,\\ Fnnc: Financials,   HltC: Health Care ,   Inds: Industrials , 
Ol\&G: Oil \& Gas, \\ Tchn:  Technology, Tlcm: Telecommunications, Utlt: Utilities }

  \end{subfigure}
\end{figure}

In the last step of our analysis, we considered the relative error of the estimates according to the Industry to which companies belong. The results for Scope 1 and Scope 2 are presented in Fig.\ref{fig:ind}. In general, for Scope 1 and Scope 2 all methods except Median and MIRF performed similarly for most industries. For Scope 1, the industries with highest variation between the methods were Technology, where Median and MIRF performed worse than other methods, as well as Telecommunication (59.8\% Obs.) and  Utilities (59.0\% Obs.) where Median performed better than the other methods. Financial industry had the highest RE, which can be explained by the metric sensitivity to small reported values ( Scope 1 median 7.9).  
\begin{table}[H]
\centering
\caption{Empirical study results: Accuracy of the 2018 estimates with respect to the 2019 reported values in the Total Sample.}
\resizebox{1.0 \textwidth}{!}{\begin{tabular}{lcccl|rrrrrrrr}

  \hline
  
  & Obs.\% & $n_{0}$ & $n_{2019}$ & Methods & CR & PI & $RE_{min}$ & $RE_{mean}$ & $RE_{max}$ & RMSE & Pearson & Spearman \\ 
  & & & & & & & & & & & Corr. & Corr. \\
  \hline
      & & &  & Median & - & - & 8.04 & 18.14 & 34.15 & 2.70 & 0.64 & 0.68 \\ 
    & & &  & LM & 70.45 & 3.80 & 6.49 & 13.48 & 24.37 & 1.86 & 0.81 & 0.81 \\ 
    & & &  &  MIPmm & 88.65 & 6.07 & 6.29 & 13.45 & 23.92 & 1.86 & 0.81 & 0.81 \\ 
\textbf{Scope1}  & 39 & 3993 & 423 & MIRF & 97.16 & 8.32 & 7.18 & 14.59 & 23.90 & 2.00 & 0.80 & 0.79 \\ 
    & & &  & Hml & 78.01 & 4.52 & 6.45 & 13.47 & 23.13 & 1.86 & 0.81 & 0.81 \\ 
     & & &  &  MIHml & 84.40 & 5.41 & 6.62 & 13.55 & 23.44 & 1.86 & 0.81 & 0.81 \\ 
    & & &  & MIH2Step & 89.83 & 6.02 & 6.17 & 13.12 & 24.48 & 1.86 & 0.81 & 0.81 \\ 
 
   \hline
   
    &  &  &  &   Median & - & - & 5.62 & 13.22 & 29.15 & 2.64 & 0.29 & 0.42 \\ 
 &  &  &  &   LM & 75.84 & 3.46 & 4.10 & 8.79 & 16.91 & 1.59 & 0.73 & 0.69 \\ 
 &  &  &  &  MIPmm & 88.59 & 5.30 & 4.22 & 8.61 & 16.76 & 1.59 & 0.73 & 0.69 \\ 
  \textbf{Scope 2}   & 39 & 4005 & 447 &  MIRF & 93.51 & 6.38 & 4.48 & 9.85 & 18.01 & 1.65 & 0.73 & 0.68 \\ 
 &  &  &  &   Hml & 83.89 & 4.28 & 4.54 & 8.93 & 16.97 & 1.58 & 0.72 & 0.69 \\ 
 &  &  &  &  MIHml & 85.91 & 4.71 & 4.36 & 8.91 & 17.04 & 1.59 & 0.73 & 0.69 \\ 
 &  &  &  &  MIH2Step & 88.81 & 5.11 & 4.35 & 8.61 & 16.84 & 1.58 & 0.73 & 0.69 \\ 

   \hline
   
 &  &  & & Median & - & - & 4.92 & 11.23 & 21.31 & 2.34 & 0.55 & 0.64 \\ 
  &  &  & & LM & 78.49 & 3.37 & 3.50 & 7.44 & 13.55 & 1.43 & 0.82 & 0.80 \\ 
 &  &  & & MIPmm & 90.07 & 4.79 & 3.51 & 7.44 & 13.60 & 1.44 & 0.82 & 0.80 \\ 
   \textbf{Scope $1+2$}   & 39 & 3997 & 423 & MIRF & 95.04 & 6.41 & 3.84 & 8.34 & 14.96 & 1.56 & 0.81 & 0.78 \\ 
  &  &  & & Hml & 85.58 & 4.18 & 3.20 & 7.64 & 13.83 & 1.44 & 0.82 & 0.80 \\ 
 &  &  & & MIHml & 88.89 & 4.53 & 3.66 & 7.51 & 13.77 & 1.44 & 0.82 & 0.79 \\ 
 &  &  & & MIH2Step & 90.78 & 4.82 & 3.63 & 7.38 & 13.58 & 1.42 & 0.82 & 0.80 \\ 
 \hline
\end{tabular}}
{\small RE =  relative error (\%), PI = Average length of the  95\% prediction interval , \\ RMSE = Root mean square error, CR = Coverage Rate (\%)}
\label{tbs:empirical}
\end{table}

\clearpage

\section{Discussion}

This paper proposed a new way to estimate missing corporate carbon emissions data to account for the potential non-randomness of the observed sample and appropriately reflect the uncertainty of the estimates of the missing values. This paper showed that if data are MNAR, the proposed multiple imputation approaches using Heckman's model can lead to a more accurate estimation of the missing corporate carbon emissions values. Moreover, reporting the length of the prediction intervals is likely to improve the transparency of the predicted values, while being simple for investors to understand. We have also shown that MI methods can reflect more accurately the uncertainty of the obtained estimates (through a wider prediction interval) than corresponding single imputation methods, and they resulted in a higher coverage rate. 

\vspace{2.5mm}

Currently used methods that predict missing emission values, such as a linear model, are typically only based on the disclosed emissions values. Therefore, these methods do not account for the uncertainty of the missing data and tend to overlook potential bias. To address these issues, we decided to derive the required linear prediction model based on an analysis of the data with multiple imputation (MI) methods for MNAR data using Heckman's model with maximum likelihood (MIHml) and two-step estimation (MIH2Step). To reflect the uncertainty of the predicted values, we proposed to report the corresponding length of the prediction interval. 

\vspace{2.5mm}

Our conclusions were drawn by examining different methods and scenarios.  In the empirical literature, emissions data are often considered to be \textit{missing not at random}. However, the non-randomness of the observed sample is not testable from the data and the performance of the proposed methods against MAR and MNAR assumption thus needed to be tested with a simulation study. In the simulation, we implemented four data missingness scenarios, namely MAR, Light MNAR (data-generation based on Heckman's approach with $\rho = -0.3$), Heavy MNAR ($\rho = -0.6$) and a Non-Heckamn scenario (MNAR mechanism where the probability of observing the data is a function of the emissions value). To compare the performance of the proposed MI methods with other techniques, we also implemented MI based on random forest (MIRF) and predictive mean matching (MIPmm) as well as single imputation (SI) using a linear model (LM) and Heckman's model with maximum likelihood (Hml). Each of the methods have been tested in terms of the accuracy of regression coefficients estimation. The resulting prediction models were then evaluated in terms of their prediction accuracy and were also compared with the results from a sector median (Median) estimation, a popular approach used by data providers.

\vspace{2.5mm}

The simulation study has shown that the proposed MIHml and MIH2Step methods appeared to improve the accuracy of parameter estimation if the data were MNAR. The estimates of coefficients obtained with Hml, MIHml amd MIH2Step seemed to be unbiased or have significantly lower bias than other methods across all missingness scenarios. MIRF, MIPmm and LM tended to result in biased estimates when data were MNAR. When the MAR mechanism was applied, all MI methods except MIRF performed equally well and their coverage rate was much higher than single imputation methods. This result was anticipated since MI methods include the between-variance when estimating standard errors (i.e., MI methods have larger standard errors than those resulting from single imputation); thus, their confidence intervals reflect the uncertainty of the estimates more accurately. 

\vspace{2.5mm}

Moreover, if the data were MNAR, the prediction models based on MIH2Step and MIHml appeared to have a smaller prediction error than other methods. In all the samples and scenarios, Median estimation was the method with the highest RMSE. As expected, MI has a greater length of the prediction interval than SI approaches due to the inclusion of the between-imputation variance. Thus, the coverage rate was higher for MI methods. However, SI Hml performed significantly better in terms of coverage than LM, which showed significant undercoverage in all simulation scenarios and samples. Therefore, neither LM nor Median (currently used by the industry) performed well in the simulation, suggesting that the proposed methods can bring improvements when being applied to corporate emissions estimation.

\vspace{2.5mm}

The methods implemented in this paper are subject to several constraints. We have shown that the relevant literature provides some evidence to support our MNAR assumption for corporate emission data; however, due to its non-testability, a MAR mechanism cannot be ruled out. Moreover, even though it is possible to approximate the distribution of the observed emissions data with a normal distribution, we cannot test Heckman's assumption about the normality of the error terms in the complete sample since the missing data cannot be observed. Furthermore, finding covariates, which would allow to satisfy the exclusion restriction for 2-Step estimation, may prove to be difficult in practice. 

\vspace{2.5mm}

In conclusion, the proposed Heckman's model based methods may result in more reliable estimates of missing carbon emissions than currently used techniques. The MI methods more accurately reflect the uncertainty of the estimation process and therefore provide investors with a better understanding of estimates’ quality.   
The limited results from the empirical study, based on only 10 \% of the missing sample that was observed in the following year, suggested that multiple imputation methods resulted in a higher coverage rate of the estimates than single imputation methods. Across all the methods, $MIH2Step$  had the lowest RMSE and relatively high coverage rate. Further research based on time-series data (with bigger sample size) should be conducted.

\clearpage

\section{Annex}

\subsection{Data}

\begin{table}[H]

 \caption{ London Stock Exchange Data (2018) Overview}
\label{tab:data}
 \resizebox*{!}{0.75\textwidth}{\begin{tabular}{|l|l|c|c|r|}
\hline

 Name &	Description & Type &  Details \\
 \hline
 Carbon Emissions Scope 1 &  Log transformed Scope 1 Carbon & Numeric & Min: 2.31     \\
 &  Emissions (MT CO2 equivalent) & &Mean: 10.95\\
  &  & & Max: 19.39\\
    &  & &Observed: 39 \%\\
 \hline
  Carbon Emissions Scope 2 &  Log transformed Scope 2 Carbon & Numeric & Min: 2.35     \\
 &  Emissions (MT CO2 equivalent) & &Mean: 11.22\\
  &  & & Max: 17.58\\
    &  & &Observed: 39 \%\\
 \hline
Country & Location of the Headquarters &	Factor & 51 levels \\
\hline
Region & Carbon Region  &	Factor  & 8 levels \\
\hline
Market & Stage of the Regional Market  &	Factor  & 5 levels\\
& Development & & \\
\hline
 IndustryName & Industry  & Factor  & 10 levels \\
 \hline
 SuperSectorName & Super-Sector  & Factor  & 20 levels \\
 \hline
 SectorName & Sector & Factor  &	40 levels  \\
 \hline
 SubSectorName & Sub-sector & Factor  &	113 levels \\
 \hline
LogRevenue  & Log Transformed Annual  & Numeric  & Min: 11.34 \\
& Revenue (\$M)&& Mean: 21.38 \\ 
&&& Max: 26.97 \\
\hline
 FirstActivity \footnotemark & Time period of Company's IPO & Factor & 4 levels   \\ 
 &   or  First Trade Recorded Date & & \\
 &&&\\
 \hline
 \textbf{SizeMarker} & Size category of the company & Factor & 4 levels: \\
 & & & \textbf{L, M, S, U} \\

\hline
\end{tabular}} 
\end{table}

\footnotetext{The first (recorded) trade of the companies have been been captured in the database since 1981.The initial public offering (IPO) dates are only recorded for the USA companies and the first trade date is available for the majority of companies. In most cases, the date of the IPO and the first trade is the same, however it is not a rule.}
 
 \newpage
 
\subsection{Evaluation Metrics}

\begin{table}[H]
    \centering
     \caption{Measure applied for methods performance in the simulation}
   \resizebox{\textwidth}{!}{\begin{tabular}{|l|l|c|c|}
    \hline
     \multicolumn{4}{|c|}{Parameter Estimation} \\
       \hline
     Measure & Abb. & Definition &  Estimate \\
    \hline
     & & & \\
   Relative bias (\%) & $Rbias$ & $\frac{\mathrm{E}[\hat{\theta}] - \theta}{\theta}$ & $\frac{1}{N} \sum_{i=1}^{N} \frac{\left(\hat{\theta}_{i}-\theta\right)}{\theta}100\%$  \\
     \hline
     & & & \\
    Root mean squared error & $RMSE$ & $\sqrt{\mathrm{E}\left((\hat{\theta}-\theta)^{2}\right)}$ & $\sqrt{\frac{1}{N-1} \sum_{i=1}^{N}\left(\hat{\theta}_{i}-\theta\right)^{2}}$ \\
         \hline
     & & & \\
    Empirical SE &  $SE_{e}$ & $\sqrt{Var(\hat{\theta})}$ & $\sqrt{\frac{1}{N-1} \sum_{i=1}^{N}\left(\hat{\theta}_{i}-\bar{\theta}\right)^{2}}$  \\
\hline
   & &  & \\
   Average model SE & $SE_{m}$ & $\sqrt{\mathrm{E}[\widehat{Var(\hat{\theta})}]}$ & $\sqrt{\frac{1}{N} \sum_{i=1}^{N} \widehat{Var}\left(\hat{\theta}_{i}\right)}$ \\
     \hline
        & &  & \\
        
  Relative error of $SE_{m}$ (\%)  &  $RE_{SE_{m}}$ & $(\frac{SE_{m}}{SE_{e}} - 1)$ & $(\frac{\widehat{SE}_{m}}{\widehat{SE}_{e}} - 1)100\%$\\
    \hline
        & &  & \\
 Coverage rate (\%) & CR &  $Pr\left(\hat{\theta}_{low} \leq \theta \leq \hat{\theta}_{upp}\right)$ & $\frac{1}{N} \sum_{i=1}^{N} \mathbf{ 1}\left(\hat{\theta}_{low,i} \leq \theta \leq \hat{\theta}_{upp, i}\right)100\%$ \\
 
 \hline
  \multicolumn{4}{|c|}{Prediction Accuracy} \\
      \hline
        Measure & Abb. & Definition &  Estimate \\
    \hline
        & &  & \\
   Average Relative Error (\%) & RE  & $ \frac{1}{n_{0}} \sum_{j=1}^{n_{0}} \frac{ \mid \hat{Y}_{j} - Y_{j} \mid }{Y_{j} }$ & $\frac{1}{N} \sum_{i=1}^{N} \left( \frac{1}{n_{0i}} \sum_{j=1}^{n_{0i}} \frac{\mid \hat{Y}_{ji} - Y_{ji} \mid}{Y_{ji} }\right) 100\%$  \\
     
        \hline
     & &  & \\
    Root mean squared error &  RMSE & $\sqrt{\mathrm{E}\left((\hat{Y}-Y)^{2}\right)}$ & $\sqrt{\frac{1}{N-1} \sum_{i=1}^{N} \left( \frac{1}{n_{0i}} \sum_{j=1}^{n_{0i}} \hat{Y_{ji}} - Y_{ji}\right)}$ \\
         \hline

     & &  & \\
 Coverage rate (\%) &  CR  &  $Pr\left(\hat{Y}_{low, j } \leq Y_{j} \leq \hat{Y}_{upp, j}\right)$ & $\frac{1}{N} \sum_{i=1}^{N} \left( \frac{1}{n_{0i}} \sum_{j=1}^{n_{0i}} \mathbf{1}\left(\hat{Y}_{low, ji} \leq Y_{ji} \leq \hat{Y}_{upp, ji}\right) \right)100\%$ \\
\hline
   & &  & \\
   Average prediction  & PI & $ \hat{Y}_{upp_{ j}} - \hat{Y}_{low_{j}} $ &  $\frac{1}{N} \sum_{i=1}^{N} ( \frac{1}{n_{0i}} \sum_{j=1}^{n_{0i}} \hat{Y}_ {upp_{j}} - \hat{Y}_ {low_{j}} )$\\
  interval length  & &  & \\
  \hline
 \end{tabular}}

{\footnotesize Formulas as presented in \cite{morris2019using}}
   
    \label{tab:evaluation}
\end{table}

\newpage

 \subsection{Imputation Methods}
 
 \begin{table}[H]
   \caption{Overview of the applied methods and corresponding predictors for each dataset}
    \centering
    \small
    \resizebox{0.74\textwidth}{!}{\begin{tabular}{|c|c|l|l|c|lllllll|} 
  \hline
  \multicolumn{5}{|c|}{  } & \multicolumn{7}{c|}{Predictors} \\ 
  \hline
 \textbf{Method} & \textbf{Abbr.} & \rot[90]{\footnotesize{Imputations}} & \rot[90]{\footnotesize{Dataset}} & \rot[90]{\footnotesize{Dependent}} & \rot[90]{\footnotesize{SubSector}}& \rot[90]{\footnotesize{Sector}} & \rot[90]{\footnotesize{LogRevenue}} & \rot[90]{\footnotesize{FirstActivity}} & \rot[90]{\footnotesize{Region}} & \rot[90]{\footnotesize{Market}} & \rot[90]{\footnotesize{Size}}\\
 \hline
 
 \multirow{10}{3.5cm}{(Sub)Sector Median}  & \multirow{10}{*}{Median}  & \multirow{10}{*}{0} & L  & $Y_{S1}$ & - & b & - & - & - & - & - \\ 
& & & M & $Y_{S1}$ & - & b & - & - & - & - & -\\
& & & S & $Y_{S1}$ & - & b & - & - & - & - &  - \\
& & & U & $Y_{S1}$ & - & b & - & - & - & - & - \\
& & & Total & $Y_{S1}$ & e & s & - & - & - & - &  - \\
& & & L & $Y_{S2}$ & - & e & - & - & - & - & - \\
& & & M & $Y_{S2}$ & - & e & - & - & - & - & - \\
& & & S & $Y_{S2}$ & - & e & - & - & - & - & -  \\
& & & U & $Y_{S2}$ & - & e & - & - & - & - & - \\
& & & Total & $Y_{S2}$ & e & s& - & - & - & - & -  \\ 
& & & Total & $Y_{S1+2}$ & e & - & - & - & - & - & -  \\ 
\hline
\multirow{3}{3.5cm}{ Linear Model} & \multirow{3}{*}{LM}  & \multirow{3}{*}{1} &  L  & $Y_{S1}$ & - & b & - b & b  & b  & - & - \\ 
& & & M & $Y_{S1}$ & - & b & b & - & - & - & - \\
& & & S & $Y_{S1}$ & - & b & b & b  & - & - & -  \\
\cline{1-3}
   \multirow{3}{3.5cm}{ MI Predictive Mean Matching }  &  \multirow{3}{*}{MIPmm}  & \multirow{3}{*}{5} &  U & $Y_{S1}$ & - & b & b & - & - & - & - \\
& & & Total & $Y_{S1}$ & e & s & b  & b & b & - & -  \\  
& & & L & $Y_{S2}$ & - & e & e & - & e  & - & - \\
& & & M & $Y_{S2}$ & - & e & e & - & e & - & - \\ \cline{1-3}
    \multirow{3}{*}{ MI Random Forest}  & \multirow{3}{*}{MIRF}  & \multirow{3}{*}{5} & S & $Y_{S2}$ & - & e & e & - & - & e & -  \\
& & & U & $Y_{S2}$ & - & e & e & - & - & - & - \\
& & & Total & $Y_{S2}$ & e & s & b  & - & b & - & b  \\ 
& & & Total & $Y_{S1+2}$ & e & - & e & - & e & - & e  \\ 
\hline
\hline
   \multirow{7}{3.5cm}{ Heckman's  ML Estimator }  &  \multirow{7}{*}{Hml}  & \multirow{7}{*}{1} &  L  & $Y_{S1}$ & - & b &  b & b  & b  & - & - \\ 
& & & M & $Y_{S1}$ & - & b & b & - & - & - & - \\
& & & S & $Y_{S1}$ & - & b & b & b  & - & - & -  \\
& & & U & $Y_{S1}$ & - & b & - & - & - & - & - \\
& & & Total & $Y_{S1}$ & e & s & - & - & - & - & -  \\
& & & L & $Y_{S2}$ & - & e & - & - & - & - & - \\
& & & M & $Y_{S2}$ & - & e & -  & - & - & - & - \\    \cline
{1-3}

      \multirow{7}{3.5cm}{ MI Heckman's  ML Estimator }  &  \multirow{7}{*}{MIHml}  & \multirow{7}{*}{1} & S & $Y_{S2}$ & - & e & - & - & - & - & -  \\
& & & U & $Y_{S2}$ & - & e & - & - & - & - & - \\
& & & Total & $Y_{S2}$ & e & s & - & - & - & - & -  \\ 
& & & Total & $Y_{S1+2}$ & e & - & - & - & - & - &  - \\
& & & L  & $R_{S1}$ & - & - &  b & b  & -  & b & - \\ 
& & & M & $R_{S1}$ & - & b & b & b & b & b &  -\\
& & & S & $R_{S1}$ & - & b & b & b  & b & - & -  \\ \cline{1-3}

 \multirow{7}{3.5cm}{ MI Heckman's  Two-Step Estimator }  &  \multirow{7}{*}{MIH2Step}  & \multirow{7}{*}{1} & U & $R_{S1}$ & -& - & b & b & b & b & - \\
& & & Total & $R_{S1}$ & e & s & b & b & b & b & b  \\
& & & L & $R_{S2}$ & - & - & e & e & e & e & - \\
& & & M & $R_{S2}$ & - & e & e & e & e & e & - \\  
& &  & S & $R_{S2}$ & - & e & e & - & e & - & -  \\
& & & U & $R_{S2}$ & - & - & e & e & e & e & - \\
& & & Total & $R_{S2}$ & e & s & b  & b & b & b & b  \\ 
& & & Total & $R_{S1+2}$ & e & - & e  & - & e & e & e  \\ 
   \hline
 \end{tabular}}

{\footnotesize Median was used only for prediction (not imputation). Otherwise, same set of predictors was used for imputation and prediction model. Each model included an intercept. $Y_{S1}$ Scope 1 outcome, $Y_{S2}$ Scope 2 outcome, $R_{S1}$ Scope 1 selection, $R_{S2}$ Scope 2 selection. Variables: $r$ selected for simulation, $e$ selected for Empirical study, $b$  selected for both.}
    \label{tab:methods}
    
 \end{table}

 \subsection{Simulation Prediction Accuracy: Total Scope 2}
  
\begin{table}[H]
\centering
\caption{$Total$:  Simulation results for the predicted missing Scope 2 emissions values.}
\label{tab:accuracy}
\resizebox{\textwidth}{!}{\begin{tabular}{|ll|rrrr|rrrr|rrrr|rrrr|}
  \hline
  && \multicolumn{4}{c|}{MAR} & \multicolumn{4}{c|}{Light MNAR} & \multicolumn{4}{c|}{Heavy MNAR} & \multicolumn{4}{c|}{NonHeckman}  \\

$\sigma^{2}_{\varepsilon}$  & Method & \small{RE} & \small{RMSE} & \small{CR} & \small{PI} & \small{RE} & \small{RMSE} & \small{CR} & \small{PI}& \small{RE} & \small{RMSE} & \small{CR} & \small{PI} & \small{RE} & \small{RMSE} & \small{CR} & \small{PI}  \\ 
  \hline
  
\multirow{7}{*}{1} & Median & 21.38 & 2.25 & - & - & 19.62 & 2.15 & - & - & 18.18 & 2.07 & - & - & 18.03 & 2.64 & - & - \\ 
 & LM & 8.74 & 1.00 & 87.68 & 3.08 & 9.06 & 1.10 & 83.21 & 3.05 & 11.21 & 1.37 & 69.54 & 2.95 & 8.54 & 1.18 & 78.63 & 2.98 \\ 
 & MIPmm & 8.76 & 1.00 & 95.16 & 3.95 & 9.03 & 1.10 & 92.12 & 3.87 & 11.15 & 1.36 & 80.49 & 3.59 & 8.57 & 1.19 & 88.22 & 3.72 \\ 
 & MIRF & 9.58 & 1.05 & 97.85 & 4.85 & 8.99 & 1.08 & 97.55 & 4.85 & 10.24 & 1.29 & 93.76 & 4.69 & 9.25 & 1.29 & 92.23 & 4.51 \\ 
 & Hml & 8.75 & 1.00 & 95.08 & 3.94 & 8.48 & 1.00 & 95.27 & 3.96 & 8.16 & 0.98 & 95.81 & 4.01 & 7.44 & 1.02 & 95.03 & 3.99 \\ 
 & MIHml & 8.73 & 1.00 & 95.39 & 3.99 & 8.48 & 1.00 & 95.60 & 4.02 & 8.16 & 0.98 & 96.29 & 4.12 & 7.44 & 1.02 & 96.00 & 4.21 \\ 
  & MIH2Step & 8.76 & 1.00 & 95.41 & 4.00 & 8.49 & 1.00 & 95.50 & 4.00 & 8.16 & 0.98 & 95.67 & 4.00 & 7.96 & 1.10 & 94.16 & 4.20 \\ 
\hline
\multirow{7}{*}{2.5} & Median & 27.07 & 2.57 & - & - & 24.89 & 2.44 & - & - & 23.18 & 2.38 & - & - & 22.17 & 3.24 & - & - \\ 
  & LM & 14.85 & 1.58 & 77.98 & 3.88 & 14.72 & 1.75 & 72.60 & 3.84 & 17.53 & 2.16 & 57.04 & 3.71 & 14.80 & 2.11 & 57.66 & 3.66 \\ 
  & MIPmm & 14.89 & 1.58 & 95.11 & 6.24 & 14.73 & 1.74 & 92.03 & 6.10 & 17.51 & 2.16 & 80.14 & 5.66 & 14.82 & 2.11 & 79.87 & 5.50 \\ 
  & MIRF & 15.79 & 1.61 & 96.48 & 6.80 & 14.80 & 1.70 & 95.26 & 6.75 & 16.60 & 2.06 & 88.34 & 6.43 & 15.44 & 2.21 & 81.58 & 5.93 \\ 
  & Hml & 14.85 & 1.58 & 85.78 & 4.64 & 14.38 & 1.58 & 86.06 & 4.67 & 13.76 & 1.57 & 86.78 & 4.72 & 11.79 & 1.68 & 83.67 & 4.71 \\ 
&  MIHml & 14.81 & 1.58 & 92.21 & 5.58 & 14.37 & 1.58 & 92.41 & 5.62 & 13.75 & 1.57 & 93.11 & 5.72 & 11.74 & 1.68 & 91.56 & 5.82 \\ 
 & MIH2Step & 14.88 & 1.58 & 95.41 & 6.32 & 14.39 & 1.58 & 95.46 & 6.32 & 13.72 & 1.56 & 95.61 & 6.31 & 12.47 & 1.79 & 92.86 & 6.55 \\ 
  \hline
\multirow{7}{*}{4} &   Median & 36.20 & 2.85 & - & - & 30.84 & 2.71 & - & - & 28.87 & 2.68 & - & - & 25.69 & 3.77 & - & - \\ 
 &  LM & 21.86 & 2.00 & 72.44 & 4.36 & 19.97 & 2.21 & 66.86 & 4.32 & 22.78 & 2.74 & 51.13 & 4.18 & 19.64 & 2.84 & 45.39 & 4.04 \\ 
 & MIPmm & 21.94 & 2.00 & 95.10 & 7.89 & 20.00 & 2.20 & 92.00 & 7.71 & 22.77 & 2.74 & 80.03 & 7.14 & 19.63 & 2.84 & 73.45 & 6.62 \\ 
 & MIRF & 23.23 & 2.03 & 95.95 & 8.30 & 20.26 & 2.16 & 94.30 & 8.21 & 21.92 & 2.62 & 86.19 & 7.77 & 20.20 & 2.92 & 74.07 & 6.89 \\ 
  & Hml & 21.83 & 2.00 & 79.59 & 5.08 & 20.12 & 2.00 & 79.79 & 5.10 & 19.27 & 1.99 & 80.43 & 5.16 & 14.93 & 2.20 & 75.26 & 5.16 \\ 
 & MIHml & 21.78 & 2.00 & 90.54 & 6.70 & 20.12 & 2.00 & 90.69 & 6.73 & 19.27 & 1.99 & 91.33 & 6.84 & 14.89 & 2.20 & 88.23 & 6.92 \\ 
 & MIH2Step & 21.89 & 2.00 & 95.41 & 8.00 & 20.18 & 1.99 & 95.47 & 8.00 & 19.23 & 1.97 & 95.59 & 7.98 & 15.87 & 2.35 & 91.39 & 8.21 \\ 
 \hline
\end{tabular}}
\newline
{\small RE =  Average relative error for all $Y_{miss}$ (\%) , PI = Average length of the prediction interval, RMSE = Root mean square error, CR = Coverage rate (\%)}
\end{table}
\newpage

\subsection{Simulation Parameter Estimation: Scope 2 Total}

\begin{figure}[H]
\centering
   \caption{$Total$ Scope 2: Boxplot over simulations over $\hat{\beta}_{Revenue}$.\\
   Red dotted lines denotes true value of the  
   coefficient.}
\begin{subfigure}{\textwidth}
\centering
    \includegraphics[width=0.6\textwidth]{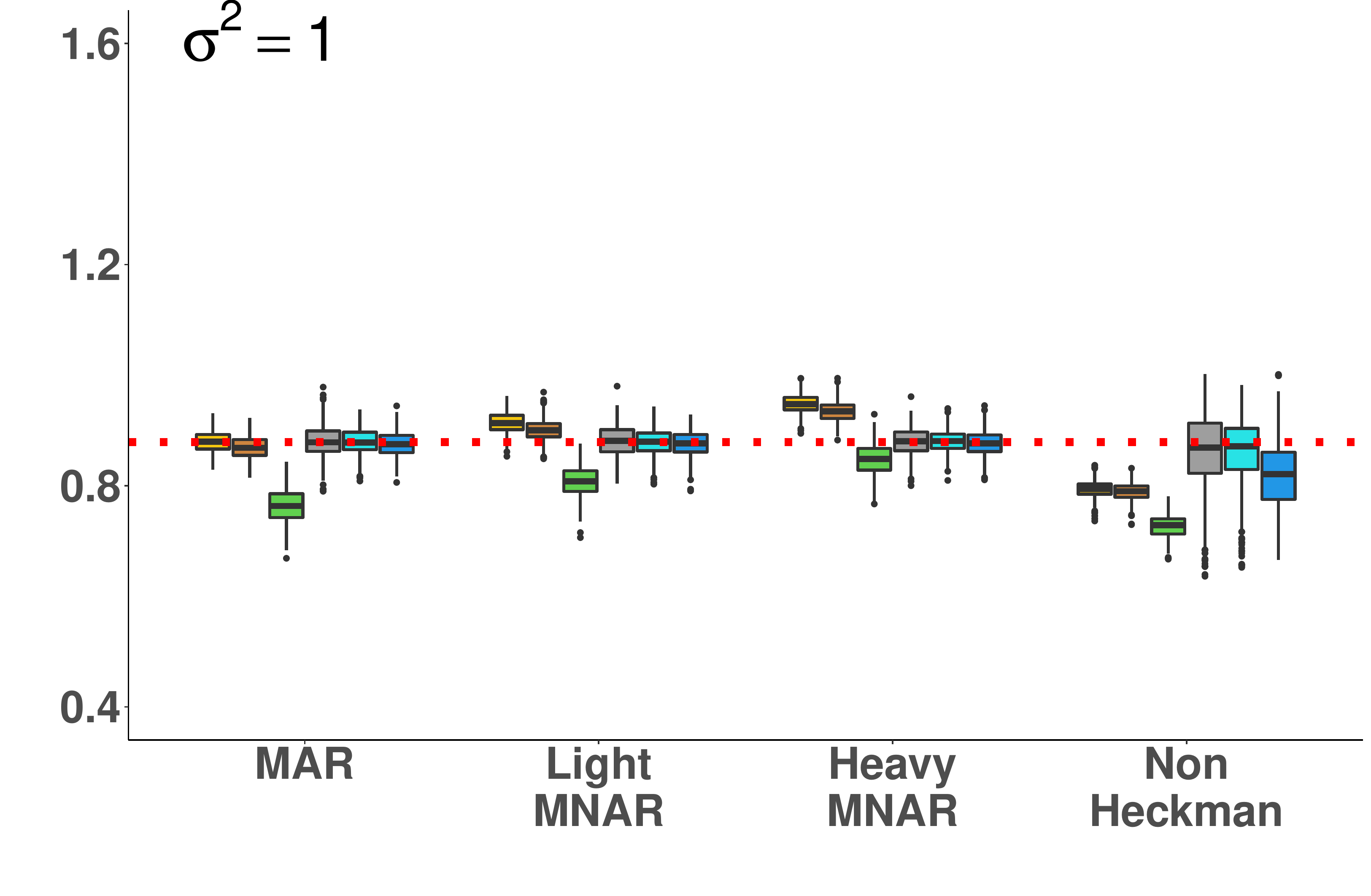}
  \end{subfigure}
  \begin{subfigure}{\textwidth}
  \centering
   \includegraphics[width=0.6\textwidth]{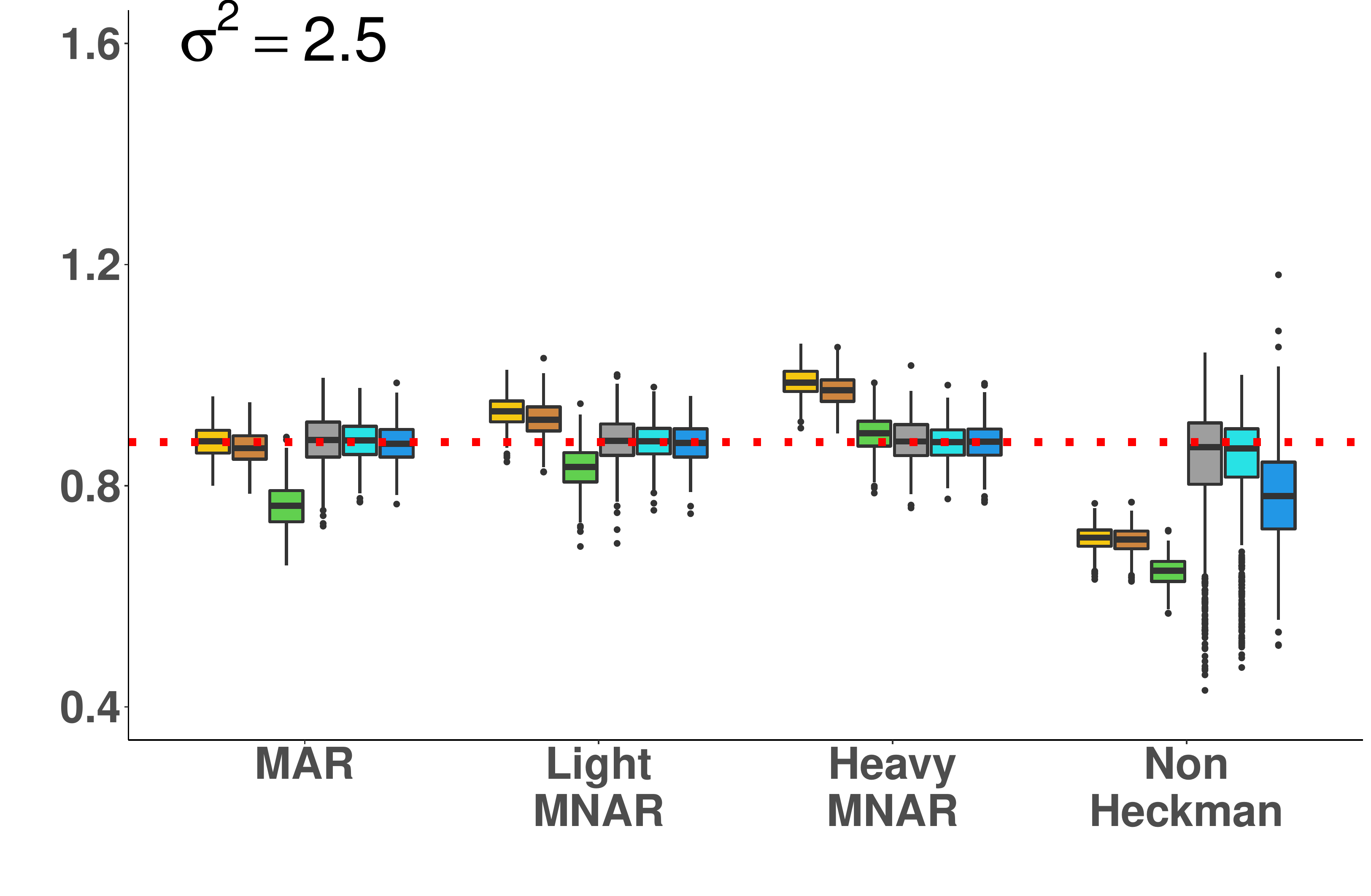}
 \end{subfigure}
  \begin{subfigure}{\textwidth}
  \centering
    \includegraphics[width=0.6\textwidth]{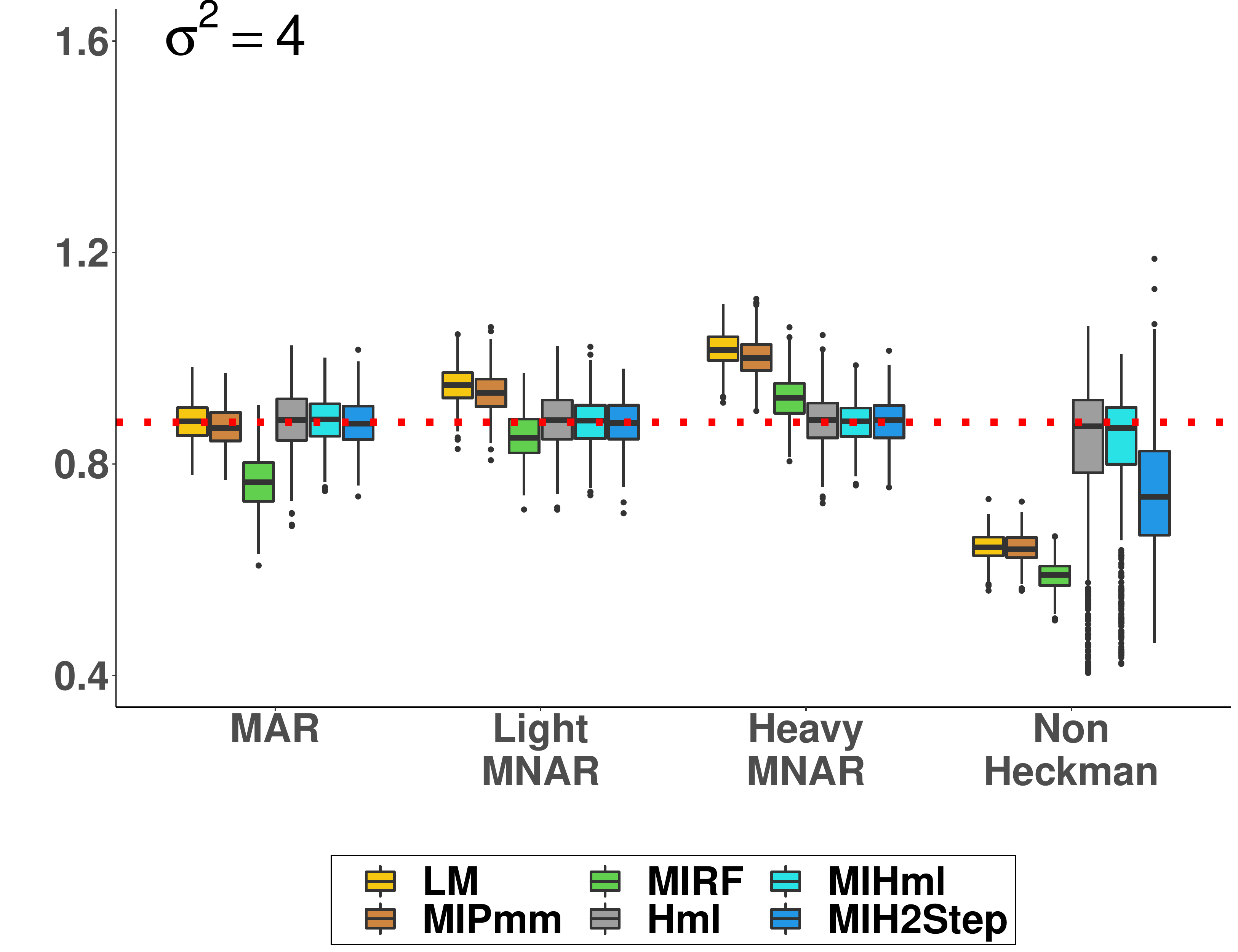}
    \end{subfigure}
    
 \label{fig:betas2}
  \end{figure}
  
  \subsection{Simulation Parameter Estimation: Sub-sets}

\begin{table}[H]
\caption{$L$ Scope 1: Simulation results for $\beta_{Revenue} = 0.875 $ estimates }


\resizebox{\textwidth}{!}{\begin{tabular}{|ll|rrrrrrr|rrrrrrr|rrrrrrr|}
  \hline
  
   &  & \multicolumn{7}{c|}{\Large \textbf{$\sigma^{2}_{\varepsilon}$ = 1}} & \multicolumn{7}{c|}{ \Large \textbf{$\sigma^{2}_{\varepsilon}$ = 2.5}} & \multicolumn{7}{c|}{ \Large \textbf{$\sigma^{2}_{\varepsilon}$ = 4}}  \\
  &&&&&&&&&&&&&&&&&&&&&& \\
 & Methods & Mean & Rbias & $SE_{m}$ & $SE_{e}$ & $RE_{SE}$ &  CR   & {\small RMSE } & Mean & Rbias & $SE_{m}$ & $SE_{e}$ & $RE_{SE}$ &  CR   & {\small RMSE } & Mean & Rbias & $SE_{m}$ & $SE_{e}$ & $RE_{SE}$ &  CR   & {\small RMSE } \\ 
  \hline

\multirow{5}{*}{\textbf{MAR}}   & LM & 0.87 & -0.56 & 2.02 & 4.74 & -57.38 & 79.20 & 3.16 & 0.87 & -0.56 & 3.15 & 6.13 & -48.61 & 79.40 & 4.85 & 0.87 & -0.56 & 3.98 & 7.07 & -43.71 & 79.80 & 6.09 \\ 
   & MIPmm & 0.86 & -1.71 & 3.06 & 4.78 & -35.98 & 88.60 & 3.81 & 0.85 & -2.85 & 4.75 & 6.33 & -24.96 & 91.40 & 5.39 & 0.85 & -2.85 & 6.01 & 7.48 & -19.65 & 92.00 & 6.68 \\ 
   & MIRF & 0.78 & -10.85 & 4.29 & 9.47 & -54.70 & 34.80 & 10.64 & 0.77 & -11.99 & 5.57 & 9.95 & -44.02 & 56.60 & 11.35 & 0.77 & -11.99 & 6.61 & 10.62 & -37.76 & 65.80 & 11.94 \\ 
   & Hml & 0.87 & -0.56 & 2.46 & 6.47 & -61.98 & 61.40 & 5.30 & 0.87 & -0.56 & 3.62 & 8.58 & -57.81 & 64.80 & 7.58 & 0.87 & -0.56 & 4.46 & 10.23 & -56.40 & 62.40 & 9.43 \\ 
   & MIHml & 0.88 & 0.58 & 4.38 & 6.29 & -30.37 & 90.40 & 4.71 & 0.87 & -0.56 & 6.36 & 8.05 & -20.99 & 90.40 & 6.74 & 0.87 & -0.56 & 7.82 & 9.38 & -16.63 & 89.00 & 8.32 \\ 
   & MIH2Step & 0.87 & -0.56 & 4.35 & 5.49 & -20.77 & 92.80 & 4.25 & 0.87 & -0.56 & 6.80 & 7.62 & -10.76 & 92.40 & 6.55 & 0.87 & -0.56 & 8.57 & 9.10 & -5.82 & 92.20 & 8.23 \\ 
   \hline
   & LM & 0.90 & 2.87 & 1.99 & 5.40 & -63.15 & 59.40 & 4.31 & 0.92 & 5.15 & 3.10 & 6.66 & -53.45 & 59.20 & 6.82 & 0.94 & 7.44 & 3.91 & 8.07 & -51.55 & 58.00 & 8.63 \\ 
\textbf{Light}   & MIPmm & 0.89 & 1.72 & 2.98 & 4.81 & -38.05 & 89.40 & 3.52 & 0.91 & 4.01 & 4.73 & 6.10 & -22.46 & 88.40 & 5.91 & 0.92 & 5.15 & 5.96 & 7.48 & -20.32 & 86.00 & 7.62 \\ 
 \textbf{MNAR}  & MIRF & 0.81 & -7.42 & 4.32 & 7.54 & -42.71 & 69.80 & 7.31 & 0.84 & -3.99 & 5.64 & 7.69 & -26.66 & 88.80 & 6.62 & 0.85 & -2.85 & 6.55 & 8.42 & -22.21 & 92.20 & 6.97 \\ 
   & Hml & 0.87 & -0.56 & 2.45 & 5.48 & -55.29 & 69.20 & 4.87 & 0.88 & 0.58 & 3.62 & 8.16 & -55.64 & 65.20 & 7.76 & 0.88 & 0.58 & 4.45 & 9.86 & -54.87 & 69.60 & 9.34 \\ 
   & MIHml & 0.88 & 0.58 & 4.04 & 5.36 & -24.63 & 89.60 & 4.50 & 0.88 & 0.58 & 5.96 & 6.89 & -13.50 & 92.00 & 6.33 & 0.88 & 0.58 & 7.30 & 8.48 & -13.92 & 90.60 & 7.89 \\ 
   & MIH2Step & 0.87 & -0.56 & 4.33 & 5.27 & -17.84 & 91.80 & 4.38 & 0.87 & -0.56 & 6.79 & 7.26 & -6.47 & 92.40 & 6.51 & 0.87 & -0.56 & 8.63 & 9.32 & -7.40 & 92.80 & 8.40 \\ 
   \hline
   & LM & 0.94 & 7.44 & 1.88 & 6.44 & -70.81 & 16.60 & 7.14 & 0.98 & 12.01 & 2.94 & 9.19 & -68.01 & 16.40 & 11.37 & 1.01 & 15.44 & 3.70 & 10.96 & -66.24 & 15.00 & 14.44 \\ 
 \textbf{Heavy}  & MIPmm & 0.92 & 5.15 & 2.87 & 5.50 & -47.82 & 56.40 & 5.72 & 0.96 & 9.72 & 4.40 & 8.21 & -46.41 & 47.60 & 9.85 & 0.99 & 13.15 & 5.56 & 9.88 & -43.72 & 43.20 & 12.80 \\ 
  \textbf{MNAR} & MIRF & 0.85 & -2.85 & 4.32 & 6.30 & -31.43 & 91.60 & 4.52 & 0.89 & 1.72 & 5.37 & 7.04 & -23.72 & 94.20 & 5.31 & 0.92 & 5.15 & 6.41 & 8.50 & -24.59 & 89.00 & 7.69 \\ 
   & Hml & 0.88 & 0.58 & 2.47 & 5.06 & -51.19 & 78.40 & 3.93 & 0.88 & 0.58 & 3.61 & 7.89 & -54.25 & 74.40 & 6.21 & 0.89 & 1.72 & 4.43 & 10.38 & -57.32 & 73.20 & 7.96 \\ 
   & MIHml & 0.88 & 0.58 & 3.59 & 4.88 & -26.43 & 94.20 & 3.57 & 0.88 & 0.58 & 5.19 & 7.63 & -31.98 & 91.20 & 5.59 & 0.89 & 1.72 & 6.28 & 9.50 & -33.89 & 90.40 & 6.83 \\ 
   & MIH2Step & 0.87 & -0.56 & 4.25 & 5.56 & -23.56 & 93.80 & 4.03 & 0.87 & -0.56 & 6.61 & 9.09 & -27.28 & 94.00 & 6.43 & 0.87 & -0.56 & 8.18 & 11.41 & -28.31 & 94.00 & 7.77 \\ 
   \hline
   & LM & 0.81 & -7.42 & 1.90 & 4.08 & -53.43 & 17.60 & 6.92 & 0.74 & -15.42 & 2.81 & 6.31 & -55.47 & 3.60 & 13.82 & 0.69 & -21.14 & 3.39 & 8.48 & -60.02 & 1.00 & 18.95 \\ 
 \textbf{Non-}  & MIPmm & 0.81 & -7.42 & 2.90 & 4.16 & -30.29 & 39.80 & 7.15 & 0.74 & -15.42 & 4.42 & 6.49 & -31.90 & 14.20 & 14.13 & 0.69 & -21.14 & 5.37 & 8.89 & -39.60 & 6.60 & 19.56 \\ 
 \textbf{Heckman}  & MIRF & 0.74 & -15.42 & 3.73 & 7.60 & -50.92 & 3.60 & 13.84 & 0.67 & -23.42 & 4.65 & 9.95 & -53.27 & 0.60 & 20.47 & 0.63 & -27.99 & 5.51 & 12.34 & -55.35 & 0.20 & 25.44 \\ 
  & Hml & 0.82 & -6.28 & 2.49 & 9.73 & -74.41 & 33.80 & 10.44 & 0.79 & -9.71 & 3.58 & 14.83 & -75.86 & 41.80 & 15.81 & 0.78 & -10.85 & 4.28 & 16.40 & -73.90 & 47.80 & 17.60 \\ 
   & MIHml & 0.82 & -6.28 & 6.02 & 9.25 & -34.92 & 67.60 & 10.07 & 0.77 & -11.99 & 7.58 & 13.11 & -42.18 & 63.60 & 15.88 & 0.76 & -13.14 & 7.87 & 16.10 & -51.12 & 62.00 & 18.60 \\ 
  & MIH2Step & 0.82 & -6.28 & 14.39 & 12.72 & 13.13 & 87.80 & 13.43 & 0.81 & -7.42 & 24.44 & 22.10 & 10.59 & 87.00 & 22.32 & 0.79 & -9.71 & 32.30 & 28.16 & 14.70 & 92.00 & 28.43 \\ 
   \hline
\end{tabular}}

\centering
{\small Mean = Average estimate $\hat{\beta}_{Revenue}$ (rounded to 2 digits), Rbias =  Average relative bias (\%), $SE_{m}$ = Root mean square of the estimated standard errors   $ \times 10^{2}$, $SE_{e}$ = Empirical standard error $ \times  10^{2}$, CR = Coverage rate (based on 95\% confidence interval), $RE_{SE}$ = relative error of $SE_{m}$(\%), RMSE = Root mean square error $ \times  10^{2}$.}
\end{table}

\begin{table}[H]
\centering
\caption{\textit{L} Scope 1: Median relative bias (\% Rbias) for variable coefficients estimates across all the categories of selected predictors}
 $\sigma^{2}_{\varepsilon}$ =1
 
\resizebox{1.0 \textwidth}{!}{\begin{tabular}{|l|rrrr|rrrr|rrrr|}
  \hline
  & \multicolumn{4}{c|}{Sector (40 categories)} & \multicolumn{4}{c|}{Region (8 categories)}& \multicolumn{4}{c|}{First Activity (4 categories)}\\
  \hline
 & MAR &\multicolumn{1}{c}{Light} & Heavy & \multicolumn{1}{c|}{Non-} & MAR &\multicolumn{1}{c}{Light} & Heavy &  \multicolumn{1}{c|}{Non-} & MAR &\multicolumn{1}{c}{Light} & Heavy &  \multicolumn{1}{c|}{Non-} \\
 & & MNAR & MNAR & Heckman & & MNAR & MNAR & Heckman & & MNAR & MNAR & Heckman\\
 
  \hline
LM & 14.06 & 15.68 & 15.87 & -8.14 & -3.42 & 0.30 & 5.88 & 1.65 & 2.14 & 16.82 & 30.57 & -7.94 \\ 
  MIPmm & 13.66 & 15.20 & 15.36 & -8.20 & -6.07 & -8.45 & -6.58 & 3.51 & 1.93 & 17.78 & 32.22 & -7.86 \\ 
  MIRF & 4.53 & 5.90 & 6.27 & -19.24 & -28.20 & -31.73 & -39.08 & -13.84 & -17.46 & -16.11 & -7.43 & -22.23 \\ 
  Hml & 14.14 & 15.20 & 15.34 & -7.83 & -1.67 & -2.01 & -0.70 & 10.93 & -3.05 & -3.72 & -3.47 & -5.66 \\ 
  MIHml & 14.18 & 15.53 & 15.57 & -7.84 & -4.71 & -4.02 & -4.54 & 5.25 & 2.11 & 2.03 & 2.28 & -7.74 \\ 
  MIH2Step & 13.87 & 15.38 & 15.48 & -8.18 & -3.18 & -3.49 & -3.18 & 2.93 & 2.54 & 2.04 & 2.26 & -6.33 \\ 
  \hline
 
\end{tabular}}

\end{table}
  
\begin{table}[H]
\centering
\caption{\textit{L} Scope 1: Median relative bias (\% Rbias) for variable coefficients estimates across all the categories of selected predictors}
 $\sigma^{2}_{\varepsilon}$ =2.5
 
\resizebox{1.0 \textwidth}{!}{\begin{tabular}{|l|rrrr|rrrr|rrrr|}
  \hline
  & \multicolumn{4}{c|}{Sector (40 categories)} & \multicolumn{4}{c|}{Region (8 categories)}& \multicolumn{4}{c|}{First Activity  (4 categories)}\\
  \hline
 & MAR &\multicolumn{1}{c}{Light} & Heavy & \multicolumn{1}{c|}{Non-} & MAR &\multicolumn{1}{c}{Light} & Heavy &  \multicolumn{1}{c|}{Non-} & MAR &\multicolumn{1}{c}{Light} & Heavy &  \multicolumn{1}{c|}{Non-} \\
 & & MNAR & MNAR & Heckman & & MNAR & MNAR & Heckman & & MNAR & MNAR & Heckman\\
  \hline
LM & 14.32 & 16.59 & 16.32 & -13.00 & -2.95 & 2.73 & 8.93 & -3.22 & 2.53 & 25.46 & 47.22 & -15.63 \\ 
  MIPmm & 13.87 & 16.16 & 15.61 & -12.88 & -4.29 & -9.34 & 1.44 & -1.78 & 2.93 & 27.15 & 48.95 & -17.47 \\ 
  MIRF & 4.33 & 5.97 & 4.66 & -22.90 & -28.30 & -36.16 & -47.68 & -19.20 & -16.41 & -14.83 & 32.97 & -28.50 \\ 
  Hml & 14.41 & 16.07 & 15.29 & -12.73 & 2.52 & 3.60 & -2.10 & 4.43 & -5.18 & -5.85 & -2.69 & -8.66 \\ 
  MIHml & 14.48 & 16.44 & 15.89 & -11.30 & -2.90 & -5.32 & -3.85 & -3.60 & 2.67 & 2.14 & 3.11 & -11.75 \\ 
  MIH2Step & 14.09 & 16.13 & 15.68 & -9.20 & -4.17 & -3.00 & -2.72 & 2.79 & 3.31 & 2.71 & 2.40 & -7.10 \\ 
   
   \hline
\end{tabular}}
\label{tab:allbeta}
\end{table}

\begin{table}[H]
\centering
\caption{\textit{L} Scope 1: Median relative bias (\% Rbias) for variable coefficients estimates across all the categories of selected predictors}
 $\sigma^{2}_{\varepsilon}$ = 4
 
\resizebox{1.0 \textwidth}{!}{\begin{tabular}{|l|rrrr|rrrr|rrrr|}
  \hline
  & \multicolumn{4}{c|}{Sector (40 categories)} & \multicolumn{4}{c|}{Region (8 categories)}& \multicolumn{4}{c|}{First Activity(4 categories)}\\
  \hline
 & MAR &\multicolumn{1}{c}{Light} & Heavy & \multicolumn{1}{c|}{Non-} & MAR &\multicolumn{1}{c}{Light} & Heavy &  \multicolumn{1}{c|}{Non-} & MAR &\multicolumn{1}{c}{Light} & Heavy &  \multicolumn{1}{c|}{Non-} \\
 & & MNAR & MNAR & Heckman & & MNAR & MNAR & Heckman & & MNAR & MNAR & Heckman\\
  \hline
LM & 14.61 & 17.41 & 16.78 & -14.31 & -2.64 & 3.04 & 12.79 & -6.68 & 2.83 & 31.77 & 59.13 & -22.21 \\ 
  MIPmm & 14.58 & 16.82 & 15.95 & -14.10 & -7.85 & -10.65 & 4.66 & -9.95 & 3.36 & 33.62 & 60.74 & -22.52 \\ 
  MIRF & 4.22 & 5.85 & 4.40 & -24.56 & -30.62 & -37.63 & -54.48 & -23.76 & -16.18 & -5.07 & 44.40 & -34.83 \\ 
  Hml & 14.66 & 17.16 & 15.61 & -12.38 & 3.22 & -0.96 & -2.62 & -7.65 & -6.50 & -4.91 & 0.59 & -10.86 \\ 
  MIHml & 14.83 & 16.74 & 16.06 & -10.74 & -2.17 & -3.98 & -5.59 & -3.38 & 2.94 & 3.34 & 3.01 & -17.15 \\ 
  MIH2Step & 14.34 & 16.78 & 16.13 & -8.80 & -4.97 & -3.17 & -2.54 & 4.06 & 3.90 & 2.61 & 3.00 & -9.22 \\ 
   
   \hline
\end{tabular}}
\label{tab:allbet}
\end{table}

\textbf{Comment:} For Sector variable, only in Non-Heckman scenario, when $\sigma^{2}_{\varepsilon} \neq 1$ MIHml, Hml and MIH2Step had lower median relative bias than other methods . For Region variable, MIHml and MIH2step tended to have lower lower median relative bias than MIRF and LM in Heavy MNAR and Non-Heckman scenario. For FirstActivity variable, in Light and Heavy MNAR scenario Hml, MIHml and MIH2Step tended to perform better than LM, MIPmm and MIRF.

\begin{table}[ht]
\caption{$M$ Scope 1: Simulation results for $\beta_{Revenue} = 1.077 $ estimates }


\resizebox{\textwidth}{!}{\begin{tabular}{|ll|rrrrrrr|rrrrrrr|rrrrrrr|}
  \hline
  
   &  & \multicolumn{7}{c|}{\Large \textbf{$\sigma^{2}_{\varepsilon}$ = 1}} & \multicolumn{7}{c|}{ \Large \textbf{$\sigma^{2}_{\varepsilon}$ = 2.5}} & \multicolumn{7}{c|}{ \Large \textbf{$\sigma^{2}_{\varepsilon}$ = 4}}  \\
  &&&&&&&&&&&&&&&&&&&&&& \\
 & Methods & Mean & Rbias & $SE_{m}$ & $SE_{e}$ & $RE_{SE}$ &  CR   & {\small RMSE } & Mean & Rbias & $SE_{m}$ & $SE_{e}$ & $RE_{SE}$ &  CR   & {\small RMSE } & Mean & Rbias & $SE_{m}$ & $SE_{e}$ & $RE_{SE}$ &  CR   & {\small RMSE } \\ 
  \hline

\multirow{5}{*}{\textbf{MAR}}  & LM & 1.08 & 0.26 & 1.61 & 8.95 & -82.01 & 66.20 & 3.34 & 1.08 & 0.26 & 2.55 & 9.56 & -73.33 & 66.40 & 5.28 & 1.08 & 0.26 & 3.23 & 10.42 & -69.00 & 66.40 & 6.67 \\ 
   & MIPmm & 1.04 & -3.45 & 3.33 & 7.85 & -57.58 & 72.80 & 5.27 & 1.04 & -3.45 & 5.17 & 8.80 & -41.25 & 84.40 & 6.80 & 1.04 & -3.45 & 6.57 & 9.86 & -33.37 & 87.60 & 7.99 \\ 
   & MIRF & 0.88 & -18.31 & 5.63 & 17.74 & -68.26 & 7.60 & 19.99 & 0.89 & -17.38 & 6.59 & 18.16 & -63.71 & 18.20 & 20.13 & 0.89 & -17.38 & 7.59 & 18.61 & -59.22 & 29.20 & 20.38 \\ 
   & Hml & 1.07 & -0.67 & 2.23 & 9.59 & -76.75 & 54.40 & 5.58 & 1.07 & -0.67 & 3.21 & 11.22 & -71.39 & 54.80 & 8.47 & 1.07 & -0.67 & 3.91 & 12.88 & -69.64 & 52.20 & 10.53 \\ 
   & MIHml & 1.08 & 0.26 & 4.49 & 9.56 & -53.03 & 92.00 & 4.52 & 1.08 & 0.26 & 6.68 & 10.78 & -38.03 & 89.60 & 7.18 & 1.09 & 1.19 & 8.26 & 12.12 & -31.85 & 89.60 & 9.07 \\ 
   & MIH2Step & 1.07 & -0.67 & 4.46 & 9.35 & -52.30 & 92.20 & 4.28 & 1.08 & 0.26 & 7.04 & 10.65 & -33.90 & 92.60 & 6.74 & 1.08 & 0.26 & 8.91 & 12.03 & -25.94 & 92.80 & 8.52 \\ 
   \hline
   & LM & 1.13 & 4.90 & 1.58 & 9.57 & -83.49 & 26.00 & 5.91 & 1.16 & 7.69 & 2.50 & 11.26 & -77.80 & 26.40 & 9.37 & 1.18 & 9.54 & 3.16 & 12.75 & -75.22 & 25.20 & 11.85 \\ 
 \textbf{Light}  & MIPmm & 1.08 & 0.26 & 3.18 & 7.29 & -56.38 & 90.40 & 3.61 & 1.11 & 3.04 & 5.13 & 8.71 & -41.10 & 84.60 & 6.56 & 1.13 & 4.90 & 6.41 & 10.12 & -36.66 & 80.00 & 8.95 \\ 
 \textbf{MNAR}  & MIRF & 0.93 & -13.67 & 5.87 & 15.31 & -61.66 & 26.40 & 15.36 & 0.96 & -10.88 & 6.68 & 14.85 & -55.02 & 54.60 & 13.40 & 0.98 & -9.02 & 7.77 & 14.64 & -46.93 & 71.20 & 12.39 \\ 
   & Hml & 1.07 & -0.67 & 2.26 & 8.10 & -72.10 & 59.80 & 5.30 & 1.07 & -0.67 & 3.23 & 10.08 & -67.96 & 58.00 & 8.06 & 1.07 & -0.67 & 3.93 & 11.88 & -66.92 & 56.80 & 9.94 \\ 
   & MIHml & 1.08 & 0.26 & 4.38 & 7.74 & -43.41 & 91.20 & 4.44 & 1.08 & 0.26 & 6.51 & 9.03 & -27.91 & 89.40 & 6.80 & 1.08 & 0.26 & 8.03 & 10.52 & -23.67 & 89.20 & 8.77 \\ 
   & MIH2Step & 1.08 & 0.26 & 4.32 & 7.91 & -45.39 & 91.80 & 4.17 & 1.08 & 0.26 & 6.93 & 9.22 & -24.84 & 92.00 & 6.71 & 1.08 & 0.26 & 8.59 & 10.61 & -19.04 & 92.80 & 8.51 \\ 
   \hline
   & LM & 1.17 & 8.61 & 1.48 & 11.18 & -86.76 & 1.60 & 10.17 & 1.23 & 14.18 & 2.34 & 14.59 & -83.96 & 2.00 & 16.12 & 1.27 & 17.90 & 2.96 & 17.33 & -82.92 & 1.40 & 20.41 \\ 
\textbf{Heavy}   & MIPmm & 1.13 & 4.90 & 3.02 & 7.76 & -61.08 & 60.60 & 6.03 & 1.18 & 9.54 & 4.71 & 10.74 & -56.15 & 42.80 & 11.62 & 1.22 & 13.26 & 5.95 & 13.25 & -55.09 & 36.40 & 15.75 \\ 
 \textbf{MNAR}   & MIRF & 0.98 & -9.02 & 5.86 & 13.55 & -56.75 & 54.80 & 11.36 & 1.03 & -4.38 & 6.79 & 12.54 & -45.85 & 86.60 & 7.88 & 1.07 & -0.67 & 7.65 & 12.47 & -38.65 & 94.00 & 7.23 \\ 
   & Hml & 1.07 & -0.67 & 2.32 & 7.73 & -69.99 & 66.60 & 4.69 & 1.07 & -0.67 & 3.30 & 10.70 & -69.16 & 65.20 & 7.05 & 1.07 & -0.67 & 3.99 & 13.63 & -70.73 & 59.80 & 8.88 \\ 
   & MIHml & 1.08 & 0.26 & 4.01 & 6.94 & -42.22 & 93.20 & 3.86 & 1.08 & 0.26 & 5.87 & 9.69 & -39.42 & 92.20 & 6.16 & 1.08 & 0.26 & 7.17 & 11.86 & -39.54 & 90.20 & 7.74 \\ 
   & MIH2Step & 1.08 & 0.26 & 4.29 & 7.22 & -40.58 & 92.40 & 4.05 & 1.09 & 1.19 & 6.53 & 9.81 & -33.44 & 92.80 & 6.54 & 1.09 & 1.19 & 8.26 & 12.09 & -31.68 & 92.80 & 8.34 \\ 
   \hline
   & LM & 0.99 & -8.10 & 1.53 & 5.75 & -73.39 & 1.00 & 9.66 & 0.89 & -17.38 & 2.28 & 7.42 & -69.27 & 0.00 & 19.46 & 0.82 & -23.88 & 2.75 & 10.15 & -72.91 & 0.00 & 26.41 \\ 
  \textbf{Non-} & MIPmm & 0.98 & -9.02 & 2.99 & 5.49 & -45.54 & 12.40 & 10.02 & 0.89 & -17.38 & 4.45 & 7.42 & -40.03 & 1.40 & 19.74 & 0.82 & -23.88 & 5.40 & 10.02 & -46.11 & 0.00 & 26.72 \\ 
  \textbf{Heckman} & MIRF & 0.88 & -18.31 & 4.79 & 11.74 & -59.20 & 1.20 & 20.10 & 0.79 & -26.66 & 5.36 & 13.14 & -59.21 & 0.60 & 28.67 & 0.74 & -31.30 & 5.93 & 15.37 & -61.42 & 0.00 & 34.69 \\ 
   & Hml & 1.02 & -5.31 & 2.35 & 14.49 & -83.78 & 29.20 & 13.99 & 0.95 & -11.81 & 3.31 & 20.55 & -83.89 & 34.60 & 22.82 & 0.94 & -12.74 & 3.94 & 24.67 & -84.03 & 40.40 & 25.79 \\ 
   & MIHml & 1.02 & -5.31 & 7.07 & 13.55 & -47.82 & 67.40 & 13.04 & 0.96 & -10.88 & 9.53 & 19.49 & -51.10 & 62.20 & 20.83 & 0.95 & -11.81 & 10.31 & 23.25 & -55.66 & 63.60 & 23.72 \\ 
   & MIH2Step & 0.99 & -8.10 & 9.48 & 10.10 & -6.14 & 78.20 & 13.03 & 0.93 & -13.67 & 14.01 & 14.91 & -6.04 & 75.00 & 19.83 & 0.88 & -18.31 & 17.07 & 17.70 & -3.56 & 71.20 & 25.16 \\ 
   \hline
\end{tabular}}

\centering
{\small Mean = Average estimate $\hat{\beta}_{Revenue}$ (rounded to 2 digits), Rbias =  Average relative bias (\%), $SE_{m}$ = Root mean square of the estimated standard errors $ \times  10^{2}$, $SE_{e}$ = Empirical standard error $\times  10^{2}$, CR = Coverage rate (based on 95\% confidence interval), $RE_{SE}$ = relative error of $SE_{m}$(\%), RMSE = Root mean square error  $ \times 10^{2}$.}
\end{table}

\begin{table}[ht]
\centering
 \caption{\textit{M}: Median relative bias (\% Rbias) for estimated coefficients across all categories}
 $\sigma^{2}_{\varepsilon}$ =1 \\
\resizebox{0.8\textwidth}{!}{\begin{tabular}{|l|rrrr|rrrr|}
  \hline
 Variable    & \multicolumn{4}{c|}{Sector (39 categories)} & \multicolumn{4}{c|}{ Market (4 categories)}\\
  \hline
 & MAR &\multicolumn{1}{c}{Light} & Heavy & \multicolumn{1}{c|}{Non-} & MAR &\multicolumn{1}{c}{Light} & Heavy &  \multicolumn{1}{c|}{Non-}  \\
Method & & MNAR & MNAR & Heckman & & MNAR & MNAR & Heckman \\
 \hline
LM & 1.40 & 1.61 & 2.12 & 1.00 & 0.76 & -8.88 & -17.84 & -7.45 \\ 
  MIPmm & 0.00 & 1.10 & 1.82 & 1.31 & -3.71 & -13.50 & -23.25 & -7.10 \\ 
  MIRF & -16.00 & -15.37 & -13.70 & -11.68 & -40.34 & -46.67 & -52.77 & -32.52 \\ 
  Hml & 0.36 & 0.26 & 0.96 & 0.29 & 1.20 & 1.04 & 0.46 & -5.16 \\ 
  MIHml & 1.61 & 1.37 & 1.75 & 0.80 & 0.76 & -0.38 & -0.52 & -4.18 \\ 
  MIH2Step & 1.17 & 1.03 & 1.69 & 0.90 & -0.79 & -0.81 & -2.42 & -7.49 \\ 
   \hline
\end{tabular}}

\vspace{3mm}
 $\sigma^{2}_{\varepsilon}$ =2.5 \\
\resizebox{0.8\textwidth}{!}{\begin{tabular}{|l|rrrr|rrrr|}
  \hline
 Variable    & \multicolumn{4}{c|}{Sector (39 categories)} & \multicolumn{4}{c|}{ Market (4 categories)}\\
  \hline
 & MAR &\multicolumn{1}{c}{Light} & Heavy & \multicolumn{1}{c|}{Non-} & MAR &\multicolumn{1}{c}{Light} & Heavy &  \multicolumn{1}{c|}{Non-}  \\
Method & & MNAR & MNAR & Heckman & & MNAR & MNAR & Heckman \\
 \hline
LM & 1.37 & 2.13 & 3.20 & 2.82 & 1.04 & -14.09 & -28.01 & -15.55 \\ 
  MIPmm & 0.54 & 0.67 & 2.35 & 2.79 & -3.40 & -18.99 & -33.53 & -14.20 \\ 
  MIRF & -15.38 & -14.07 & -10.37 & -10.68 & -38.11 & -47.90 & -54.56 & -37.88 \\ 
  Hml & 0.32 & 0.01 & 1.34 & 0.59 & 2.47 & 1.17 & 0.91 & -10.71 \\ 
  MIHml & 1.09 & 0.72 & 1.38 & 1.43 & 1.10 & 0.22 & -0.95 & -7.99 \\ 
  MIH2Step & 0.51 & 0.30 & 1.37 & 1.93 & -1.34 & -2.19 & -4.68 & -11.44 \\ 
   \hline
\end{tabular}}

\vspace{3mm}

 $\sigma^{2}_{\varepsilon}$ =4 \\
\resizebox{0.8\textwidth}{!}{\begin{tabular}{|l|rrrr|rrrr|}
  \hline
 Variable    & \multicolumn{4}{c|}{Sector (39 categories)} & \multicolumn{4}{c|}{ Market (4 categories)}\\
  \hline
 & MAR &\multicolumn{1}{c}{Light} & Heavy & \multicolumn{1}{c|}{Non-} & MAR &\multicolumn{1}{c}{Light} & Heavy &  \multicolumn{1}{c|}{Non-}  \\
Method & & MNAR & MNAR & Heckman & & MNAR & MNAR & Heckman \\
 \hline
LM & 0.90 & 2.38 & 3.48 & 3.83 & 1.24 & -18.32 & -34.98 & -22.44 \\ 
  MIPmm & 0.38 & 1.35 & 2.84 & 3.62 & -2.85 & -23.33 & -40.05 & -21.20 \\ 
  MIRF & -15.04 & -12.84 & -9.72 & -11.02 & -36.03 & -48.47 & -56.28 & -44.12 \\ 
  Hml & 0.46 & 0.52 & 1.01 & 1.17 & 5.86 & 2.30 & 1.48 & -12.62 \\ 
  MIHml & 0.78 & 0.86 & 1.59 & 1.19 & 0.32 & -1.09 & -0.58 & -11.61 \\ 
  MIH2Step & 0.47 & 0.59 & 1.19 & 2.95 & -1.73 & -2.18 & -4.43 & -15.83 \\ 
   \hline
\end{tabular}}
\label{tab:medianM}

\end{table}

\textbf{Comment:} For Sector variable, Hml, MIHml and MIH2Step had lower median relative bias than other methods across all MNAR scenarios  and values of $\sigma^{2}_{\varepsilon}$. For Market, MIHml and MIH2step performed better than other methods across all the MNAR scenarios and values of $\sigma^{2}_{\varepsilon}$. MIH2Step had lower median relative bias than other methods in Heavy MNAR  and Non-Heckman scenarios.


\begin{table}[ht]
\caption{$S$ Scope 1: Simulation results for $\beta_{Revenue} = 1.165 $ estimates }
\label{tab:BetaTot}


\resizebox{\textwidth}{!}{\begin{tabular}{|ll|rrrrrrr|rrrrrrr|rrrrrrr|}
  \hline
  
   &  & \multicolumn{7}{c|}{\Large \textbf{$\sigma^{2}_{\varepsilon}$ = 1}} & \multicolumn{7}{c|}{ \Large \textbf{$\sigma^{2}_{\varepsilon}$ = 2.5}} & \multicolumn{7}{c|}{ \Large \textbf{$\sigma^{2}_{\varepsilon}$ = 4}}  \\
  &&&&&&&&&&&&&&&&&&&&&& \\
 & Methods & Mean & Rbias & $SE_{m}$ & $SE_{e}$ & $RE_{SE}$ &  CR   & {\small RMSE } & Mean & Rbias & $SE_{m}$ & $SE_{e}$ & $RE_{SE}$ &  CR   & {\small RMSE } & Mean & Rbias & $SE_{m}$ & $SE_{e}$ & $RE_{SE}$ &  CR   & {\small RMSE } \\ 
  \hline

\multirow{5}{*}{\textbf{MAR}}   & LM & 1.06 & -9.05 & 0.89 & 10.45 & -91.48 & 1.80 & 11.84 & 1.16 & -0.47 & 1.40 & 13.11 & -89.32 & 28.80 & 7.09 & 1.16 & -0.47 & 1.78 & 13.97 & -87.26 & 28.80 & 8.97 \\ 
   & MIPmm & 1.01 & -13.34 & 4.51 & 9.54 & -52.73 & 10.80 & 15.94 & 1.12 & -3.90 & 7.22 & 12.21 & -40.87 & 82.60 & 8.76 & 1.11 & -4.76 & 9.08 & 13.31 & -31.78 & 83.80 & 10.57 \\ 
  3 & MIRF & 0.83 & -28.79 & 6.31 & 21.21 & -70.25 & 0.20 & 34.55 & 0.90 & -22.78 & 7.21 & 24.26 & -70.28 & 7.60 & 27.48 & 0.91 & -21.92 & 7.72 & 24.38 & -68.33 & 14.00 & 27.34 \\ 
  & Hml & 1.05 & -9.91 & 1.86 & 11.11 & -83.26 & 10.80 & 12.93 & 1.16 & -0.47 & 2.50 & 14.09 & -82.26 & 37.20 & 9.69 & 1.16 & -0.47 & 2.94 & 15.45 & -80.97 & 34.80 & 12.20 \\ 
   & MIHml & 1.06 & -9.05 & 4.88 & 10.75 & -54.60 & 39.00 & 12.07 & 1.16 & -0.47 & 7.49 & 13.33 & -43.81 & 90.00 & 7.63 & 1.16 & -0.47 & 9.42 & 14.30 & -34.13 & 89.60 & 9.63 \\ 
   & MIH2Step & 1.05 & -9.91 & 4.90 & 10.41 & -52.93 & 35.60 & 12.22 & 1.16 & -0.47 & 7.75 & 13.26 & -41.55 & 92.40 & 7.50 & 1.17 & 0.39 & 9.81 & 14.30 & -31.40 & 92.40 & 9.48 \\ 
   \hline
   & LM & 1.21 & 3.82 & 0.88 & 12.92 & -93.19 & 16.80 & 6.39 & 1.24 & 6.39 & 1.38 & 14.49 & -90.48 & 16.20 & 10.13 & 1.26 & 8.11 & 1.75 & 16.10 & -89.13 & 16.40 & 12.74 \\ 
\textbf{Light}   & MIPmm & 1.17 & 0.39 & 4.50 & 11.10 & -59.46 & 88.00 & 4.77 & 1.19 & 2.10 & 6.99 & 12.33 & -43.31 & 87.20 & 7.61 & 1.20 & 2.96 & 8.67 & 13.65 & -36.48 & 83.40 & 9.88 \\ 
\textbf{MNAR}   & MIRF & 0.93 & -20.20 & 6.37 & 22.97 & -72.27 & 7.00 & 24.06 & 0.96 & -17.63 & 7.19 & 21.94 & -67.23 & 23.40 & 21.89 & 0.98 & -15.91 & 8.04 & 21.52 & -62.64 & 34.60 & 20.96 \\ 
   & Hml & 1.16 & -0.47 & 1.88 & 11.48 & -83.62 & 45.60 & 6.14 & 1.16 & -0.47 & 2.53 & 12.76 & -80.17 & 40.40 & 9.55 & 1.16 & -0.47 & 2.97 & 14.66 & -79.74 & 37.20 & 11.67 \\ 
   & MIHml & 1.17 & 0.39 & 4.85 & 11.19 & -56.66 & 89.40 & 5.14 & 1.16 & -0.47 & 7.51 & 11.78 & -36.25 & 91.60 & 7.74 & 1.15 & -1.33 & 9.10 & 12.84 & -29.13 & 86.80 & 9.51 \\ 
   & MIH2Step & 1.16 & -0.47 & 5.06 & 11.04 & -54.17 & 90.00 & 4.97 & 1.17 & 0.39 & 7.83 & 12.27 & -36.19 & 91.00 & 7.91 & 1.17 & 0.39 & 9.69 & 13.02 & -25.58 & 92.80 & 9.39 \\ 
   \hline
   & LM & 1.26 & 8.11 & 0.84 & 14.43 & -94.18 & 2.20 & 10.33 & 1.31 & 12.40 & 1.33 & 18.21 & -92.70 & 2.20 & 16.20 & 1.35 & 15.83 & 1.68 & 20.94 & -91.98 & 2.40 & 20.42 \\ 
\textbf{Heavy}   & MIPmm & 1.21 & 3.82 & 4.28 & 11.47 & -62.69 & 76.60 & 6.27 & 1.25 & 7.25 & 6.65 & 14.21 & -53.20 & 68.60 & 10.97 & 1.28 & 9.83 & 8.10 & 16.18 & -49.94 & 65.60 & 14.35 \\ 
\textbf{MNAR}  & MIRF & 0.97 & -16.77 & 6.48 & 21.77 & -70.23 & 14.80 & 20.76 & 1.01 & -13.34 & 7.40 & 20.08 & -63.15 & 42.60 & 17.11 & 1.04 & -10.77 & 7.98 & 19.61 & -59.31 & 61.60 & 14.85 \\ 
   & Hml & 1.16 & -0.47 & 1.94 & 10.95 & -82.28 & 50.20 & 5.55 & 1.15 & -1.33 & 2.62 & 12.93 & -79.74 & 46.80 & 8.55 & 1.14 & -2.19 & 3.08 & 15.51 & -80.14 & 41.20 & 10.99 \\ 
   & MIHml & 1.17 & 0.39 & 4.67 & 10.18 & -54.13 & 91.20 & 4.61 & 1.15 & -1.33 & 6.98 & 11.67 & -40.19 & 88.60 & 7.16 & 1.15 & -1.33 & 8.65 & 13.85 & -37.55 & 84.80 & 9.29 \\ 
   & MIH2Step & 1.16 & -0.47 & 4.78 & 10.44 & -54.21 & 92.00 & 4.72 & 1.16 & -0.47 & 7.38 & 12.07 & -38.86 & 92.00 & 7.21 & 1.17 & 0.39 & 9.40 & 13.89 & -32.33 & 89.20 & 9.25 \\ 
   \hline
   & LM & 1.04 & -10.77 & 0.82 & 7.55 & -89.14 & 0.20 & 12.89 & 0.92 & -21.06 & 1.26 & 10.18 & -87.62 & 0.00 & 25.55 & 0.83 & -28.79 & 1.55 & 14.96 & -89.64 & 0.00 & 34.46 \\ 
 \textbf{Non-}  & MIPmm & 1.02 & -12.48 & 3.65 & 7.88 & -53.68 & 4.60 & 14.71 & 0.90 & -22.78 & 5.71 & 11.02 & -48.19 & 0.80 & 27.25 & 0.81 & -30.50 & 6.31 & 15.54 & -59.40 & 0.00 & 35.93 \\ 
 \textbf{Heckman}  & MIRF & 0.89 & -23.64 & 6.64 & 17.27 & -61.55 & 6.40 & 28.30 & 0.79 & -32.22 & 6.62 & 19.17 & -65.47 & 0.40 & 38.28 & 0.71 & -39.08 & 6.55 & 22.89 & -71.38 & 0.00 & 45.92 \\ 
   & Hml & 1.08 & -7.33 & 2.04 & 22.27 & -90.84 & 16.00 & 22.49 & 1.03 & -11.62 & 2.74 & 31.28 & -91.24 & 15.00 & 32.27 & 1.06 & -9.05 & 3.24 & 38.90 & -91.67 & 18.00 & 34.43 \\ 
   & MIHml & 1.09 & -6.48 & 11.48 & 20.03 & -42.69 & 66.20 & 19.79 & 1.03 & -11.62 & 16.18 & 29.40 & -44.97 & 66.40 & 29.73 & 1.01 & -13.34 & 15.60 & 34.89 & -55.29 & 63.20 & 34.58 \\ 
   & MIH2Step & 1.07 & -8.19 & 14.72 & 16.56 & -11.11 & 81.20 & 17.24 & 0.99 & -15.06 & 23.91 & 25.21 & -5.16 & 77.20 & 28.46 & 0.93 & -20.20 & 28.24 & 30.21 & -6.52 & 76.40 & 36.17 \\ 
   \hline
   \hline
\end{tabular}}
\centering
{\small Mean = Average estimate $\hat{\beta}_{Revenue}$ (rounded to 2 digits), Rbias =  Average relative bias (\%), $SE_{m}$ = Root mean square of the estimated standard errors $ \times 10^{2}$, $SE_{e}$ = Empirical standard error $ \times  10^{2}$, CR = Coverage rate (based on 95\% confidence interval), $RE_{SE}$ = relative error of $SE_{m}$(\%), RMSE = Root mean square error  $ \times 10^{2}$.}
\end{table}

\clearpage

\begin{table}[H]
\centering
 \caption{\textit{S}: Median relative bias (\% Rbias) for estimated coefficients across all categories}
 $\sigma^{2}_{\varepsilon}$ =1 \\
\resizebox{0.8\textwidth}{!}{\begin{tabular}{|l|rrrr|rrrr|}
  \hline
 Variable    & \multicolumn{4}{c|}{Sector (39 categories)} & \multicolumn{4}{c|}{FirstActivity (4 categories)}\\
  \hline
 & MAR &\multicolumn{1}{c}{Light} & Heavy & \multicolumn{1}{c|}{Non-} & MAR &\multicolumn{1}{c}{Light} & Heavy &  \multicolumn{1}{c|}{Non-} \\
Method & & MNAR & MNAR & Heckman & & MNAR & MNAR & Heckman\\
\hline
LM & -22.49 & 0.62 & 1.29 & 0.06 & -22.41 & -4.71 & -7.80 & -12.15 \\ 
  MIPmm & -23.24 & 0.02 & -0.14 & 0.11 & -23.71 & -7.64 & -10.38 & -12.77 \\ 
  MIRF & -51.92 & -37.02 & -37.00 & -49.31 & -45.10 & -49.25 & -51.86 & -24.70 \\ 
  Hml & -21.84 & 0.19 & -1.16 & -1.13 & -15.91 & 8.26 & 7.19 & -0.73 \\ 
  MIHml & -21.99 & 0.59 & 0.38 & -0.22 & -17.52 & -0.91 & -1.06 & -5.56 \\ 
  MIH2Step & -22.09 & -0.15 & -0.20 & -0.11 & -15.79 & -0.12 & -0.55 & -4.99 \\ 
   \hline
   \end{tabular}}
 
 \vspace{2mm}
 
  $\sigma^{2}_{\varepsilon}$ =2.5 \\
  \resizebox{0.8\textwidth}{!}{\begin{tabular}{|l|rrrr|rrrr|}
  \hline
 Variable    & \multicolumn{4}{c|}{Sector (37 categories)} & \multicolumn{4}{c|}{FirstActivity (4 categories)}\\
  \hline
 & MAR &\multicolumn{1}{c}{Light} & Heavy & \multicolumn{1}{c|}{Non-} & MAR &\multicolumn{1}{c}{Light} & Heavy &  \multicolumn{1}{c|}{Non-} \\
Method & & MNAR & MNAR & Heckman & & MNAR & MNAR & Heckman\\
\hline
LM & 0.15 & 1.34 & 2.39 & -0.25 & -2.67 & -8.36 & -12.18 & -24.73 \\ 
  MIPmm & -0.21 & 0.22 & 0.65 & -0.26 & -5.29 & -10.42 & -15.05 & -24.77 \\ 
  MIRF & -38.08 & -37.36 & -35.94 & -44.64 & -20.11 & -48.88 & -51.98 & -34.22 \\ 
  Hml & -1.38 & 2.08 & 1.98 & -1.57 & 8.87 & 6.93 & 9.30 & -15.24 \\ 
  MIHml & 0.78 & 0.60 & 0.68 & -1.00 & -2.25 & -3.69 & -3.17 & -11.36 \\ 
  MIH2Step & -0.23 & 1.43 & -0.01 & -2.17 & -0.98 & -1.28 & -0.01 & -16.33 \\ 
   \hline
   \end{tabular}}
 
 \vspace{2mm}
 
  $\sigma^{2}_{\varepsilon}$ =4 \\
  \resizebox{0.8 \textwidth}{!}{\begin{tabular}{|l|rrrr|rrrr|}
  \hline
 Variable    & \multicolumn{4}{c|}{Sector (37 categories)} & \multicolumn{4}{c|}{FirstActivity (4 categories)}\\
  \hline
 & MAR &\multicolumn{1}{c}{Light} & Heavy & \multicolumn{1}{c|}{Non-} & MAR &\multicolumn{1}{c}{Light} & Heavy &  \multicolumn{1}{c|}{Non-} \\
Method & & MNAR & MNAR & Heckman & & MNAR & MNAR & Heckman\\
\hline
LM & 0.18 & 2.17 & 2.95 & -0.60 & -3.37 & -10.57 & -15.34 & -33.54 \\ 
  MIPmm & -1.00 & 0.32 & 1.17 & -2.21 & -5.85 & -13.66 & -17.24 & -31.21 \\ 
  MIRF & -37.48 & -35.27 & -35.39 & -39.94 & -18.15 & -39.24 & -51.84 & -38.80 \\ 
  Hml & 1.22 & 3.36 & 1.96 & -3.40 & 9.11 & 12.63 & 8.01 & -15.15 \\ 
  MIHml & 0.54 & 0.62 & 0.32 & -2.53 & -2.42 & -2.61 & -2.50 & -18.65 \\ 
  MIH2Step & -0.36 & 0.32 & 0.99 & -1.40 & -1.22 & -1.26 & -1.95 & -24.82 \\ 
   \hline
   
\end{tabular}}
\label{tab:medianS}
\end{table}

\textbf{Comment:} For Sector variable, MIHml and MIH2Step tended to have lower median relative bias than other methods except MIPmm  in Light MNAR and Heavy MNAR scenario. For Non-Heckman scenario, LM does performed better than all the other methods. For the FirstActivity variable, MIHStep2 and MIHml had lower median relative bias than other methods under Light MNAR and Heavy MNAR scenario. In the Non-Heckman scenario, Hml, MIHml and MIH2Step had lower relative biased than LM, MIPmm and MIRF across all values of $\sigma^{2}_{\varepsilon}$.
\newpage

\clearpage


\begin{table}[ht]
\caption{$U$ Scope 1: Simulation results for $\beta_{Revenue} = 0.837 $ estimates }


\resizebox{\textwidth}{!}{\begin{tabular}{|ll|rrrrrrr|rrrrrrr|rrrrrrr|}
  \hline
  
   &  & \multicolumn{7}{c|}{\Large \textbf{$\sigma^{2}_{\varepsilon}$ = 1}} & \multicolumn{7}{c|}{ \Large \textbf{$\sigma^{2}_{\varepsilon}$ = 2.5}} & \multicolumn{7}{c|}{ \Large \textbf{$\sigma^{2}_{\varepsilon}$ = 4}}  \\
  &&&&&&&&&&&&&&&&&&&&&& \\
 & Methods & Mean & Rbias & $SE_{m}$ & $SE_{e}$ & $RE_{SE}$ &  CR   & {\small RMSE } & Mean & Rbias & $SE_{m}$ & $SE_{e}$ & $RE_{SE}$ &  CR   & {\small RMSE } & Mean & Rbias & $SE_{m}$ & $SE_{e}$ & $RE_{SE}$ &  CR   & {\small RMSE } \\ 
  \hline

\multirow{5}{*}{\textbf{MAR}}  & LM & 0.84 & 0.38 & 0.55 & 5.68 & -90.32 & 17.60 & 5.58 & 0.84 & 0.38 & 0.87 & 8.93 & -90.26 & 17.40 & 8.82 & 0.84 & 0.38 & 1.10 & 11.30 & -90.27 & 17.40 & 11.17 \\ 
   & MIPmm & 0.82 & -2.01 & 5.80 & 6.11 & -5.07 & 88.60 & 6.08 & 0.82 & -2.01 & 8.93 & 9.40 & -5.00 & 86.00 & 9.43 & 0.82 & -2.01 & 11.25 & 11.82 & -4.82 & 86.40 & 11.84 \\ 
   & MIRF & 0.83 & -0.82 & 10.31 & 10.05 & 2.59 & 90.00 & 10.01 & 0.83 & -0.82 & 10.57 & 11.29 & -6.38 & 88.20 & 11.28 & 0.84 & 0.38 & 11.46 & 12.85 & -10.82 & 86.60 & 12.80 \\ 
   & Hml & 0.84 & 0.38 & 1.69 & 7.15 & -76.36 & 34.00 & 7.10 & 0.84 & 0.38 & 2.20 & 11.27 & -80.48 & 28.20 & 11.21 & 0.84 & 0.38 & 2.54 & 14.18 & -82.09 & 27.80 & 14.11 \\ 
   & MIHml & 0.84 & 0.38 & 5.64 & 6.06 & -6.93 & 87.80 & 5.97 & 0.83 & -0.82 & 8.63 & 9.47 & -8.87 & 88.20 & 9.44 & 0.83 & -0.82 & 10.82 & 11.92 & -9.23 & 87.20 & 11.89 \\ 
   & MIH2Step & 0.83 & -0.82 & 6.11 & 6.36 & -3.93 & 87.20 & 6.33 & 0.83 & -0.82 & 9.64 & 10.04 & -3.98 & 87.40 & 10.00 & 0.83 & -0.82 & 12.15 & 12.68 & -4.18 & 87.40 & 12.64 \\ 
  \hline
   & LM & 0.85 & 1.57 & 0.54 & 5.67 & -90.48 & 13.80 & 5.62 & 0.86 & 2.77 & 0.86 & 8.83 & -90.26 & 13.00 & 8.87 & 0.86 & 2.77 & 1.09 & 11.16 & -90.23 & 13.40 & 11.21 \\ 
  \textbf{Light} & MIPmm & 0.83 & -0.82 & 5.73 & 5.75 & -0.35 & 87.00 & 5.68 & 0.83 & -0.82 & 9.06 & 8.92 & 1.57 & 88.80 & 8.83 & 0.83 & -0.82 & 11.37 & 11.72 & -2.99 & 89.00 & 11.64 \\ 
 \textbf{MNAR}  & MIRF & 0.83 & -0.82 & 10.59 & 9.66 & 9.63 & 91.60 & 9.61 & 0.84 & 0.38 & 10.75 & 11.05 & -2.71 & 88.40 & 10.95 & 0.84 & 0.38 & 11.26 & 12.46 & -9.63 & 88.80 & 12.40 \\ 
   & Hml & 0.84 & 0.38 & 1.71 & 6.81 & -74.89 & 35.80 & 6.79 & 0.84 & 0.38 & 2.23 & 11.24 & -80.16 & 28.00 & 11.25 & 0.84 & 0.38 & 2.58 & 13.65 & -81.10 & 28.40 & 13.65 \\ 
   & MIHml & 0.84 & 0.38 & 5.43 & 5.79 & -6.22 & 87.40 & 5.73 & 0.84 & 0.38 & 8.25 & 9.41 & -12.33 & 85.80 & 9.35 & 0.84 & 0.38 & 10.29 & 11.07 & -7.05 & 87.00 & 11.03 \\ 
   & MIH2Step & 0.83 & -0.82 & 6.25 & 6.11 & 2.29 & 87.80 & 6.03 & 0.84 & 0.38 & 9.90 & 9.48 & 4.43 & 89.00 & 9.41 & 0.84 & 0.38 & 12.17 & 11.78 & 3.31 & 89.20 & 11.72 \\ 
  \hline
   & LM & 0.86 & 2.77 & 0.52 & 5.64 & -90.78 & 14.60 & 5.73 & 0.87 & 3.96 & 0.83 & 8.68 & -90.44 & 15.60 & 8.90 & 0.88 & 5.16 & 1.04 & 11.00 & -90.55 & 14.40 & 11.13 \\ 
 \textbf{Heavy}  & MIPmm & 0.84 & 0.38 & 5.36 & 5.64 & -4.96 & 87.40 & 5.58 & 0.85 & 1.57 & 8.63 & 8.80 & -1.93 & 88.40 & 8.85 & 0.84 & 0.38 & 10.87 & 11.15 & -2.51 & 87.40 & 11.09 \\ 
  \textbf{MNAR}  & MIRF & 0.83 & -0.82 & 10.15 & 9.63 & 5.40 & 90.20 & 9.57 & 0.84 & 0.38 & 10.92 & 10.94 & -0.18 & 91.00 & 10.84 & 0.85 & 1.57 & 11.32 & 11.90 & -4.87 & 90.20 & 11.79 \\ 
   & Hml & 0.84 & 0.38 & 1.78 & 6.81 & -73.86 & 42.40 & 6.78 & 0.84 & 0.38 & 2.33 & 10.89 & -78.60 & 30.00 & 10.75 & 0.84 & 0.38 & 2.70 & 13.16 & -79.48 & 31.60 & 12.99 \\ 
   & MIHml & 0.84 & 0.38 & 5.02 & 5.80 & -13.45 & 85.80 & 5.73 & 0.84 & 0.38 & 7.68 & 8.85 & -13.22 & 85.80 & 8.74 & 0.83 & -0.82 & 9.80 & 11.19 & -12.42 & 85.20 & 11.03 \\ 
   & MIH2Step & 0.84 & 0.38 & 5.76 & 5.85 & -1.54 & 89.40 & 5.72 & 0.84 & 0.38 & 8.92 & 9.05 & -1.44 & 90.60 & 8.89 & 0.84 & 0.38 & 11.29 & 11.10 & 1.71 & 89.40 & 10.90 \\ 
  \hline
   & LM & 0.73 & -12.77 & 0.45 & 11.71 & -96.16 & 1.40 & 11.78 & 0.63 & -24.72 & 0.73 & 9.73 & -92.50 & 0.00 & 21.94 & 0.56 & -33.08 & 0.94 & 8.92 & -89.46 & 0.00 & 28.56 \\ 
  \textbf{Non-} & MIPmm & 0.69 & -17.55 & 5.25 & 9.84 & -46.65 & 23.60 & 15.41 & 0.60 & -28.30 & 7.14 & 8.99 & -20.58 & 8.60 & 24.69 & 0.54 & -35.47 & 8.90 & 9.19 & -3.16 & 9.60 & 30.74 \\ 
 \textbf{Heckman} & MIRF & 0.53 & -36.67 & 8.98 & 18.54 & -51.56 & 17.40 & 32.77 & 0.48 & -42.64 & 8.17 & 15.85 & -48.45 & 6.80 & 37.30 & 0.45 & -46.23 & 7.93 & 15.56 & -49.04 & 4.60 & 40.67 \\ 
   & Hml & 0.63 & -24.72 & 2.44 & 34.65 & -92.96 & 4.80 & 39.53 & 0.57 & -31.89 & 3.05 & 44.22 & -93.10 & 5.80 & 51.46 & 0.56 & -33.08 & 3.34 & 44.12 & -92.43 & 7.80 & 51.88 \\ 
   & MIHml & 0.60 & -28.30 & 14.09 & 33.37 & -57.78 & 36.00 & 39.73 & 0.56 & -33.08 & 18.78 & 41.49 & -54.74 & 41.40 & 49.69 & 0.55 & -34.28 & 18.43 & 41.91 & -56.02 & 41.60 & 50.58 \\ 
   & MIH2Step & 0.72 & -13.96 & 19.37 & 20.50 & -5.51 & 82.20 & 21.01 & 0.62 & -25.91 & 28.32 & 25.89 & 9.39 & 78.20 & 33.08 & 0.58 & -30.69 & 26.80 & 27.15 & -1.29 & 75.00 & 36.96 \\ 
   \hline
\end{tabular}}

\centering
{\small Mean = Average estimate $\hat{\beta}_{Revenue}$ (rounded to 2 digits), Rbias =  Average relative bias (\%), $SE_{m}$ = Root mean square of the estimated standard errors  $ \times 10^{2}$, $SE_{e}$ = Empirical standard error  $ \times 10^{2}$, CR = Coverage rate (based on 95\% confidence interval), $RE_{SE}$ = relative error of $SE_{m}$(\%), RMSE = Root mean square error  $ \times 10^{2}$.}
\end{table}

   \begin{table}[H]
\centering
\caption{\textit{U}: Median relative bias (\% Rbias) for estimated coefficients across all categories}
 $\sigma^{2}_{\varepsilon}$ =1 \\
\resizebox{0.5\textwidth}{!}{\begin{tabular}{|l|rrrr|}
  \hline
 Variable    & \multicolumn{4}{c|}{Sector (32 categories)}\\
  \hline
 & MAR &\multicolumn{1}{c}{Light} & Heavy & \multicolumn{1}{c|}{Non-}  \\
Method & & MNAR & MNAR & Heckman \\
\hline
LM & 0.74 & -0.29 & -1.71 & -2.17 \\ 
  MIPmm & -1.07 & -1.28 & -4.06 & -8.58 \\ 
  MIRF & -59.05 & -59.18 & -61.41 & -54.64 \\ 
  Hml & 0.49 & 0.79 & 0.09 & -0.94 \\ 
  MIHml & 0.43 & 0.78 & 0.51 & -0.62 \\ 
  MIH2Step & -2.14 & -2.64 & -0.88 & -2.41 \\ 
   \hline
\end{tabular}} 
\vspace{2mm}

 $\sigma^{2}_{\varepsilon}$ =2.5 \\
\resizebox{0.5\textwidth}{!}{\begin{tabular}{|l|rrrr|}
  \hline
 Variable    & \multicolumn{4}{c|}{Sector (32 categories)}\\
  \hline
 & MAR &\multicolumn{1}{c}{Light} & Heavy & \multicolumn{1}{c|}{Non-}  \\
Method & & MNAR & MNAR & Heckman \\
\hline
LM & 0.69 & -0.31 & -4.31 & -2.91 \\ 
  MIPmm & -3.07 & -3.22 & -4.29 & -6.66 \\ 
  MIRF & -59.67 & -57.71 & -60.11 & -52.69 \\ 
  Hml & 2.40 & 2.71 & 0.77 & -0.39 \\ 
  MIHml & 0.52 & 0.57 & 0.37 & -1.69 \\ 
  MIH2Step & -2.73 & -1.68 & -0.93 & -3.60 \\ 
       \hline
\end{tabular}} 

\vspace{2mm}

 $\sigma^{2}_{\varepsilon}$ =4 \\
\resizebox{0.5\textwidth}{!}{\begin{tabular}{|l|rrrr|}
  \hline
 Variable    & \multicolumn{4}{c|}{Sector (32 categories)}\\
  \hline
 & MAR &\multicolumn{1}{c}{Light} & Heavy & \multicolumn{1}{c|}{Non-} \\
Method & & MNAR & MNAR & Heckman \\
\hline
LM & 0.51 & -0.88 & -3.78 & -3.84 \\ 
  MIPmm & -4.13 & -5.02 & -7.40 & -8.35 \\ 
  MIRF & -59.91 & -58.60 & -62.25 & -42.96 \\ 
  Hml & 2.43 & 4.77 & 1.57 & -1.00 \\ 
  MIHml & 0.75 & 0.53 & 0.73 & -2.44 \\ 
  MIH2Step & -3.41 & -1.92 & 0.17 & -5.94 \\ 
   \hline
\end{tabular}}
\label{tab:medianU}
\end{table}

\textbf{Comment:} For Sector variable, Hml, MIHml and MIH2Step performed better than other methods under Heavy MNAR scenario across all values of $\sigma^{2}_{\varepsilon}$. Hml and MIHml  performed better than other methods in Non-Heckman scenario. 

\newpage

\begin{figure}[H]
   \caption{Boxplot over simulations over $\hat{\beta}_{Revenue}$. Red dotted lines denotes true value of the coefficient}
\begin{subfigure}{0.5\textwidth}
\caption{$L$}
    \includegraphics[width=\textwidth]{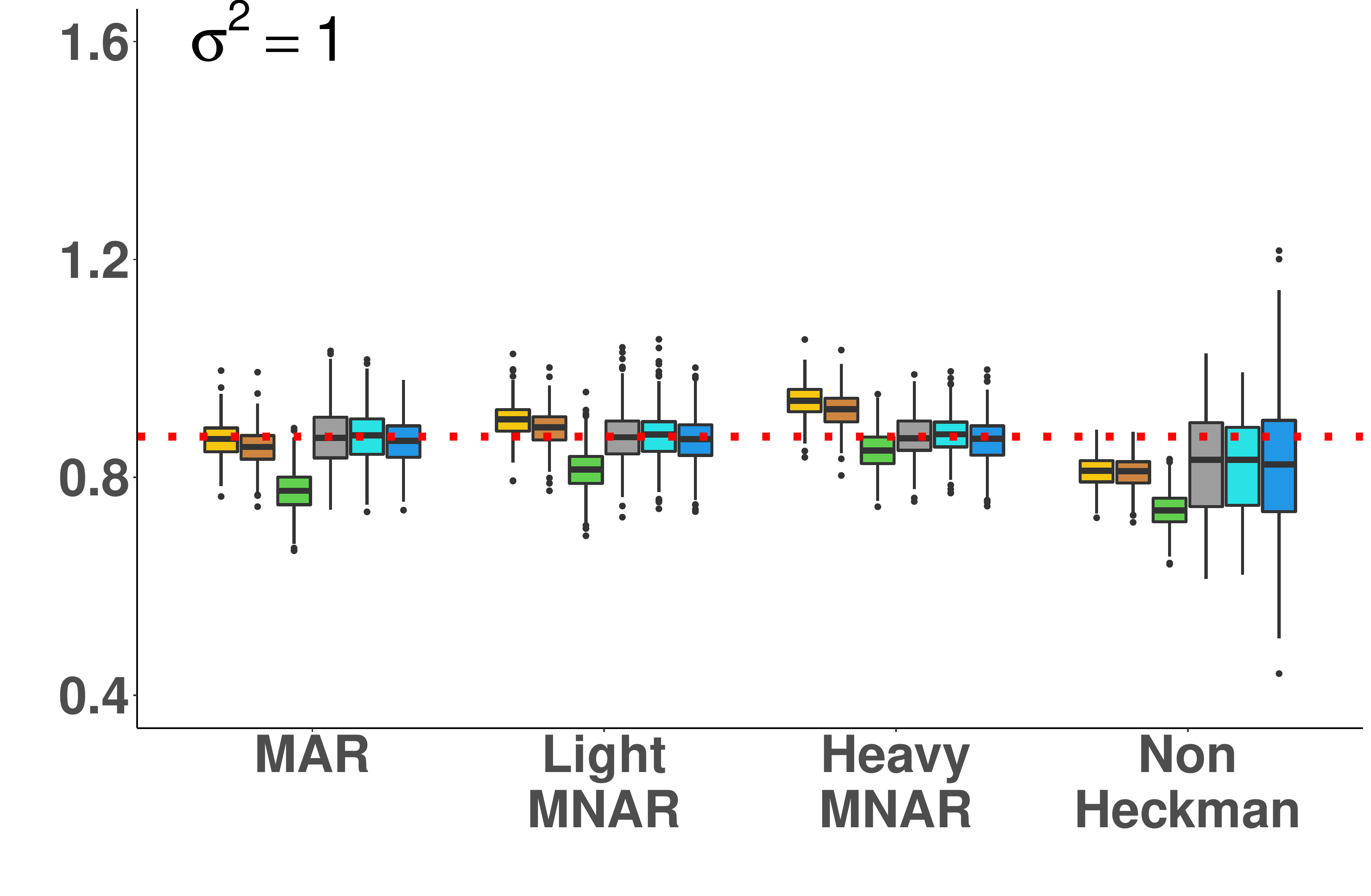}
      \end{subfigure}%
  \begin{subfigure}{0.5\textwidth}
\caption{$M$}
    \includegraphics[width=\textwidth]{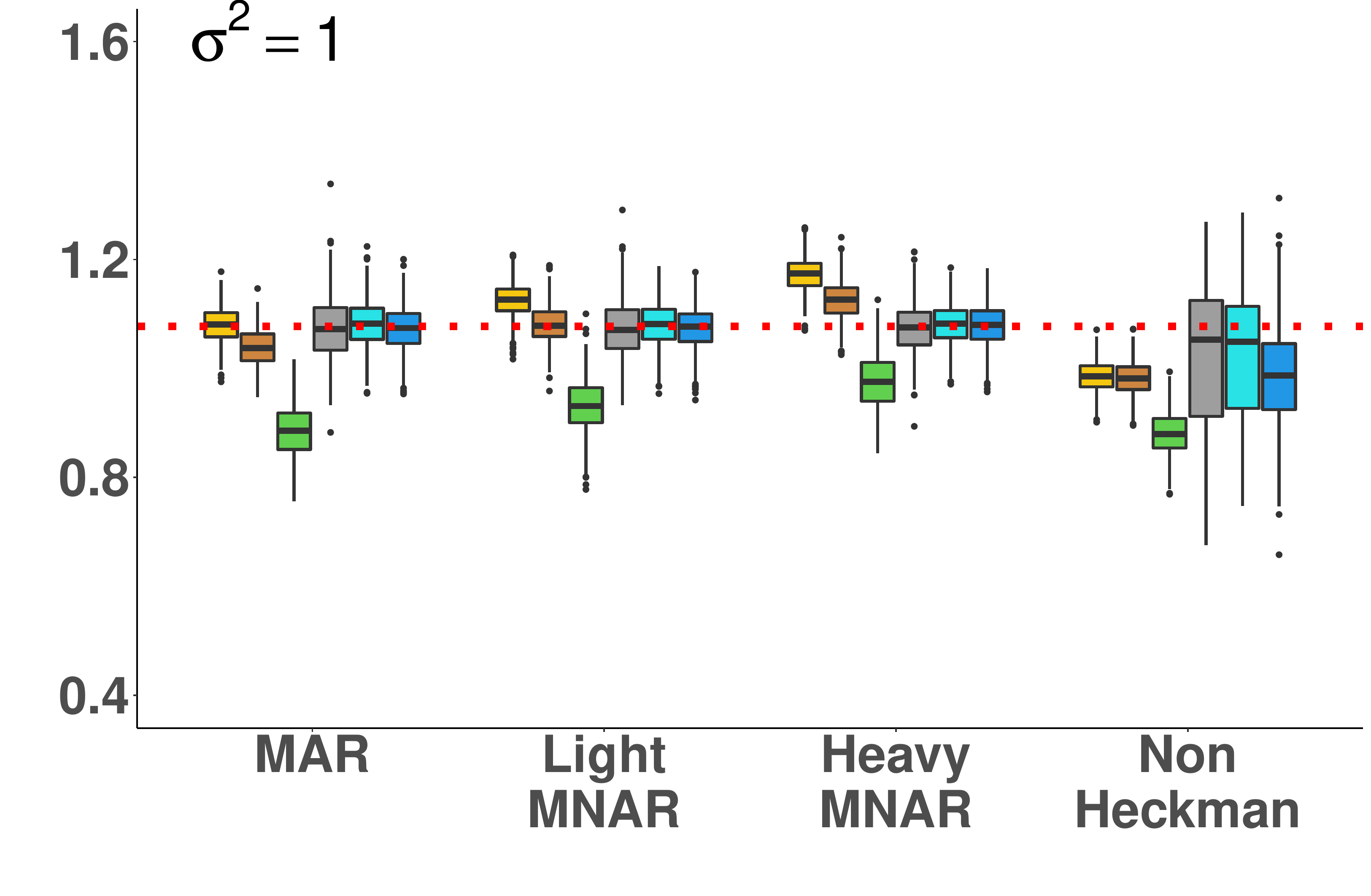}
    \end{subfigure}
  \end{figure}

\begin{figure}[H]
\begin{subfigure}{0.5\textwidth}
     \includegraphics[width=\textwidth]{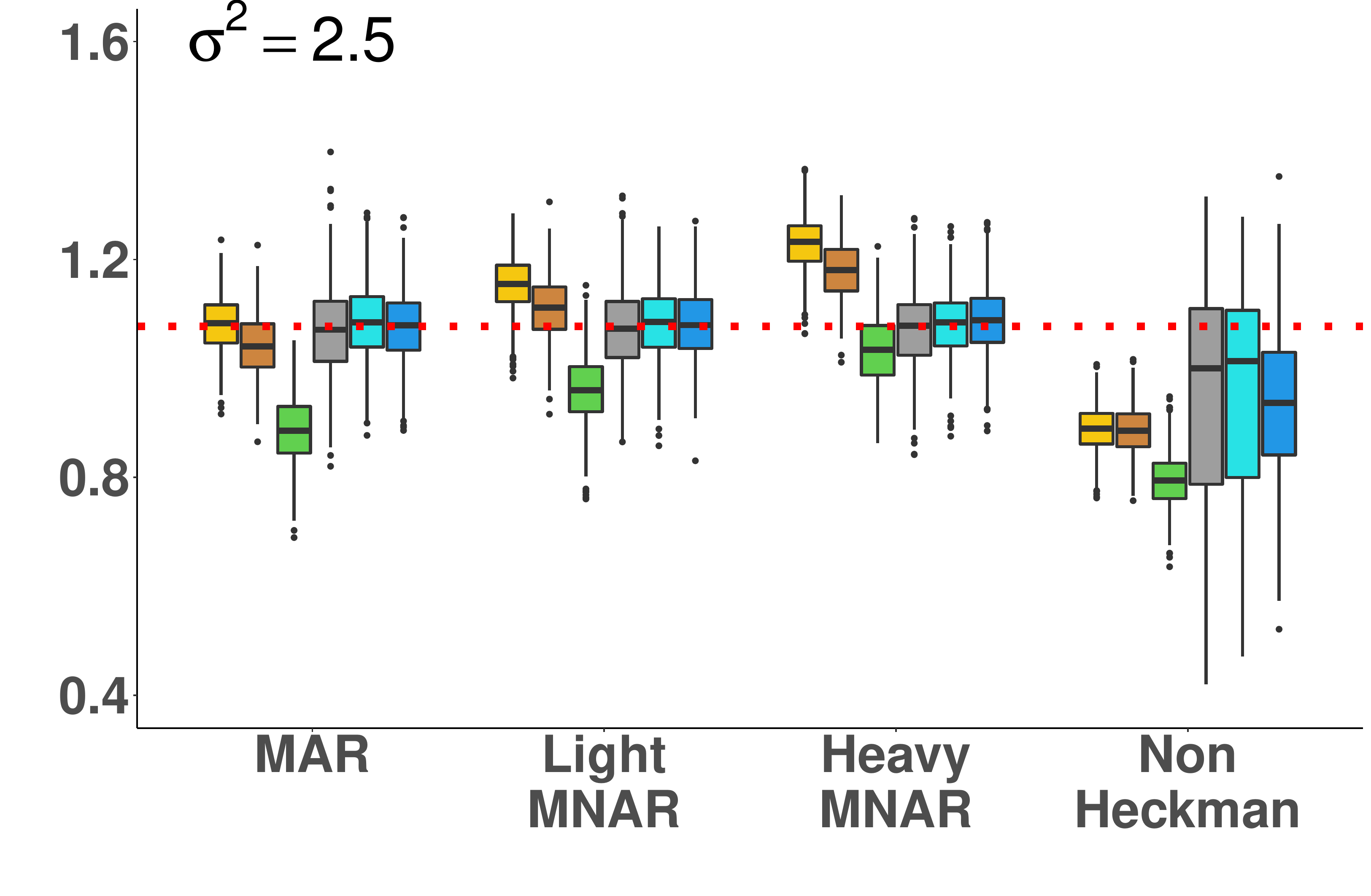}
      \end{subfigure}%
  \begin{subfigure}{0.5\textwidth}
      \includegraphics[width=\textwidth]{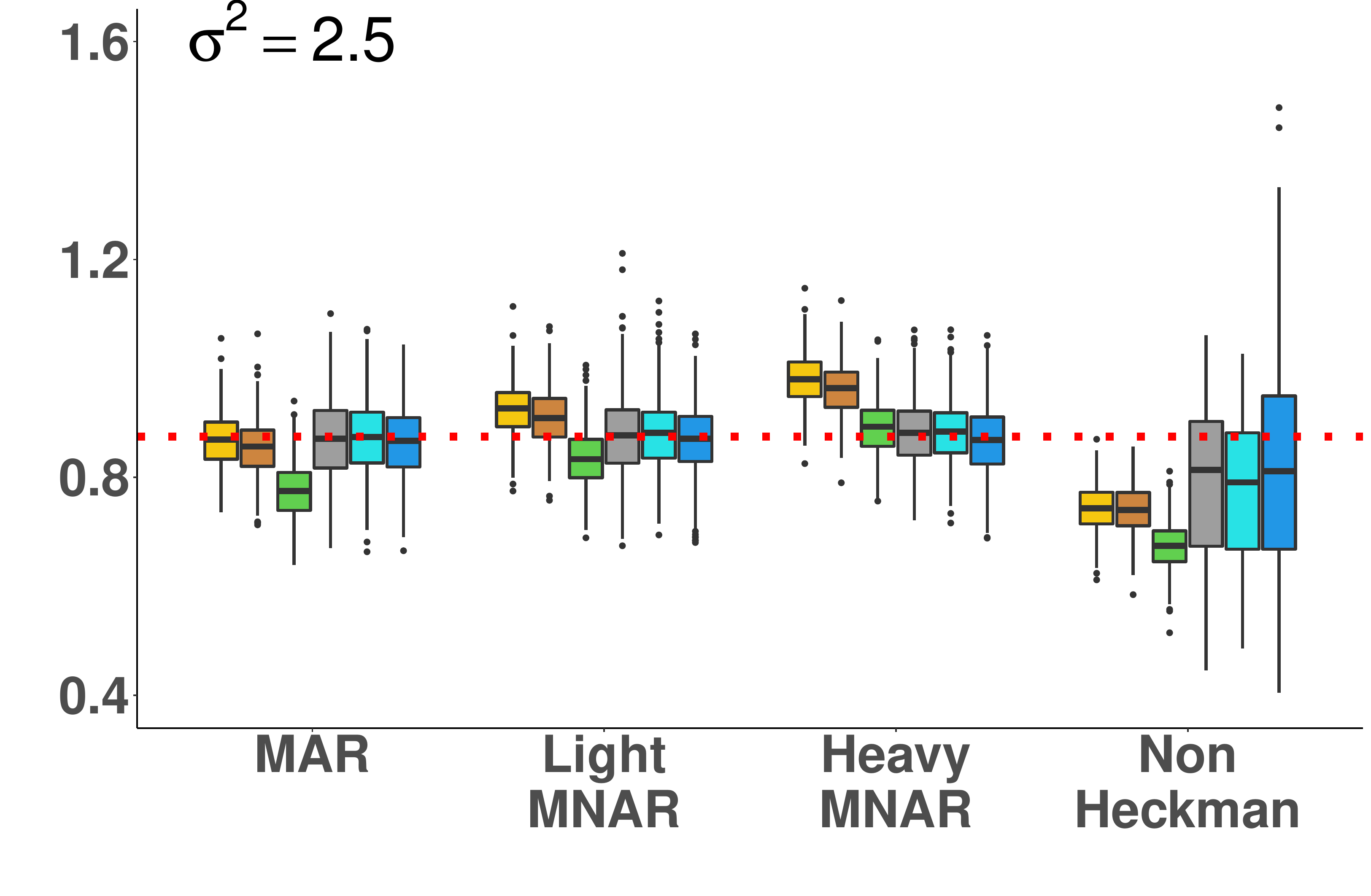}
    \end{subfigure}
  \end{figure}  

\begin{figure}[H]
\begin{subfigure}{0.5\textwidth}
    \includegraphics[width=\textwidth]{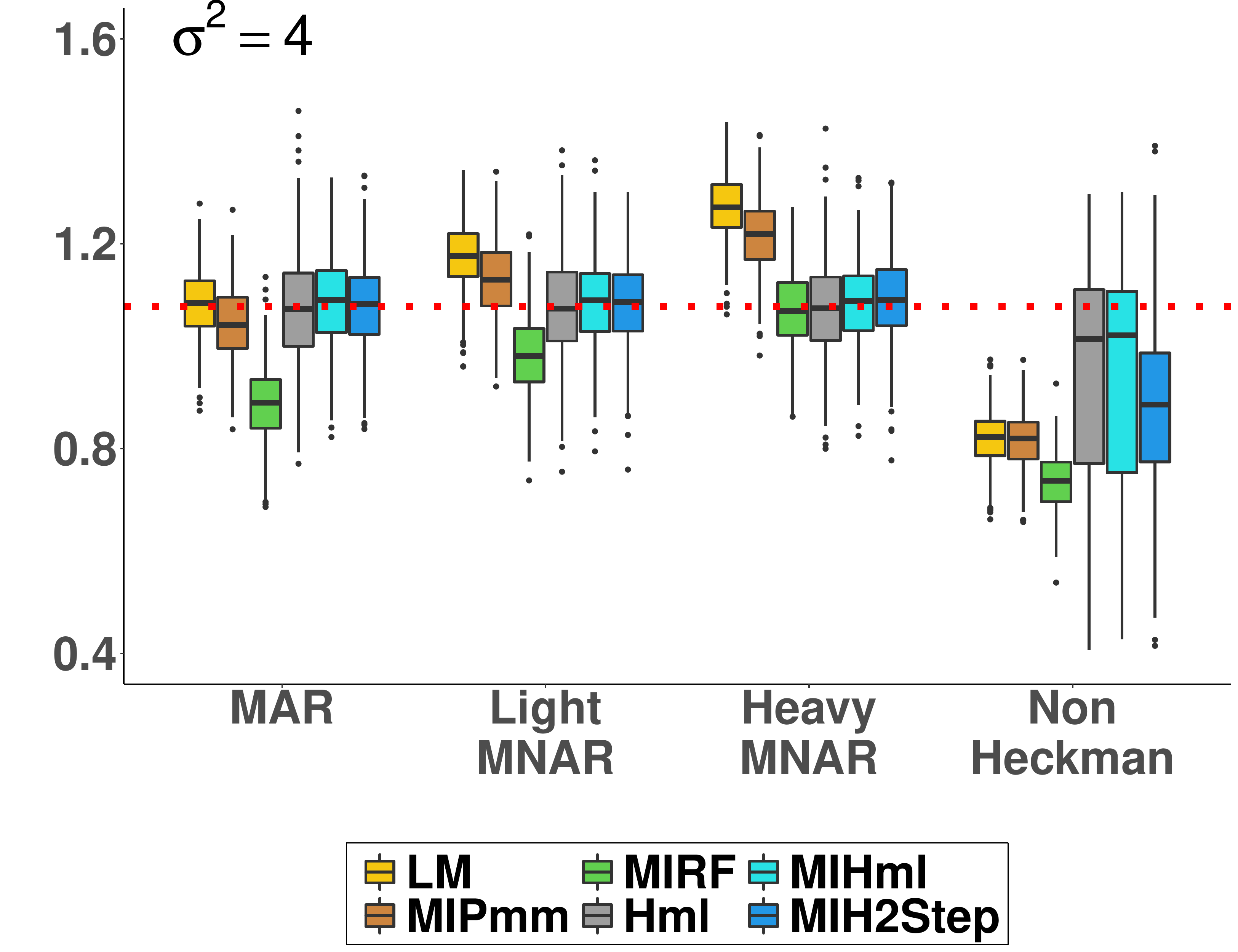}
      \end{subfigure}%
  \begin{subfigure}{0.5\textwidth}
         \includegraphics[width=\textwidth]{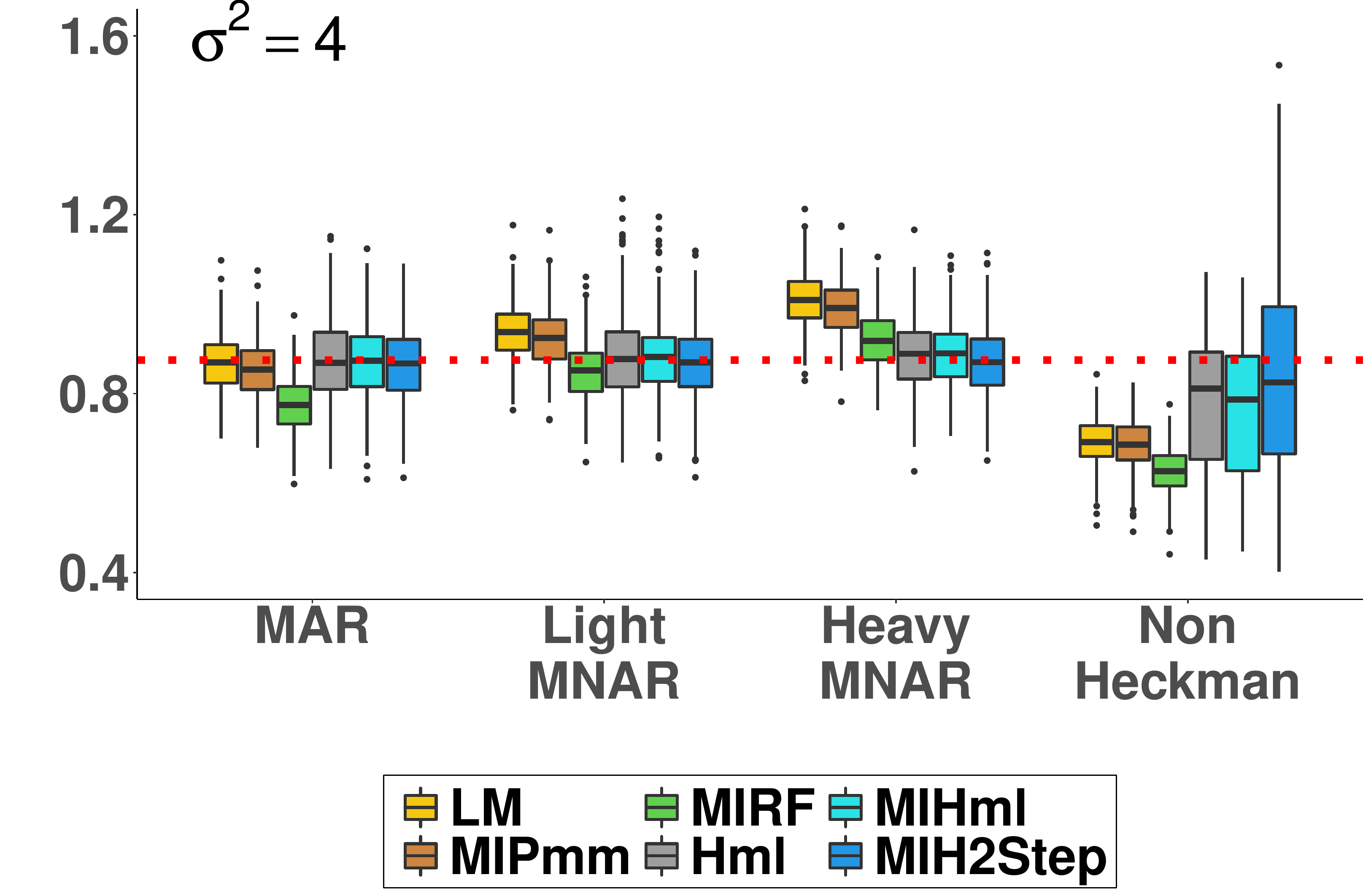}
    \end{subfigure}
  \end{figure} 

\clearpage

\begin{figure}[H]

   \caption{Boxplot over simulations over $\hat{\beta}_{Revenue}$. Red dotted lines denotes true value of the coefficient }
\begin{subfigure}{0.5\textwidth}
\caption{$S$}
    \includegraphics[width=\textwidth]{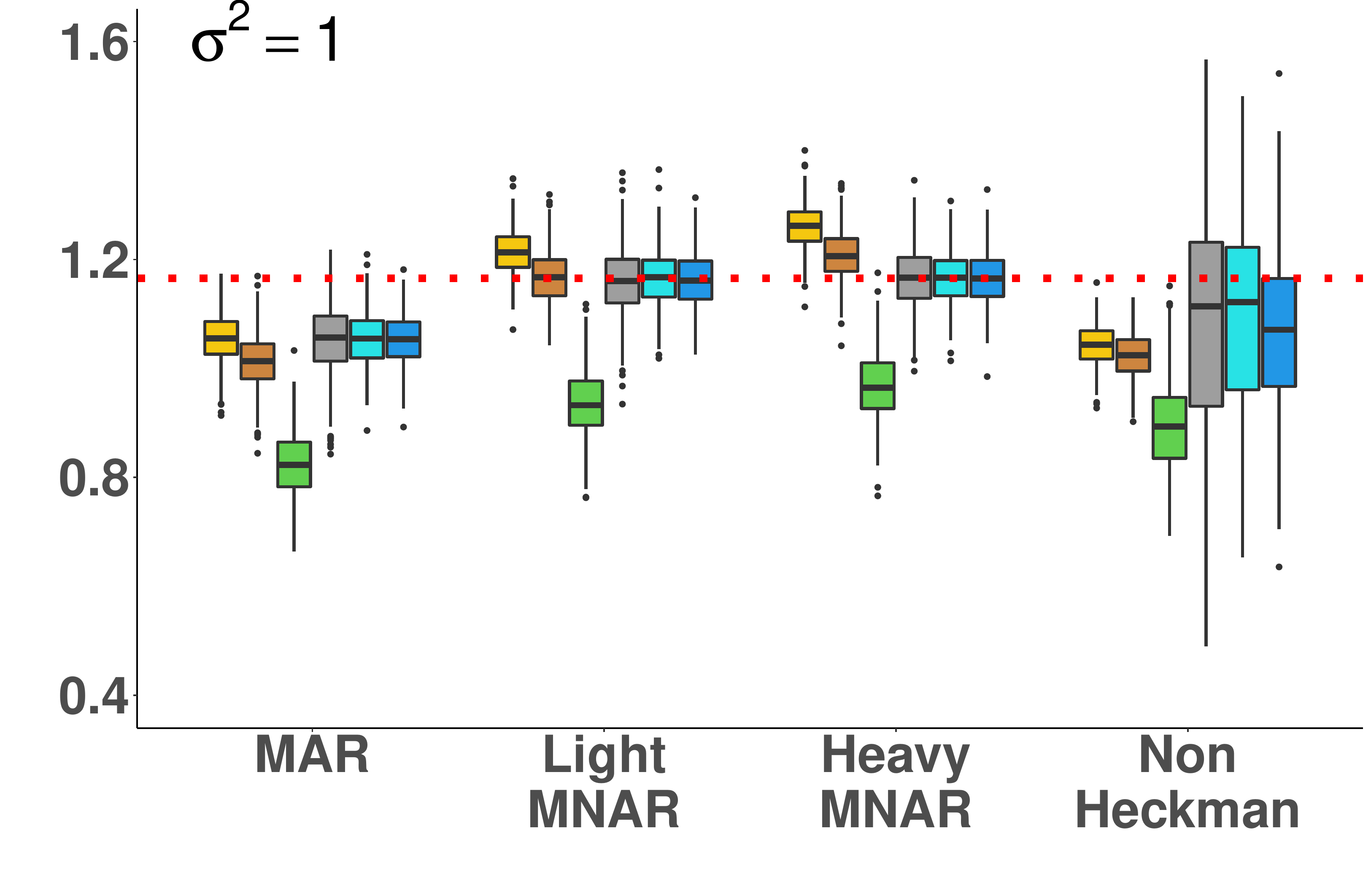}
  \end{subfigure}%
  \begin{subfigure}{0.5\textwidth}
\caption{$U$}
    \includegraphics[width=\textwidth]{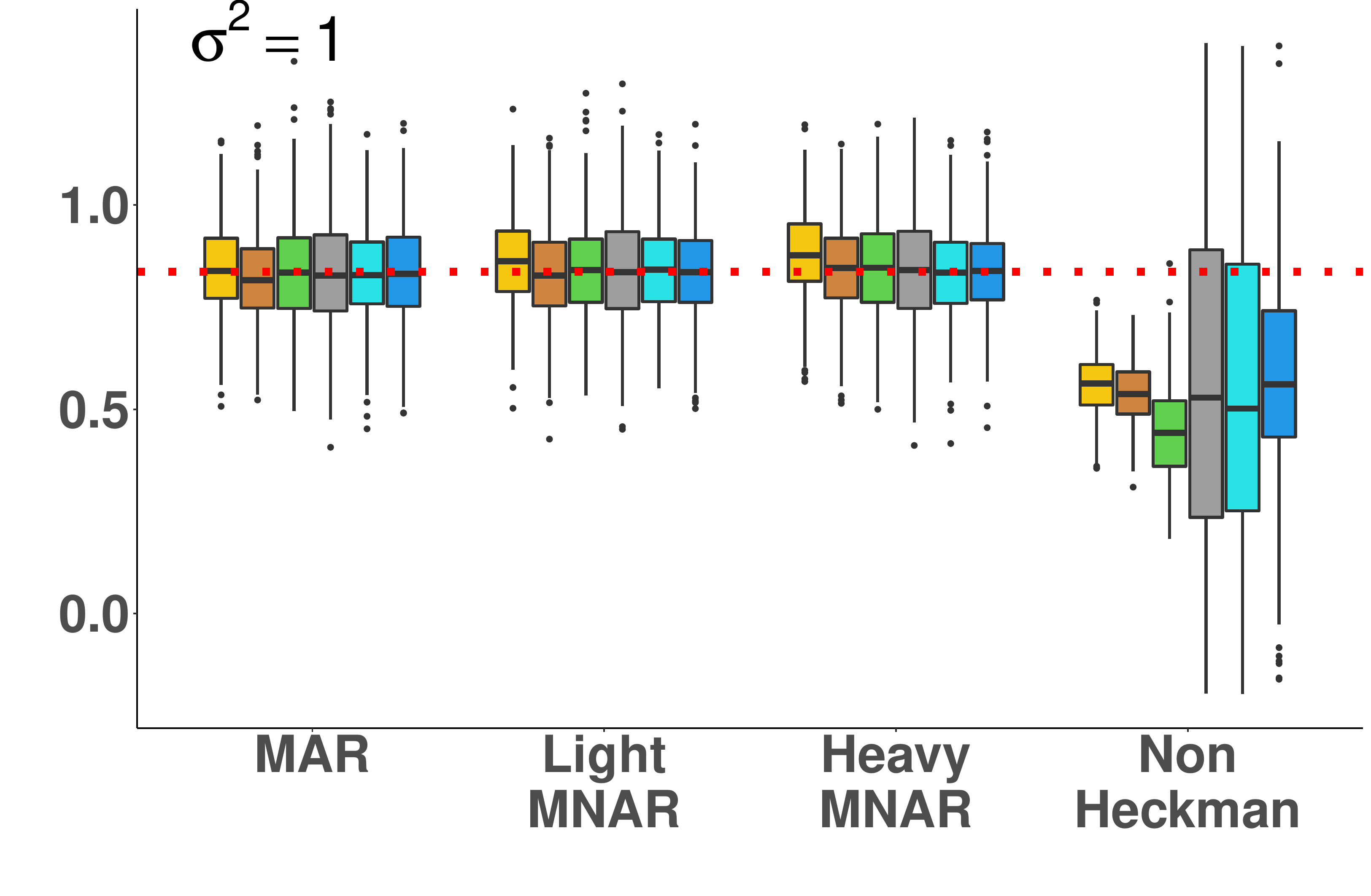}
  \end{subfigure}
  
\end{figure}

\begin{figure}[H]

\begin{subfigure}{0.5\textwidth}
 \includegraphics[width=\textwidth]{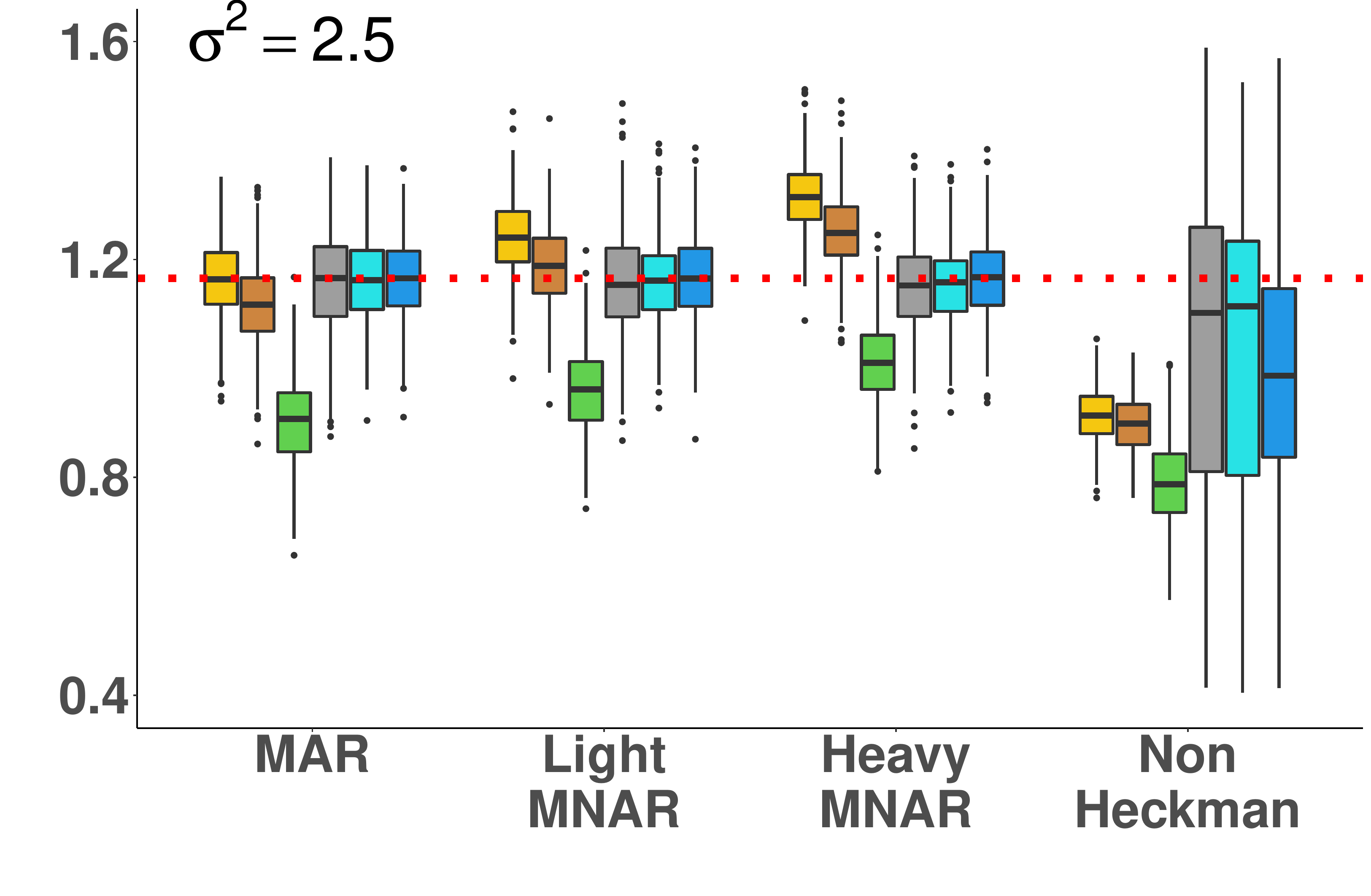}
  \end{subfigure}%
  \begin{subfigure}{0.5\textwidth}
   \includegraphics[width=\textwidth]{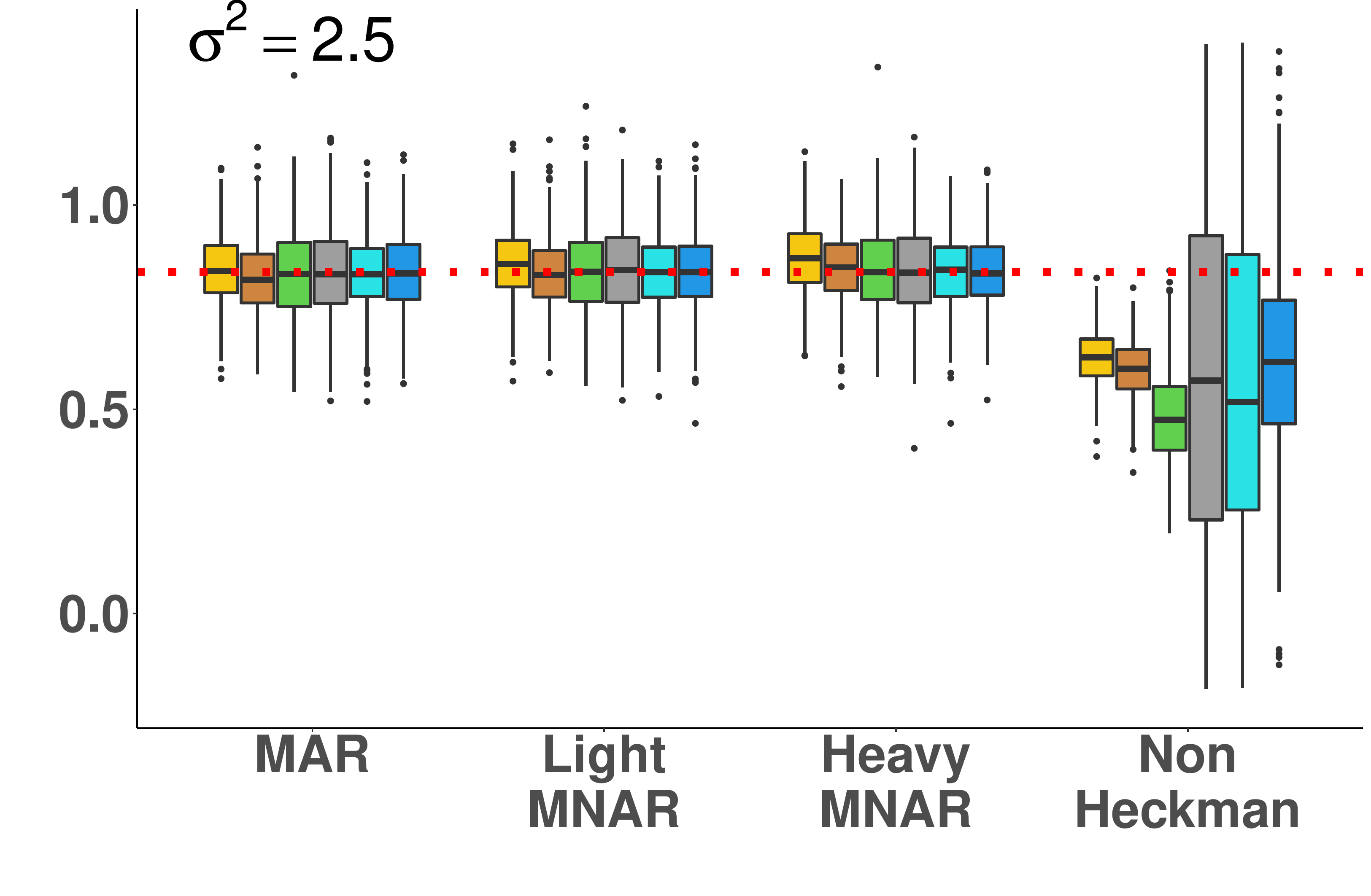}
  \end{subfigure}
  
\end{figure}

\begin{figure}[H]

\begin{subfigure}{0.5\textwidth}
     \includegraphics[width=\textwidth]{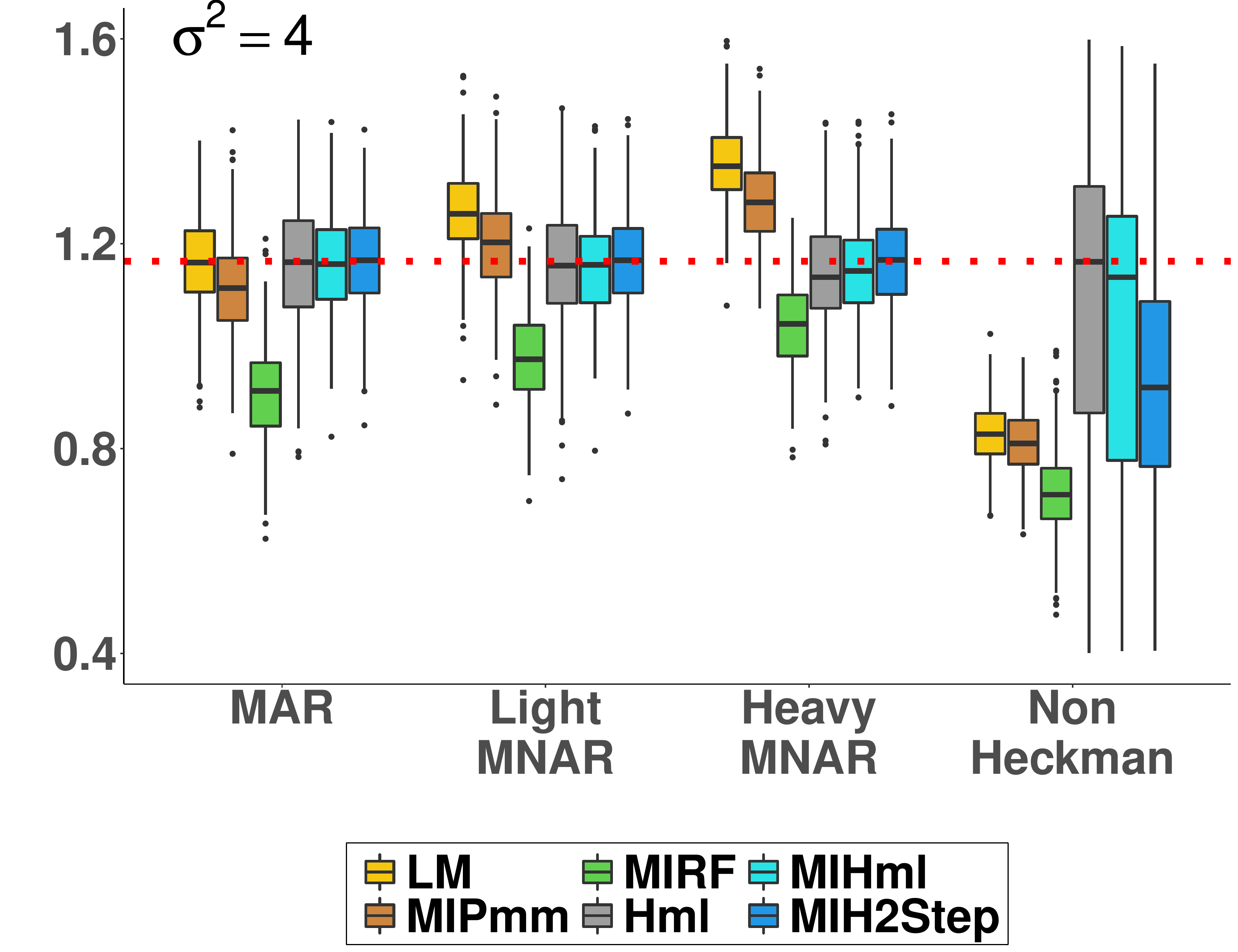}
  \end{subfigure}%
  \begin{subfigure}{0.5\textwidth}
    \includegraphics[width=\textwidth]{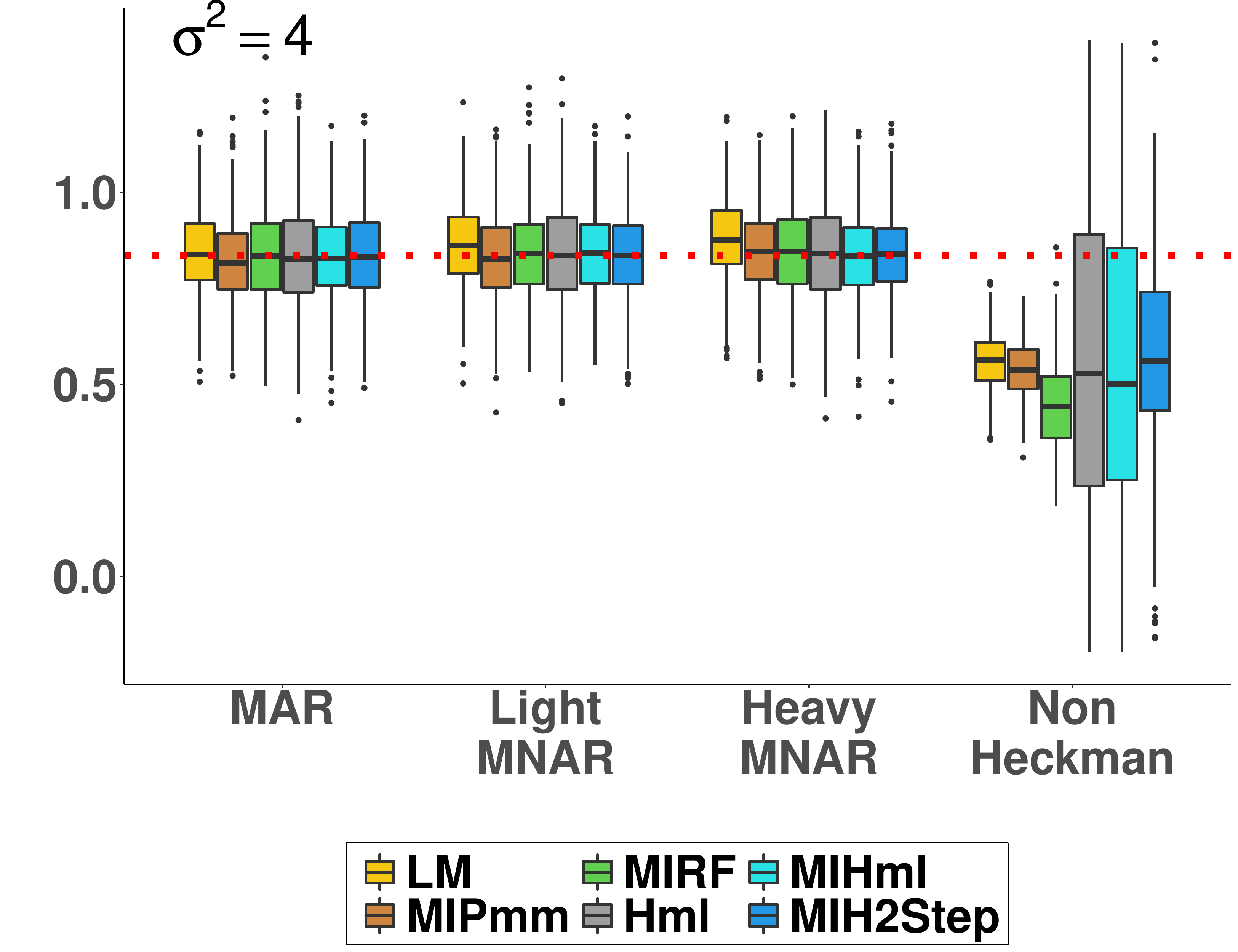}
  \end{subfigure}
  
\end{figure}

  \clearpage

\subsection{Simulation study results: prediction accuracy}

\begin{table}[ht]
\centering
\caption{$L$: Simulation results for the predicted missing emissions values.}
\label{tab:accuracyU}
\resizebox{\textwidth}{!}{\begin{tabular}{|ll|rrrr|rrrr|rrrr|rrrr|}
  \hline
  && \multicolumn{4}{c|}{MAR} & \multicolumn{4}{c|}{Light MNAR} & \multicolumn{4}{c|}{Heavy MNAR} & \multicolumn{4}{c|}{NonHeckman}  \\

$\sigma^{2}_{\varepsilon}$  & Method & \small{RE} & \small{RMSE} & \small{CR} & \small{PI} & \small{RE} & \small{RMSE} & \small{CR} & \small{PI}& \small{RE} & \small{RMSE} & \small{CR} & \small{PI} & \small{RE} & \small{RMSE} & \small{CR} & \small{PI}  \\ 
  \hline
\multirow{7}{*}{1} & 
Median & 27.21 & 2.18 & - & - & 23.44 & 2.12 & - & - & 21.27 & 2.11 & - & - & 16.24 & 2.28 & - & - \\ 
&  LM & 12.10 & 1.07 & 91.30 & 3.65 & 11.38 & 1.15 & 88.48 & 3.62 & 12.68 & 1.35 & 80.12 & 3.52 & 8.87 & 1.16 & 87.15 & 3.54 \\ 
 & MIPmm & 12.23 & 1.07 & 94.23 & 4.11 & 11.37 & 1.15 & 92.08 & 4.04 & 12.54 & 1.34 & 84.63 & 3.84 & 8.88 & 1.16 & 90.45 & 3.87 \\ 
 & MIRF & 13.14 & 1.11 & 97.61 & 5.17 & 11.60 & 1.15 & 97.42 & 5.17 & 12.09 & 1.31 & 95.39 & 5.07 & 9.35 & 1.24 & 94.55 & 4.67 \\ 
 & Hml & 12.09 & 1.07 & 93.88 & 4.03 & 11.11 & 1.10 & 93.28 & 4.02 & 10.82 & 1.15 & 92.15 & 4.04 & 8.67 & 1.13 & 92.66 & 4.05 \\ 
 & MIHml & 12.05 & 1.07 & 95.09 & 4.29 & 11.12 & 1.10 & 94.36 & 4.25 & 10.89 & 1.16 & 93.16 & 4.23 & 8.74 & 1.14 & 94.42 & 4.36 \\ 
 & MIH2Step & 12.14 & 1.07 & 95.08 & 4.28 & 11.11 & 1.09 & 94.64 & 4.28 & 10.77 & 1.14 & 93.54 & 4.26 & 8.72 & 1.14 & 97.21 & 5.33 \\ 
  \hline
\multirow{7}{*}{2.5} &   Median.1 & 45.01 & 2.53 & - & - & 36.75 & 2.46 & - & - & 29.31 & 2.52 & - & - & 20.74 & 2.92 & - & - \\ 
 & LM & 23.43 & 1.63 & 83.93 & 4.56 & 20.56 & 1.75 & 80.25 & 4.53 & 19.40 & 2.08 & 68.29 & 4.40 & 14.88 & 2.04 & 68.21 & 4.31 \\ 
 & MIPmm & 23.75 & 1.63 & 94.91 & 6.40 & 20.71 & 1.74 & 92.85 & 6.30 & 19.37 & 2.06 & 84.96 & 5.96 & 14.91 & 2.04 & 84.31 & 5.81 \\ 
&  MIRF & 25.50 & 1.65 & 96.80 & 7.14 & 21.63 & 1.74 & 95.96 & 7.12 & 19.24 & 2.03 & 91.61 & 6.87 & 15.35 & 2.11 & 87.08 & 6.34 \\ 
&  Hml & 23.54 & 1.63 & 86.69 & 4.89 & 21.02 & 1.67 & 85.81 & 4.89 & 17.68 & 1.78 & 82.80 & 4.88 & 13.90 & 1.91 & 78.55 & 4.85 \\ 
&  MIHml & 23.36 & 1.63 & 94.42 & 6.27 & 20.99 & 1.67 & 93.68 & 6.24 & 17.74 & 1.79 & 91.66 & 6.18 & 14.20 & 1.95 & 88.75 & 6.18 \\ 
&  MIH2Step & 23.48 & 1.63 & 95.80 & 6.69 & 21.12 & 1.66 & 95.42 & 6.69 & 17.59 & 1.74 & 94.35 & 6.66 & 13.48 & 1.86 & 97.15 & 8.88 \\ 
  \hline
 \multirow{7}{*}{4} &  Median.2 & 57.77 & 2.83 & - & - & 58.71 & 2.78 & - & - & 40.54 & 2.89 & - & - & 24.46 & 3.49 & - & - \\ 
&  LM & 35.41 & 2.04 & 79.19 & 5.13 & 31.28 & 2.19 & 75.10 & 5.08 & 24.74 & 2.61 & 62.08 & 4.94 & 19.51 & 2.74 & 55.15 & 4.73 \\ 
&  MIPmm & 35.81 & 2.04 & 95.14 & 8.07 & 31.66 & 2.18 & 93.06 & 7.94 & 24.86 & 2.60 & 85.03 & 7.50 & 19.55 & 2.74 & 79.37 & 7.07 \\ 
 & MIRF & 38.05 & 2.06 & 96.42 & 8.66 & 34.09 & 2.17 & 95.29 & 8.61 & 25.35 & 2.56 & 89.98 & 8.27 & 19.93 & 2.80 & 81.57 & 7.50 \\ 
&  Hml & 35.53 & 2.04 & 81.69 & 5.43 & 33.44 & 2.09 & 80.45 & 5.42 & 24.45 & 2.24 & 76.52 & 5.40 & 17.56 & 2.50 & 68.46 & 5.31 \\ 
&  MIHml & 35.35 & 2.04 & 94.02 & 7.72 & 33.31 & 2.09 & 93.14 & 7.66 & 24.48 & 2.25 & 90.69 & 7.56 & 17.89 & 2.54 & 85.12 & 7.37 \\ 
&  MIH2Step & 35.44 & 2.04 & 96.01 & 8.44 & 33.84 & 2.07 & 95.62 & 8.44 & 24.52 & 2.19 & 94.36 & 8.36 & 16.90 & 2.42 & 97.29 & 11.78 \\ 
   \hline
\end{tabular}}
\end{table}

\clearpage

\begin{table}[ht]
\centering
\caption{$M$:Simulation results for the predicted missing emissions values.}
\label{tab:accuracyM}
\resizebox{\textwidth}{!}{\begin{tabular}{|ll|rrrr|rrrr|rrrr|rrrr|}
  \hline
  && \multicolumn{4}{c|}{MAR} & \multicolumn{4}{c|}{Light MNAR} & \multicolumn{4}{c|}{Heavy MNAR} & \multicolumn{4}{c|}{NonHeckman}  \\

$\sigma^{2}_{\varepsilon}$  & Method & \small{RE} & \small{RMSE} & \small{CR} & \small{PI} & \small{RE} & \small{RMSE} & \small{CR} & \small{PI}& \small{RE} & \small{RMSE} & \small{CR} & \small{PI} & \small{RE} & \small{RMSE} & \small{CR} & \small{PI}  \\ 
  \hline
\multirow{7}{*}{1} & Median & 48.89 & 2.03 & - & - & 29.49 & 1.98 & - & - & 26.60 & 1.99 & - & - & 17.28 & 2.23 & - & - \\ 
  & LM & 15.39 & 1.00 & 90.94 & 3.38 & 12.32 & 1.08 & 87.80 & 3.34 & 13.75 & 1.30 & 77.66 & 3.24 & 9.71 & 1.15 & 84.49 & 3.29 \\ 
  & MIPmm & 16.44 & 1.00 & 95.63 & 4.05 & 12.36 & 1.07 & 93.56 & 3.97 & 13.43 & 1.28 & 85.44 & 3.72 & 9.72 & 1.15 & 89.91 & 3.78 \\ 
  & MIRF & 22.25 & 1.07 & 98.71 & 5.36 & 13.68 & 1.06 & 98.81 & 5.35 & 12.95 & 1.19 & 97.54 & 5.21 & 10.28 & 1.24 & 96.15 & 5.01 \\ 
  & Hml & 15.62 & 1.00 & 95.31 & 3.97 & 12.15 & 1.00 & 95.42 & 3.99 & 11.19 & 1.00 & 95.65 & 4.05 & 9.14 & 1.08 & 94.08 & 4.07 \\ 
   & MIHml & 15.31 & 1.00 & 96.06 & 4.12 & 12.10 & 1.00 & 95.95 & 4.14 & 11.22 & 1.01 & 96.16 & 4.22 & 9.18 & 1.08 & 95.48 & 4.36 \\ 
  & MIH2Step & 15.49 & 1.00 & 96.12 & 4.14 & 12.11 & 1.01 & 95.82 & 4.12 & 11.27 & 1.02 & 95.15 & 4.07 & 9.68 & 1.15 & 93.91 & 4.35 \\ 
  \hline
 \multirow{7}{*}{2.5} &  Median & 50.28 & 2.39 & - & - & 53.12 & 2.35 & - & - & 35.25 & 2.43 & - & - & 22.66 & 2.88 & - &  -\\ 
 & LM & 26.29 & 1.58 & 82.12 & 4.25 & 23.34 & 1.71 & 77.90 & 4.21 & 21.85 & 2.05 & 65.03 & 4.07 & 16.47 & 2.03 & 64.71 & 4.02 \\ 
  & MIPmm & 26.77 & 1.58 & 95.49 & 6.35 & 23.77 & 1.70 & 93.26 & 6.22 & 21.41 & 2.03 & 84.72 & 5.81 & 16.49 & 2.03 & 83.10 & 5.65 \\ 
  & MIRF & 30.94 & 1.63 & 97.36 & 7.25 & 26.51 & 1.66 & 96.99 & 7.19 & 21.28 & 1.90 & 93.52 & 6.91 & 16.99 & 2.11 & 87.73 & 6.44 \\ 
  & Hml & 26.50 & 1.58 & 86.82 & 4.77 & 24.65 & 1.59 & 86.80 & 4.78 & 19.39 & 1.61 & 86.50 & 4.83 & 15.04 & 1.87 & 79.73 & 4.83 \\ 
  & MIHml & 26.20 & 1.58 & 94.02 & 5.96 & 24.48 & 1.59 & 93.80 & 5.96 & 19.38 & 1.63 & 93.39 & 6.01 & 14.89 & 1.85 & 90.16 & 6.12 \\ 
  & MIH2Step & 26.19 & 1.58 & 96.12 & 6.54 & 24.48 & 1.59 & 95.79 & 6.50 & 19.31 & 1.63 & 94.92 & 6.40 & 15.39 & 1.91 & 91.69 & 6.63 \\ 
  \hline
  \multirow{7}{*}{4} & Median & 292.60 & 2.71 & - & - & 63.42 & 2.68 & - & - & 55.15 & 2.81 & - & - & 26.49 & 3.44 & - & - \\ 
 &  LM & 172.17 & 2.00 & 76.81 & 4.78 & 33.42 & 2.16 & 72.31 & 4.73 & 32.00 & 2.59 & 58.74 & 4.57 & 21.49 & 2.73 & 51.93 & 4.41 \\ 
  & MIPmm & 179.01 & 2.00 & 95.46 & 8.01 & 34.16 & 2.15 & 93.13 & 7.84 & 31.04 & 2.57 & 84.48 & 7.33 & 21.51 & 2.74 & 77.50 & 6.83 \\ 
  & MIRF & 214.96 & 2.04 & 96.77 & 8.73 & 38.31 & 2.10 & 96.03 & 8.64 & 31.24 & 2.42 & 91.55 & 8.25 & 21.87 & 2.80 & 81.23 & 7.45 \\ 
  & Hml & 173.29 & 2.00 & 81.15 & 5.26 & 36.93 & 2.01 & 80.96 & 5.27 & 30.52 & 2.06 & 80.12 & 5.31 & 18.55 & 2.40 & 70.97 & 5.27 \\ 
  & MIHml & 170.73 & 2.00 & 93.01 & 7.26 & 36.61 & 2.02 & 92.64 & 7.25 & 30.39 & 2.07 & 91.97 & 7.28 & 18.41 & 2.38 & 87.16 & 7.27 \\ 
  & MIH2Step & 170.44 & 2.00 & 96.12 & 8.27 & 36.59 & 2.02 & 95.73 & 8.21 & 30.25 & 2.06 & 94.94 & 8.10 & 19.69 & 2.53 & 89.18 & 8.18 \\ 
   \hline
\end{tabular}}
\end{table}

\clearpage

\begin{table}[ht]
\centering
\caption{$S$: Simulation results for the predicted missing emissions values.}
\label{tab:accuracyS}
\resizebox{\textwidth}{!}{\begin{tabular}{|ll|rrrr|rrrr|rrrr|rrrr|}
  \hline
  && \multicolumn{4}{c|}{MAR} & \multicolumn{4}{c|}{Light MNAR} & \multicolumn{4}{c|}{Heavy MNAR} & \multicolumn{4}{c|}{NonHeckman}  \\

$\sigma^{2}_{\varepsilon}$  & Method & \small{RE} & \small{RMSE} & \small{CR} & \small{PI} & \small{RE} & \small{RMSE} & \small{CR} & \small{PI}& \small{RE} & \small{RMSE} & \small{CR} & \small{PI} & \small{RE} & \small{RMSE} & \small{CR} & \small{PI}  \\ 
  \hline
\multirow{7}{*}{1} & Median & 22.03 & 2.19 & - & - & 21.34 & 2.11 & - & - & 18.90 & 2.07 & - & - & 18.74 & 2.71 & - & - \\ 
 & LM & 8.64 & 1.00 & 82.64 & 2.72 & 8.99 & 1.07 & 79.39 & 2.70 & 10.20 & 1.24 & 70.12 & 2.64 & 8.95 & 1.22 & 70.58 & 2.62 \\ 
 & MIPmm & 8.79 & 1.00 & 95.54 & 4.07 & 8.73 & 1.06 & 94.12 & 4.02 & 9.81 & 1.22 & 88.73 & 3.88 & 9.00 & 1.23 & 87.13 & 3.77 \\ 
 & MIRF & 12.67 & 1.28 & 98.92 & 6.76 & 11.47 & 1.23 & 99.33 & 6.80 & 10.68 & 1.27 & 99.29 & 6.79 & 10.78 & 1.53 & 97.89 & 6.87 \\ 
 & Hml & 8.63 & 1.00 & 95.11 & 3.94 & 8.45 & 0.99 & 95.34 & 3.96 & 8.12 & 0.97 & 96.05 & 4.02 & 8.20 & 1.12 & 93.33 & 4.10 \\ 
 & MIHml & 8.64 & 1.00 & 95.89 & 4.12 & 8.46 & 0.99 & 96.23 & 4.16 & 8.12 & 0.97 & 96.99 & 4.25 & 8.08 & 1.10 & 96.75 & 4.85 \\ 
 & MIH2Step & 8.63 & 1.00 & 96.23 & 4.20 & 8.45 & 0.99 & 96.35 & 4.20 & 8.12 & 0.97 & 96.66 & 4.19 & 8.34 & 1.14 & 96.92 & 5.16 \\ 
 \hline
\multirow{7}{*}{2.5} & Median & 28.10 & 2.55 & - & - & 26.62 & 2.45 & - & - & 25.36 & 2.45 & - & -  & 23.30 & 3.32 & - & - \\ 
  & LM & 15.53 & 1.58 & 72.03 & 3.42 & 15.08 & 1.69 & 68.49 & 3.40 & 16.63 & 1.96 & 58.66 & 3.33 & 15.68 & 2.19 & 49.42 & 3.24 \\ 
 & MIPmm & 15.83 & 1.58 & 95.38 & 6.37 & 14.96 & 1.67 & 93.80 & 6.29 & 16.37 & 1.93 & 88.17 & 6.05 & 15.74 & 2.20 & 77.49 & 5.50 \\ 
& MIRF & 19.53 & 1.78 & 97.96 & 8.30 & 17.33 & 1.74 & 98.31 & 8.34 & 17.02 & 1.88 & 97.36 & 8.26 & 16.76 & 2.40 & 89.36 & 7.60 \\ 
 & Hml & 15.50 & 1.58 & 85.18 & 4.57 & 14.80 & 1.57 & 85.60 & 4.59 & 14.52 & 1.55 & 86.96 & 4.67 & 13.08 & 1.84 & 79.79 & 4.75 \\ 
 & MIHml & 15.55 & 1.58 & 92.16 & 5.61 & 14.82 & 1.57 & 92.60 & 5.66 & 14.55 & 1.54 & 93.75 & 5.80 & 13.00 & 1.83 & 91.72 & 6.52 \\ 
 &  MIH2Step & 15.54 & 1.58 & 96.24 & 6.64 & 14.81 & 1.57 & 96.33 & 6.64 & 14.51 & 1.54 & 96.66 & 6.62 & 13.55 & 1.91 & 95.43 & 8.10 \\ 
\hline
\multirow{7}{*}{4}&  Median & 197.52 & 2.86 & - & - & 33.51 & 2.76 & - & - & 29.47 & 2.81 & - &  - & 52.98 & 3.66 & - & - \\ 
 & LM & 118.65 & 2.00 & 66.35 & 3.85 & 20.97 & 2.13 & 62.82 & 3.82 & 21.22 & 2.48 & 53.08 & 3.74 & 44.96 & 2.93 & 34.96 & 3.28 \\ 
&  MIPmm & 125.56 & 2.00 & 95.33 & 8.03 & 21.16 & 2.12 & 93.70 & 7.92 & 21.07 & 2.44 & 87.98 & 7.62 & 45.09 & 2.94 & 74.35 & 7.13 \\ 
 & MIRF & 165.72 & 2.16 & 97.31 & 9.58 & 24.23 & 2.15 & 97.44 & 9.61 & 21.86 & 2.35 & 95.78 & 9.48 & 46.06 & 3.08 & 82.57 & 8.45 \\ 
&  Hml & 117.13 & 2.00 & 78.39 & 4.95 & 21.47 & 1.99 & 78.83 & 4.97 & 19.56 & 1.96 & 80.46 & 5.07 & 44.05 & 2.36 & 83.87 & 6.79 \\ 
 & MIHml & 119.57 & 2.00 & 90.03 & 6.65 & 21.60 & 1.99 & 90.48 & 6.69 & 19.61 & 1.96 & 91.79 & 6.86 & 43.93 & 2.44 & 92.11 & 9.71 \\ 
&  MIH2Step & 118.16 & 2.00 & 96.24 & 8.40 & 21.58 & 1.99 & 96.33 & 8.39 & 19.54 & 1.95 & 96.65 & 8.38 & 44.21 & 2.65 & 98.27 & 19.55 \\ 
   \hline
\end{tabular}}
\end{table}

\clearpage

\begin{table}[ht]
\centering
\caption{$U$: Simulation results for the predicted missing emissions values.}
\resizebox{\textwidth}{!}{\begin{tabular}{|ll|rrrr|rrrr|rrrr|rrrr|}
  \hline
  && \multicolumn{4}{c|}{MAR} & \multicolumn{4}{c|}{Light MNAR} & \multicolumn{4}{c|}{Heavy MNAR} & \multicolumn{4}{c|}{NonHeckman}  \\

$\sigma^{2}_{\varepsilon}$  & Method & \small{RE} & \small{RMSE} & \small{CR} & \small{PI} & \small{RE} & \small{RMSE} & \small{CR} & \small{PI}& \small{RE} & \small{RMSE} & \small{CR} & \small{PI} & \small{RE} & \small{RMSE} & \small{CR} & \small{PI}  \\ 
  \hline
\multirow{7}{*}{1} & Median & 270.74 & 2.14 & - & - & 233.78 & 2.15 & - & - & 224.45 & 2.22 & - & - & 228.64 & 2.73 &  &  \\ 
 & LM & 176.00 & 1.12 & 72.06 & 2.23 & 132.11 & 1.19 & 68.28 & 2.21 & 138.28 & 1.37 & 57.92 & 2.17 & 136.97 & 1.40 & 52.45 & 2.01 \\ 
 & MIPmm & 175.61 & 1.12 & 94.62 & 4.28 & 130.26 & 1.19 & 92.85 & 4.23 & 133.30 & 1.37 & 87.04 & 4.08 & 133.12 & 1.43 & 83.95 & 3.99 \\ 
 & MIRF & 293.04 & 1.81 & 98.65 & 9.37 & 200.47 & 1.86 & 98.66 & 9.35 & 161.41 & 1.98 & 98.21 & 9.25 & 176.12 & 2.09 & 97.78 & 9.19 \\ 
 & Hml & 175.05 & 1.12 & 93.32 & 3.90 & 139.97 & 1.12 & 93.57 & 3.93 & 132.87 & 1.10 & 94.47 & 4.01 & 226.25 & 2.00 & 72.75 & 4.61 \\ 
 & MIHml & 175.89 & 1.12 & 95.11 & 4.35 & 140.59 & 1.12 & 95.26 & 4.36 & 132.48 & 1.10 & 95.89 & 4.41 & 246.97 & 2.14 & 89.76 & 6.79 \\ 
 & MIH2Step & 175.66 & 1.12 & 96.13 & 4.67 & 139.24 & 1.12 & 96.06 & 4.64 & 130.32 & 1.10 & 96.33 & 4.62 & 140.12 & 1.42 & 97.76 & 7.76 \\ 
  \hline
\multirow{7}{*}{2.5} & Median & 390.32 & 2.55 & - & - & 1429.69 & 2.59 & - & - & 260.01 & 2.73 & - & - & 267.27 & 3.40 &  - & - \\ 
& LM & 265.90 & 1.66 & 61.41 & 2.80 & 412.29 & 1.78 & 57.64 & 2.78 & 272.91 & 2.07 & 47.74 & 2.73 & 206.62 & 2.40 & 34.44 & 2.56 \\ 
&  MIPmm & 265.53 & 1.66 & 94.78 & 6.63 & 420.19 & 1.78 & 92.95 & 6.56 & 264.88 & 2.08 & 86.76 & 6.30 & 203.72 & 2.43 & 71.79 & 5.55 \\ 
 & MIRF & 336.82 & 2.18 & 98.14 & 10.47 & 387.11 & 2.27 & 97.77 & 10.42 & 258.19 & 2.51 & 96.23 & 10.29 & 143.50 & 2.78 & 91.29 & 9.39 \\ 
 & Hml & 264.01 & 1.67 & 82.91 & 4.46 & 519.65 & 1.66 & 83.37 & 4.48 & 270.51 & 1.62 & 85.13 & 4.58 & 240.29 & 2.74 & 57.98 & 5.12 \\ 
 & MIHml & 265.75 & 1.66 & 91.96 & 5.92 & 523.00 & 1.66 & 92.16 & 5.92 & 268.09 & 1.62 & 93.16 & 5.97 & 242.79 & 2.77 & 87.53 & 8.68 \\ 
 & MIH2Step & 267.58 & 1.66 & 96.65 & 7.37 & 518.49 & 1.66 & 96.69 & 7.36 & 266.16 & 1.62 & 96.93 & 7.29 & 208.90 & 2.43 & 95.60 & 11.20 \\ 
 \hline
 \multirow{7}{*}{4}& Median & 418.90 & 2.91 & - & - & 291.53 & 2.96 & - & - & 408.72 & 3.18 & - & - & 342.51 & 4.00 &  - & - \\ 
 & LM & 292.85 & 2.07 & 56.12 & 3.15 & 214.88 & 2.21 & 52.50 & 3.13 & 385.41 & 2.59 & 43.04 & 3.07 & 273.31 & 3.19 & 26.75 & 2.91 \\ 
 & MIPmm & 290.18 & 2.07 & 94.92 & 8.33 & 210.82 & 2.21 & 93.11 & 8.24 & 371.34 & 2.59 & 86.82 & 7.93 & 270.66 & 3.22 & 63.58 & 6.51 \\ 
  & MIRF & 331.47 & 2.50 & 97.69 & 11.49 & 221.93 & 2.63 & 96.96 & 11.39 & 269.21 & 2.94 & 94.57 & 11.20 & 191.09 & 3.45 & 82.94 & 9.57 \\ 
  & Hml & 290.86 & 2.07 & 75.88 & 4.79 & 237.70 & 2.06 & 76.48 & 4.83 & 397.08 & 2.01 & 78.40 & 4.93 & 280.76 & 3.26 & 51.19 & 5.37 \\ 
  & MIHml & 291.03 & 2.07 & 90.02 & 6.99 & 238.75 & 2.06 & 90.29 & 6.99 & 396.74 & 2.02 & 91.34 & 7.02 & 279.91 & 3.26 & 82.10 & 9.18 \\ 
  & MIH2Step & 292.18 & 2.07 & 96.91 & 9.32 & 237.37 & 2.06 & 96.95 & 9.29 & 392.57 & 2.02 & 97.17 & 9.20 & 256.52 & 3.05 & 92.75 & 12.07 \\ 
   \hline
   
\end{tabular}}
\end{table}

\clearpage

\subsection{Empirical Study Results}
\begin{table}[h!]
\centering
\caption{Empirical study results: Accuracy of the 2018 estimates with respect to the 2019 reported values in Total Sample.}
\resizebox{0.8 \textwidth}{!}{\begin{tabular}{lcccl|rrrrrrrr}

  \hline
  
  & Obs.\% & $n_{0}$ & $n_{2019}$ & Methods & CR & PI & $RE_{min}$ & $RE_{mean}$ & $RE_{max}$ & RMSE & Pearson & Spearman \\ 
  & & & & & & & & & & & Corr. & Corr. \\
  \hline

   &  &  &  &   Median & - & - & 5.96 & 17.86 & 34.64 & 2.89 & 0.58 & 0.62 \\ 
 \textbf{ L}     &  &  &  &  LM & 80.00 & 4.54 & 7.25 & 13.76 & 23.01 & 1.90 & 0.82 & 0.82 \\ 
      &  &  &  &   MIPmm & 91.76 & 6.34 & 7.36 & 13.68 & 22.30 & 1.90 & 0.82 & 0.81 \\ 
   \textbf{Scope 1}   & 71 & 393 &  85 & MIRF & 94.12 & 7.18 & 8.57 & 13.23 & 22.81 & 1.97 & 0.81 & 0.81 \\ 
     &  &  &  &   Hml & 82.35 & 4.83 & 5.62 & 14.57 & 23.12 & 1.90 & 0.82 & 0.82 \\ 
   &  &  &  &   MIHml & 90.59 & 6.01 & 8.47 & 14.05 & 21.70 & 1.90 & 0.81 & 0.82 \\ 
     &  &  &  &  MIH2Step & 92.94 & 6.48 & 5.49 & 14.18 & 24.02 & 1.91 & 0.82 & 0.81 \\ 
 
   \hline
   
   &  &  &  &    Median & - & - & 4.18 & 9.28 & 24.79 & 2.60 & 0.32 & 0.56 \\ 
   &  &  &  &   LM & 81.82 & 3.96 & 3.91 & 7.28 & 13.28 & 1.56 & 0.80 & 0.75 \\ 
     &  &  &  &   MIPmm & 85.23 & 4.97 & 3.98 & 7.45 & 13.80 & 1.61 & 0.80 & 0.74 \\ 
  \textbf{Scope 2} & 71 & 395 &  88 & MIRF & 86.36 & 5.3 & 4.09 & 7.87 & 14.18 & 1.64 & 0.79 & 0.73 \\ 
    &  &  &  &   Hml & 81.82 & 4.39 & 2.91 & 6.59 & 14.47 & 1.66 & 0.79 & 0.76 \\ 
     &  &  &  &   MIHml & 85.23 & 4.85 & 3.10 & 7.12 & 14.32 & 1.60 & 0.80 & 0.75 \\ 
    &  &  &  &   MIH2Step & 85.23 & 5.05 & 3.41 & 7.15 & 14.83 & 1.67 & 0.80 & 0.75 \\ 

   \hline
   
   &  &  &  &  Median & - & - & 7.91 & 19.53 & 32.93 & 2.57 & 0.69 & 0.71 \\ 
   \textbf{ M}     &  &  &  &  LM & 62.99 & 4.34 & 6.02 & 15.79 & 28.55 & 2.08 & 0.78 & 0.78 \\ 
   &  &  &  &  MIPmm & 88.19 & 6.59 & 6.46 & 15.65 & 29.46 & 2.08 & 0.78 & 0.78 \\ 
 \textbf{Scope 1}  & 55 & 767 & 127 & MIRF & 96.06 & 7.57 & 6.04 & 15.56 & 29.85 & 2.09 & 0.78 & 0.79 \\ 
     &  &  &  &  Hml & 67.72 & 4.83 & 6.22 & 16.42 & 31.18 & 2.12 & 0.78 & 0.78 \\ 
   &  &  &  &  MIHml & 84.25 & 6.07 & 7.25 & 15.36 & 30.87 & 2.12 & 0.78 & 0.78 \\ 
    &  &  &  &   MIH2Step & 90.55 & 6.84 & 7.05 & 15.40 & 31.10 & 2.15 & 0.78 & 0.78 \\ 
 
   \hline
   
   &  &  &  &    Median & - & - & 4.88 & 9.36 & 22.98 & 2.18 & 0.38 & 0.50 \\ 
   &  &  &  &    LM & 77.17 & 3.95 & 3.64 & 9.14 & 17.89 & 1.58 & 0.63 & 0.62 \\ 
     &  &  &  &   MIPmm & 92.13 & 5.63 & 3.73 & 9.52 & 18.63 & 1.58 & 0.63 & 0.62 \\ 
 \textbf{Scope 2} & 54 & 771 & 127 & MIRF & 92.91 & 6.06 & 3.61 & 9.53 & 19.03 & 1.61 & 0.63 & 0.64 \\ 
    &  &  &  &   Hml & 81.1 & 4.51 & 3.83 & 9.00 & 18.88 & 1.59 & 0.63 & 0.62 \\ 
   &  &  &  &    MIHml & 91.34 & 5.32 & 4.20 & 9.95 & 18.11 & 1.58 & 0.63 & 0.62 \\ 
   &  &  &  &    MIH2Step & 92.91 & 5.78 & 4.32 & 9.90 & 18.01 & 1.56 & 0.63 & 0.62 \\ 
   \hline
      &  &  &  &   Median & - & - & 5.99 & 15.38 & 31.67 & 2.51 & 0.58 & 0.65 \\ 
  \textbf{S}     &  &  &   & LM & 64.56 & 3.57 & 6.46 & 14.18 & 24.10 & 1.89 & 0.75 & 0.75 \\ 
   &  &  &  &   MIPmm & 92.41 & 6.78 & 6.66 & 14.44 & 24.62 & 1.89 & 0.75 & 0.75 \\ 
   \textbf{Scope 1}  & 24 & 1577 & 158 & MIRF & 96.84 & 8.5 & 6.11 & 15.11 & 24.93 & 2.00 & 0.72 & 0.72 \\ 
     &  &  &  &  Hml & 74.68 & 4.63 & 7.31 & 14.34 & 24.98 & 1.96 & 0.74 & 0.73 \\ 
    &  &  &  &  MIHml & 86.08 & 5.86 & 5.89 & 14.41 & 27.12 & 1.90 & 0.75 & 0.74 \\ 
     &  &  &  &  MIH2Step & 93.04 & 7.12 & 6.45 & 13.68 & 25.35 & 1.92 & 0.75 & 0.74 \\ 
 
   \hline
   
     &  &  &  &   Median & - & - & 7.63 & 13.34 & 22.54 & 2.21 & 0.32 & 0.36 \\ 
    &  &  &  &   LM & 68.18 & 3.32 & 4.76 & 9.63 & 18.46 & 1.65 & 0.60 & 0.57 \\ 
   &  &  &  &   MIPmm & 92.61 & 6.15 & 4.60 & 9.79 & 18.80 & 1.66 & 0.59 & 0.56 \\ 
 \textbf{Scope 2} & 24 & 1581 & 176 & MIRF & 96.02 & 6.8 & 5.59 & 11.82 & 18.45 & 1.63 & 0.59 & 0.54 \\ 
   &  &  &  &    Hml & 78.41 & 4.41 & 5.77 & 10.44 & 20.13 & 1.69 & 0.58 & 0.54 \\ 
     &  &  &  &   MIHml & 89.2 & 5.4 & 5.64 & 10.04 & 18.39 & 1.66 & 0.59 & 0.55 \\ 
   &  &  &  &   MIH2Step & 93.75 & 6.17 & 5.71 & 10.67 & 19.30 & 1.67 & 0.59 & 0.55 \\ 

   \hline

    &  &  &  &  Median & - & - & 12.89 & 21.26 & 30.38 & 3.44 & 0.55 & 0.66 \\ 
   \textbf{U}     &  &  &   & LM & 38.78 & 2.95 & 10.69 & 17.28 & 27.84 & 2.21 & 0.81 & 0.82 \\ 
    &  &  &  &  MIPmm & 89.8 & 7.2 & 11.49 & 18.55 & 32.68 & 2.24 & 0.81 & 0.80 \\ 
  \textbf{Scope 1} & 12 & 1256 &  49 & MIRF & 97.96 & 9.74 & 9.57 & 21.32 & 32.66 & 2.71 & 0.73 & 0.74 \\ 
   &  &  &  &  Hml & 59.18 & 4.57 & 9.99 & 17.27 & 29.32 & 2.29 & 0.78 & 0.80 \\ 
     &  &  &  &  MIHml & 77.55 & 6.23 & 9.71 & 17.73 & 27.02 & 2.19 & 0.80 & 0.81 \\ 
    &  &  &  &  MIH2Step & 89.8 & 7.9 & 10.82 & 15.73 & 31.92 & 2.19 & 0.82 & 0.84 \\ 
 
   \hline

   &  &  &  &   Median & - & - & 6.82 & 22.15 & 29.07 & 3.58 & 0.19 & 0.36 \\ 
   &  &  &  &    LM & 52.94 & 2.67 & 7.40 & 12.31 & 17.85 & 1.95 & 0.72 & 0.66 \\ 
    &  &  &  &   MIPmm & 88.24 & 6.17 & 5.06 & 12.97 & 22.73 & 2.05 & 0.68 & 0.64 \\ 
\textbf{Scope 2}  & 12 & 1258 &  51 & MIRF & 98.04 & 7.23 & 7.02 & 13.15 & 19.39 & 1.80 & 0.78 & 0.76 \\ 
   &  &  &  &    Hml & 72.55 & 4.41 & 6.36 & 13.54 & 22.61 & 1.96 & 0.78 & 0.75 \\ 
     &  &  &  &   MIHml & 80.39 & 5.93 & 5.92 & 16.17 & 22.23 & 2.13 & 0.72 & 0.67 \\ 
   &  &  &  &    MIH2Step & 86.27 & 7.23 & 5.95 & 15.81 & 25.83 & 2.23 & 0.70 & 0.64 \\ 

   \hline
\end{tabular}}
\newline
{\footnotesize RE =  relative error (\%), PI = Average length of the  95\% prediction interval , \\  RMSE = Root mean square error, CR = Coverage Rate (\%)}
\label{tbs:empiricalsub}
\end{table}

\clearpage
 
\bibliographystyle{elsarticle-harv} 
\bibliography{elsarticle-template-harv}





\end{document}